\documentclass[showpacs,pra,onecolumn]{revtex4-1}
\usepackage{amsmath}
\usepackage{graphicx}
\usepackage{subfigure}
\usepackage{epstopdf}

\begin{document}

\title{Domain walls and vortices in linearly coupled systems}
\author{Nir Dror, Boris A. Malomed, and Jianhua Zeng}

\begin{abstract}
We investigate one-dimensional (1D) and 2D radial domain-wall (DW)
states in the system of two
nonlinear-Schr\"{o}dinger/Gross-Pitaevskii (NLS/GP) equations, which
are coupled by the linear mixing and by the nonlinear XPM
(cross-phase-modulation). The system has straightforward
applications to two-component Bose-Einstein condensates, and to the
bimodal light propagation in nonlinear optics. In the former case,
the two components represent different hyperfine atomic states,
while in the latter setting they correspond to orthogonal
polarizations of light. Conditions guaranteeing the stability of
flat continuous wave (CW) asymmetric bimodal states are established,
followed by the study of families of the corresponding DW patterns.
Approximate analytical solutions for the DWs are found near the
point of the symmetry-breaking bifurcation of the CW states. An
\emph{exact} DW solution is produced for ratio $3:1$ of the XPM and
SPM coefficients. The DWs between flat asymmetric states, which are
mirror images to each other, are completely stable, and all other
species of the DWs, with zero crossings in one or two components,
are fully unstable. Interactions between two DWs are considered too,
and an effective potential accounting for the attraction between
them is derived analytically. Direct simulations demonstrate merger
and annihilation of the interacting DWs. The analysis is extended
for the system including single- and double-peak external
potentials. Generic solutions for trapped DWs are obtained in a
numerical form, and their stability is investigated. An exact stable
solution is found for the DW trapped by a single-peak potential. In
the 2D geometry, stable two-component vortices are found, with
topological charges $s=1,2,3$. Radial oscillations of annular
DW-shaped pulsons, with $s=0,1,2$, are studied too. A linear
relation between the period of the oscillations and the mean radius
of the DW ring is derived analytically.
\end{abstract}

\maketitle

\affiliation{Department of Physical Electronics, School of
Electrical Engineering, Faculty of Engineering, Tel Aviv University,
Tel Aviv 69978, Israel}

\section{Introduction}

\label{sec:Introduction}

A ubiquitous type of topologically protected patterns in binary
(two-component) nonlinear systems is represented by domain walls (DWs),
alias ``grain boundaries". A commonly known origin of DWs is in the theory
of media with a vectorial local order parameter, such as magnetics \cite%
{magnetic}, ferroelectrics \cite{electric}, and liquid crystals \cite{liquid}%
. In systems described by binary wave functions (at the fundamental or
phenomenological level), the DW represents a transient layer between
semi-infinite domains carrying different components, or distinct
combinations of the components.

A simple but physically significant example is a rectilinear border between
two regions occupied by spontaneously emerging roll structures with
different orientations. This is, for instance, a generic defect observed in
patterns on the surface of thermal-convection layers \cite%
{Pomeau,convection,Alik}. The theoretical description of such patterns is
based on systems of coupled Ginzburg-Landau equations, each equation
governing a slowly varying amplitude of a plane wave whose superposition
forms the DW \cite{we-old}. Formally similar systems of coupled
nonlinear-Schr\"{o}dinger (NLS) equations and Gross-Pitaevskii (GP)
equations describe, respectively, the co-propagation of electromagnetic
waves with orthogonal polarizations in nonlinear optical fibers \cite%
{Agrawal}, and binary mixtures of Bose-Einstein condensates (BECs)
in cigar-shaped traps \cite{Pit}. While the Ginzburg-Landau systems
are dissipative ones, on the contrary to the conservative NLS/GP
systems, their stationary versions essentially coincide, hence DW
patterns, generated by the stationary equations, are ubiquitous,
playing a fundamental role in sundry physical media, both
dissipative and conservative ones. Of course, the dynamics of
perturbed DW structures may be different in the dissipative and
conservative systems.

The basic models which give rise to DWs formed by two fields feature
nonlinear XPM\ (cross-phase-modulation) interaction between the fields. In
this context, one-dimensional (1D) solutions for DWs were reported in Ref.
\cite{we-old} for grain boundaries in thermal-convection patterns, in Refs.
\cite{Haelt,me} for temporal-domain DWs between electromagnetic waves with
orthogonal circular polarizations in bimodal optical fibers, and (in a
number of different forms) in two-component BECs \cite{Warsaw}. Various
extensions of these settings were investigated too, including grain
boundaries between domains filled by different cellular \cite{cellular} and
quasi-periodic \cite{Horacio} patterns, and between traveling-wave domains
in the model of the oscillatory thermal convection \cite{HS}. Optical DWs
between polarized waves with different wavelengths were also studied \cite%
{Wabnitz}. Moreover, the analysis was performed for DW states in the
discrete version of the nonlinearly-coupled NLS system, which describes
arrays of parallel bimodal optical waveguides \cite{Panos} or BEC\
fragmented in a deep optical-lattice potential (in the latter case, the
system takes the form of the Bose-Hubbard model) \cite{Mering}. Also
considered were the transition to immiscibility in binary BEC with
long-range dipole-dipole interactions \cite{DD}, and various forms of DWs in
the three-component spinor BEC \cite{spinor}.

As concerns the experiment, linear grain boundaries between patches filled
by rolls with different orientations had been well documented in many
observations of the thermal convection \cite{rolls,convection}. Similar
linear defects were reported in laser cavities \cite{cavityDW} and in
experimental studies of BEC\ \cite{BEC-experiment}. Well-pronounced DW
structures have also been created in bimodal optical fibers \cite{fiber-exp}
and in fiber lasers \cite{FiberLaserDW}.

In many physically relevant settings, the nonlinear interaction between
coexisting waves competes with the \emph{linear interconversion} between
them, which strongly affects the DW patterns in such binary systems. In
particular, the linear coupling between the orthogonal polarizations induced
by the twist or elliptic deformation of the optical fiber gives rise to an
effective force accelerating the corresponding DW \cite{me}. The linear
interconversion between two components of a binary BEC representing
different atomic hyperfine states, coupled by a resonant radio-frequency
wave, affects the formation of the DW in immiscible binary condensates \cite%
{BEClin-coupling}. The objective of the present work is to develop a
comprehensive study of DW states in systems of NLS/GP equations coupled by
both the linear and nonlinear (XPM) terms, in 1D and 2D geometries. Related
two-component vortical states in 2D are studied too.

The paper is organized as follows. The model is formulated in Section \ref%
{sec:model}. In Section \ref{sec:CW}, we consider flat continuous-wave (CW)
states, which represent backgrounds supporting DWs. Both symmetric and
asymmetric CW states are found, and the symmetry-breaking bifurcation (SBB),
which gives rise to asymmetric CW backgrounds, is identified. The stability
of the flat states is also investigated in this section. Basic types of DWs
in the free space (without an external potential) are considered in Section %
\ref{sec:free-space}, where both analytical and numerical solutions for the
DWs are reported, and the stability of the DW patterns is established.
Section \ref{DWwithPotential} deals with one and two DWs interacting with an
external potential, in the form of one or two peaks. In particular, an
analytical stable solution is produced for a DW pinned to a potential peak.
Two-dimensional axisymmetric patterns are considered in Section \ref%
{sec:2DModel}. Stable vortices with topological charges $s=1,2,3$, supported
\ by the asymmetric CW background, are found, and radial oscillations of
annular pulsons in the form of circular DWs are studied too. The paper is
concluded by Section \ref{sec:Conclusion}.

\section{The model}

\label{sec:model} Our starting point is the system of scaled one-dimensional
NLS/GP equations for two wave functions $\psi _{1,2}$, coupled by the linear
and nonlinear (XPM)\ terms:%
\begin{eqnarray}
i\left( \psi _{1}\right) _{t} &=&-\left( 1/2\right) \left( \psi _{1}\right)
_{xx}+\sigma \left\vert \psi _{1}\right\vert ^{2}\psi _{1}+g\left\vert \psi
_{2}\right\vert ^{2}\psi _{1}-\kappa \psi _{2},  \notag \\
&&  \label{GPE} \\
i\left( \psi _{2}\right) _{t} &=&-\left( 1/2\right) \left( \psi _{2}\right)
_{xx}+\sigma \left\vert \psi _{2}\right\vert ^{2}\psi _{2}+g\left\vert \psi
_{1}\right\vert ^{2}\psi _{2}-\kappa \psi _{1},  \notag
\end{eqnarray}%
where $\kappa $ is the rate of the linear interconversion between the two
atomic states, if the system is interpreted in terms of BEC \cite%
{BEClin-coupling}. The same equations, with time $t$ replaced by the
propagation distance, $z$, and $x$ replaced by the reduced time, $\tau
\equiv t-z/V_{\mathrm{gr}}$, where $V_{\mathrm{gr}}$ is the group velocity
of the carrier wave, may be realized in optics as the model of the light
propagation in an ordinary or photonic-crystal fiber \cite{Agrawal}. In the
fiber-optic model, amplitudes $\psi _{1,2}$ represent two mutually
orthogonal polarizations of light, with the linear interconversion induced
by the birefringence or twist of the fiber, for the circular or linear
polarizations, respectively. Coefficients $\sigma $ and $g$ in Eqs. (\ref%
{GPE}) account for the SPM (self-phase-modulation) and XPM nonlinearities,
respectively (in the ordinary optical fiber, the XPM/SPM\ ratio is $g/\sigma
=2$ and $4/3$ for the of circular and linear polarizations, respectively).

To secure the modulational stability of CW states supporting DWs,
coefficient $g$ will be kept positive, which corresponds to the
repulsive XPM nonlinearity, while SPM coefficient $\sigma $ may have
any sign. In optics, opposite signs of the XPM and SPM coefficients
is an exotic
situation, which is, nevertheless, possible in photonic-crystal fibers \cite%
{Agrawal}. In BEC, the sign of either coefficient may be switched by means
of the Feshbach-resonance effect \cite{Pit}.

The value of $g$ may be fixed by dint of an obvious rescaling [for instance,
it is possible to set $g\equiv 3$, which is a natural choice in view of the
existence of an exact DW solution in the form of Eqs. (\ref{ansatz}) and (%
\ref{exact}), see below, which requires $g=3\sigma $]. In addition, the
rescaling allows one to fix $|\kappa |~\equiv 1$, thus we will assume $%
\kappa =\pm 1$. Nevertheless, in the analytical expressions written below,
we keep $g$ and $\kappa $ as free parameters, as it is easier to analyze the
results in such a form.

Stationary solutions to Eqs. (\ref{GPE}) with chemical potential $\mu $ are
sought for as%
\begin{equation}
\psi _{1,2}\left( x,t\right) =e^{-i\mu t}\phi _{1,2}(x),  \label{stationary}
\end{equation}%
with real functions $\phi _{1,2}(x)$ satisfying equations%
\begin{eqnarray}
\mu \phi _{1}+\left( 1/2\right) \phi _{1}^{\prime \prime }-\sigma \phi
_{1}^{3}-g\phi _{2}^{2}\phi _{1}+\kappa \phi _{2} &=&0,  \notag \\
&&  \label{Phi12_model} \\
\mu \phi _{2}+\left( 1/2\right) \phi _{2}^{\prime \prime }-\sigma \phi
_{2}^{3}-g\phi _{1}^{2}\phi _{2}+\kappa \phi _{1} &=&0,  \notag
\end{eqnarray}%
with the prime standing for $d/dx$. The energy (Hamiltonian) corresponding
to stationary states (\ref{stationary}) is $H=\int_{-\infty }^{+\infty }%
\mathcal{H}dx,$ with density%
\begin{gather}
\mathcal{H}=(1/2)\left[ \left( \phi _{1}^{\prime }\right) ^{2}+\left( \phi
_{2}^{\prime }\right) ^{2}\right]  \notag \\
+\left( \sigma /2\right) \left( \phi _{1}^{4}+\phi _{2}^{4}\right) +g\phi
_{1}^{2}\phi _{2}^{2}-2\kappa \phi _{1}\phi _{2}.  \label{H}
\end{gather}

\section{Flat (continuous-wave)\ states}

\label{sec:CW}

Symmetric flat ($x$-independent) solutions to Eqs. (\ref{Phi12_model}), with
equal amplitudes of both components, are%
\begin{equation}
\phi _{1}=\phi _{2}\equiv A_{0}=\sqrt{\left( \mu +\kappa \right) /\left(
\sigma +g\right) }.  \label{symm}
\end{equation}%
The Hamiltonian density (\ref{H}) for this solution is%
\begin{equation}
\mathcal{H}_{\mathrm{symm}}=\left( \mu ^{2}-\kappa ^{2}\right) /\left(
g+\sigma \right) .  \label{Hsymm}
\end{equation}%
Antisymmetric CW solutions are equivalent to Eq. (\ref{symm}) with $\kappa $
replaced by $-\kappa $. In view of this relation, we will define the
symmetric solutions as those for $\kappa =+1$, while the antisymmetric ones
will be replaced by symmetric states for $\kappa =-1$.

Asymmetric flat states, which are generated by the SBB, can also be found in
the exact form:%
\begin{eqnarray}
\phi _{1}^{2} &=&\frac{\mu }{2\sigma }\pm \sqrt{\frac{\mu ^{2}}{4\sigma ^{2}}%
-\frac{\kappa ^{2}}{\left( g-\sigma \right) ^{2}}}\equiv A_{1}^{2},  \notag
\\
&&  \label{phi12_asymm} \\
\phi _{2}^{2} &=&\frac{\mu }{2\sigma }\mp \sqrt{\frac{\mu ^{2}}{4\sigma ^{2}}%
-\frac{\kappa ^{2}}{\left( g-\sigma \right) ^{2}}}\equiv A_{2}^{2},  \notag
\end{eqnarray}%
with the signs of $\phi _{1}$ and $\phi _{2}$ determined by relation%
\begin{equation}
\phi _{1}\phi _{2}=\kappa \left( g-\sigma \right) ^{-1}\,,  \label{21}
\end{equation}%
which is compatible with Eqs. (\ref{phi12_asymm}). The Hamiltonian density (%
\ref{H}) for these states is%
\begin{equation}
\mathcal{H}_{\mathrm{asymm}}=\mu ^{2}/\left( 2\sigma \right) -\kappa
^{2}/\left( g-\sigma \right) .  \label{Hasymm}
\end{equation}

Taking into account the above convention, $\phi _{1}\phi _{2}>0$, and the
condition that $\phi _{1,2}^{2}$ must be positive, it follows from Eqs. (\ref%
{phi12_asymm}) that the asymmetric CW state emerges, i.e., the SBB takes
place, at a specific value of the chemical potential,%
\begin{equation}
\mu =2\sigma \kappa \left( g-\sigma \right) ^{-1}\equiv \mu _{\mathrm{cr}},
\label{mu-crit}
\end{equation}%
the corresponding value of $A_{0}^{2}$ at the bifurcation point being
obtained by the substitution of this value into Eq. (\ref{symm}):%
\begin{equation}
A_{\mathrm{cr}}^{2}\equiv \kappa \left( g-\sigma \right) ^{-1}.  \label{crit}
\end{equation}

The asymmetric solution (\ref{21}), (\ref{phi12_asymm}) exists at $\mu
^{2}>\mu _{\mathrm{cr}}^{2}$. Slightly above the bifurcation point, i.e., at
\begin{equation}
\mu =\mu _{\mathrm{cr}}+\delta \mu ,~\mathrm{with}~\ \left\vert \delta \mu
\right\vert \ll \left\vert \mu _{\mathrm{cr}}\right\vert ,  \label{over}
\end{equation}%
one can expand Eqs. (\ref{phi12_asymm}) and (\ref{21}) in powers of small $%
\delta \mu $, which yields%
\begin{equation}
\phi _{1,2}=\sqrt{\frac{\kappa }{g-\sigma }}\pm \frac{1}{2}\sqrt{\frac{%
\delta \mu }{\sigma }}+\mathcal{O}\left( \delta \mu \right) .  \label{delta}
\end{equation}

\subsection{The modulational stability of the symmetric solution at the
bifurcation point}

A crucial condition necessary for the existence of stable DWs is the\ absence%
\emph{\ }of the modulational instability (MI) of the corresponding
CW background. Here we explicitly consider the MI of the symmetric
CW, and, in particular, we will find conditions providing for the
stability of the symmetric state exactly at the SBB point. Examining
this case secures that the asymmetric CW states are not subject to
the MI -- at least, close enough to the bifurcation point.

To analyze the MI, we look for perturbed solutions to Eqs.
(\ref{GPE}) in the well-known general form, i.e., as
\begin{equation}
\psi _{1,2}(x,t)=\left[ A_{0}+a_{1,2}(x,t)\right] \exp \left[ -i\mu t+i\chi
_{1,2}\left( x,t)\right) \right] ,  \label{pert}
\end{equation}%
where amplitude $A_{0}$ is the same as in Eq. (\ref{symm}), while
$a_{1,2}$ and $\chi _{1,2}$ are infinitesimal perturbations of the
amplitudes and phases of the two components (this form of the
solution implies the incorporation of the small complex
perturbations into the unperturbed CW state). Then, eigenmodes of
the perturbations are
sought for as%
\begin{equation}
\left\{ a_{1,2}(x,t),\chi _{1,2}(x,t)\right\} =\left\{ a_{1,2}^{(0)},\chi
_{1,2}^{(0)}\right\} \exp \left( \gamma t+ipx\right) ,  \label{mode}
\end{equation}%
where $p$ is an arbitrary real wavenumber of the perturbation, and $\gamma $
the corresponding instability gain, which may be complex. The stability
condition is that \textrm{Re}$\left\{ \gamma (p)\right\} \equiv 0$ for all
real $p$.

The substitution of expressions (\ref{pert}) and (\ref{mode}) into Eqs. (\ref%
{GPE}) and linearization with respect to infinitesimal perturbations yields
the following system of equations, which actually splits into two separate
subsystems for $\left( a_{1}^{(0)}+a_{2}^{(0)}\right) ,$ $\left( \chi
_{1}^{(0)}+\chi _{2}^{(0)}\right) $ and $\left(
a_{1}^{(0)}-a_{2}^{(0)}\right) ,$ $\left( \chi _{1}^{(0)}-\chi
_{2}^{(0)}\right) $:%
\begin{eqnarray}
\gamma \left( a_{1}^{(0)}+a_{2}^{(0)}\right) -\left( 1/2\right)
A_{0}p^{2}\left( \chi _{1}^{(0)}+\chi _{2}^{(0)}\right) &=&0,  \notag \\
&&  \label{lin+} \\
\left[ 2A^{2}\left( g+\sigma \right) +\left( 1/2\right) p^{2}\right] \left(
a_{1}^{(0)}+a_{2}^{(0)}\right) +\gamma A_{0}\left( \chi _{1}^{(0)}+\chi
_{2}^{(0)}\right) &=&0;  \notag
\end{eqnarray}%
\begin{eqnarray}
\gamma \left( a_{1}^{(0)}-a_{2}^{(0)}\right) -\left[ \left( 1/2\right)
p^{2}+2\kappa \right] A_{0}\left( \chi _{1}^{(0)}-\chi _{2}^{(0)}\right)
&=&0,  \notag \\
&&  \label{lin-} \\
\left[ 2A^{2}\left( g-\sigma \right) -2\kappa -\left( 1/2\right) p^{2}\right]
\left( a_{1}^{(0)}-a_{2}^{(0)}\right) -\gamma A_{0}\left( \chi
_{1}^{(0)}-\chi _{2}^{(0)}\right) &=&0.  \notag
\end{eqnarray}%
The resolvability conditions for Eqs. (\ref{lin+}) and (\ref{lin-}) yield,
respectively, the following expressions for $\gamma (p)$:%
\begin{eqnarray}
\gamma _{+}^{2} &=&-\left( 1/2\right) p^{2}\left[ 2\left( g+\sigma \right)
A_{0}^{2}+\left( 1/2\right) p^{2}\right] ,  \label{+} \\
\gamma _{-}^{2} &=&\left( 2\kappa +\left( 1/2\right) p^{2}\right) \left[
2\left( g-\sigma \right) A_{0}^{2}-2\kappa -\left( 1/2\right) p^{2}\right] .
\label{-}
\end{eqnarray}%
The stability condition ensuing from Eq. (\ref{+}), i.e., $\gamma _{+}^{2}<0$%
, is obvious:%
\begin{equation}
g+\sigma >0.  \label{g+s}
\end{equation}%
Expression (\ref{-}) simplifies at the bifurcation point (\ref{mu-crit}),
where $A_{0}^{2}=\kappa /\left( g-\sigma \right) $, as per Eq. (\ref{crit}):%
\begin{equation}
\gamma _{-}^{2}=-\left( 1/2\right) p^{2}\left[ 2\kappa +\left( 1/2\right)
p^{2}\right] .  \label{-simple}
\end{equation}%
Evidently, the stability condition following from Eq. (\ref{-simple}) is $%
\kappa >0$, which, as a matter of fact, means that only the symmetric flat
solution may be stable at the SBB point, while its antisymmetric
counterpart, that (as defined above) corresponds to $\kappa <0$, is
unstable. Further, from $\kappa >0$ and the positiveness of expression (\ref%
{crit}) for $A_{\mathrm{cr}}^{2}$, condition $g-\sigma >0$ follows. Combined
with Eq. (\ref{g+s}), it gives rise to the following relation between the
XPM and SPM coefficients necessary for the existence and stability of the
asymmetric CW states:%
\begin{equation}
\left\vert \sigma \right\vert <g,  \label{<}
\end{equation}%
while $\sigma $ may be positive or negative. In fact, it will be
demonstrated below that DW solutions do not exists for $\sigma <0$.

Note that condition (\ref{<}) does not hold for $g=0$, hence the DWs that
can be found in the system with the solely linear coupling are always
unstable, as demonstrated in Ref. \cite{Rich}. In that work, it was shown
that the unstable DW gives rise to an expanding layer filled with turbulent
waves.

\subsection{The modulational stability of the asymmetric background}

In the general case of the asymmetric CW background, the perturbed solution
is taken as [cf. Eq. (\ref{pert})]%
\begin{equation}
\psi _{1,2}(x,t)=A_{1,2}\left[ 1+b_{1,2}(x,t)\right] \exp \left[ -i\mu
t+i\chi _{1,2}\left( x,t)\right) \right] ,  \label{pert-general}
\end{equation}%
where $A_{1,2}$ are given by Eqs. (\ref{phi12_asymm}), and infinitesimal
perturbations are taken as [cf. Eqs. (\ref{mode})]%
\begin{equation}
\left\{ b_{1,2}(x,t),\chi _{1,2}(x,t)\right\} =\left\{ b_{1,2}^{(0)},\chi
_{1,2}^{(0)}\right\} \exp \left( \gamma t+ipx\right) .  \label{bchi}
\end{equation}%
The substitution of expressions (\ref{pert-general}) and (\ref{bchi}) into
Eqs. (\ref{GPE}), and the subsequent linearization, lead to a system of four
linear equations:%
\begin{eqnarray}
\gamma b_{1}^{(0)}-\frac{1}{2}p^{2}\chi _{1}^{(0)}-\kappa \frac{A_{2}}{A_{1}}%
\left( \chi _{1}^{(0)}-\chi _{2}^{(0)}\right) &=&0,  \notag \\
\gamma b_{2}^{(0)}-\frac{1}{2}p^{2}\chi _{2}^{(0)}+\kappa \frac{A_{1}}{A_{2}}%
\left( \chi _{1}^{(0)}-\chi _{2}^{(0)}\right) &=&0,  \notag \\
\gamma \chi _{1}^{(0)}+\frac{1}{2}p^{2}b_{1}^{(0)}+2\sigma
A_{1}^{2}b_{1}^{(0)}+2gA_{2}^{2}b_{2}^{(0)}+\kappa \frac{A_{2}}{A_{1}}\left(
b_{1}^{(0)}-b_{2}^{(0)}\right) &=&0,  \notag \\
\gamma \chi _{2}^{(0)}+\frac{1}{2}p^{2}b_{2}^{(0)}+2\sigma
A_{2}^{2}b_{2}^{(0)}+2gA_{1}^{2}b_{1}^{(0)}-\kappa \frac{A_{1}}{A_{2}}\left(
b_{1}^{(0)}-b_{2}^{(0)}\right) &=&0.  \label{4}
\end{eqnarray}%
In the special case of $A_{1}=A_{2}\equiv A_{0}$, it is easy to check that
Eqs. (\ref{4}) are tantamount to Eqs. (\ref{lin+}) and (\ref{lin-}) written
above.

The dispersion relation between $\gamma $ and $p^{2}$ is determined by the
resolvability condition of system (\ref{4}):%
\begin{equation}
\left\vert
\begin{array}{cccc}
\gamma & 0 & -\left( \kappa \frac{\phi _{2}}{\phi _{1}}+\frac{1}{2}%
p^{2}\right) & \kappa \frac{\phi _{2}}{\phi _{1}} \\
0 & \gamma & \kappa \frac{\phi _{1}}{\phi _{2}} & -\left( \kappa \frac{\phi
_{1}}{\phi _{2}}+\frac{1}{2}p^{2}\right) \\
2\sigma \phi _{1}^{2}+\kappa \frac{\phi _{2}}{\phi _{1}}+\frac{1}{2}p^{2} &
2g\phi _{2}^{2}-\kappa \frac{\phi _{2}}{\phi _{1}} & \gamma & 0 \\
2g\phi _{1}^{2}-\kappa \frac{\phi _{1}}{\phi _{2}} & 2\sigma \phi
_{2}+\kappa \frac{\phi _{1}}{\phi _{2}}+\frac{1}{2}p^{2} & 0 & \gamma%
\end{array}%
\right\vert =0.  \label{det}
\end{equation}

Equation (\ref{det}) was solved numerically, yielding four roots $\gamma
(p^{2})$. It has been verified that, with $A_{1}$ and $A_{2}$ taken as per
Eqs. (\ref{phi12_asymm}) and (\ref{21}), and for all $p^{2}\geq 0$, all the
roots satisfy condition $\mathrm{Re}\left\{ \gamma \left( p^{2}\right)
\right\} =0$, if $\kappa $ is positive and inequality (\ref{<}) holds. Thus,
as long as the symmetric CW are stable at the SBB point, its asymmetric
counterparts are stable too.

\section{Domain-wall solutions in the free space}

\label{sec:free-space}

\subsection{Numerical results}

Stationary solutions to equations (\ref{Phi12_model}) were constructed by
means of the Newton-Raphson method for the corresponding nonlinear
boundary-value problem. In particular, the boundary conditions were fixed as
zero values of the derivatives at both edges of the integration domain.

\begin{figure}[tbp]
\subfigure[]{\includegraphics[width=2.3in]{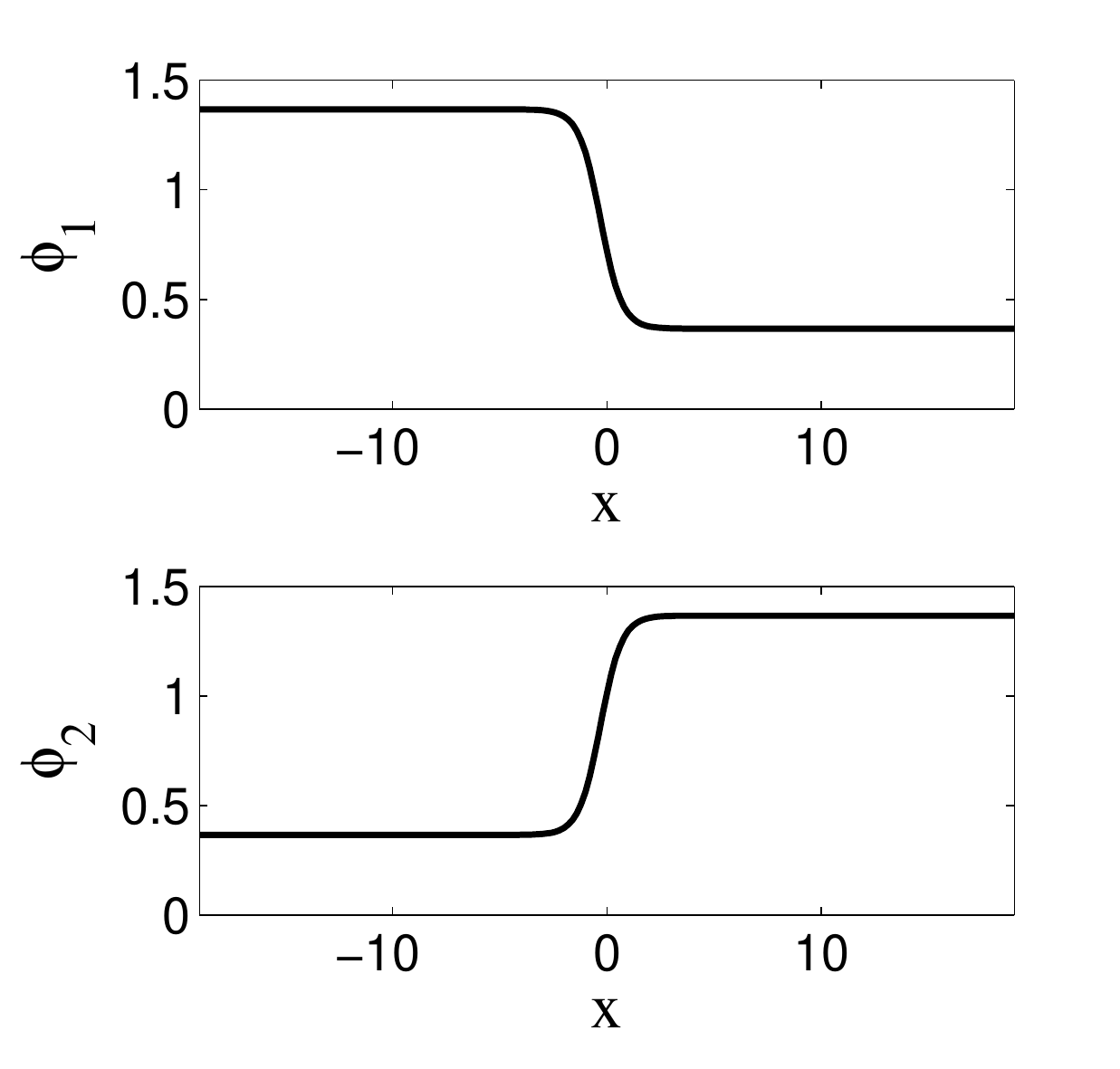}%
\label{Phi12_profiles-a}} 
\subfigure[]{\includegraphics[width=2.3in]{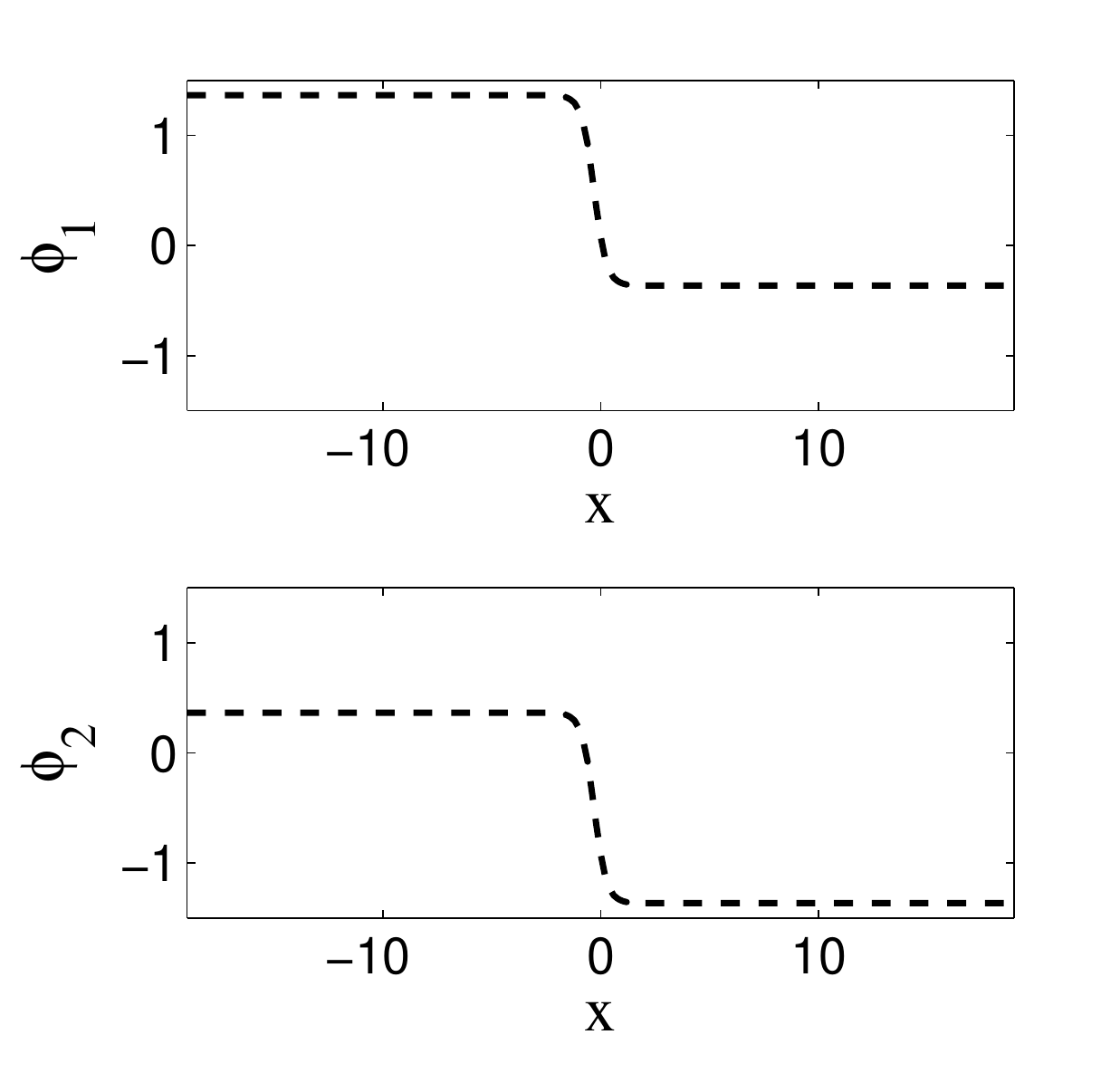}%
\label{Phi12_profiles-b}} \\
\subfigure[]{\includegraphics[width=2.3in]{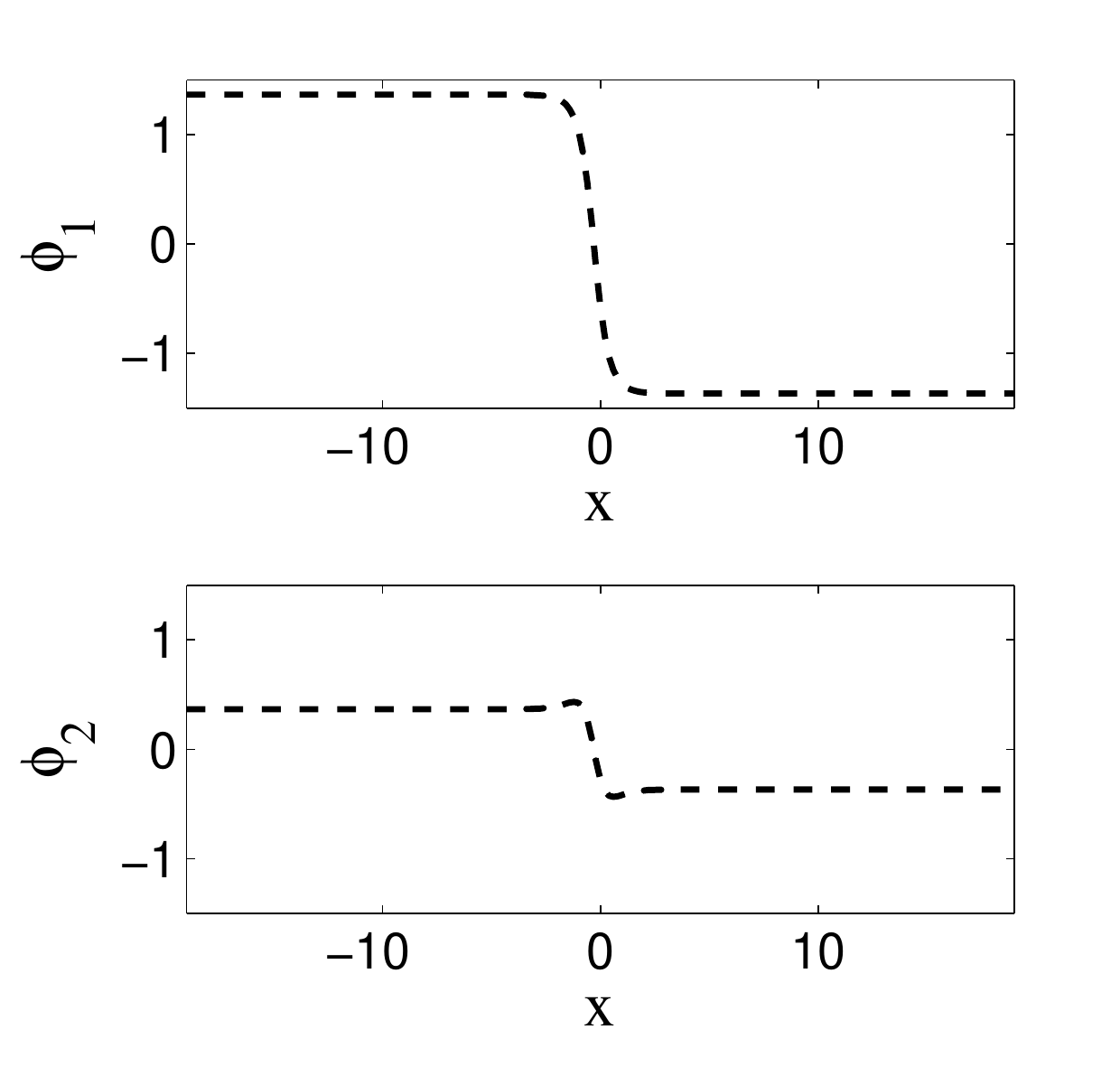}%
\label{Phi12_profiles-c}} 
\subfigure[]{\includegraphics[width=2.3in]{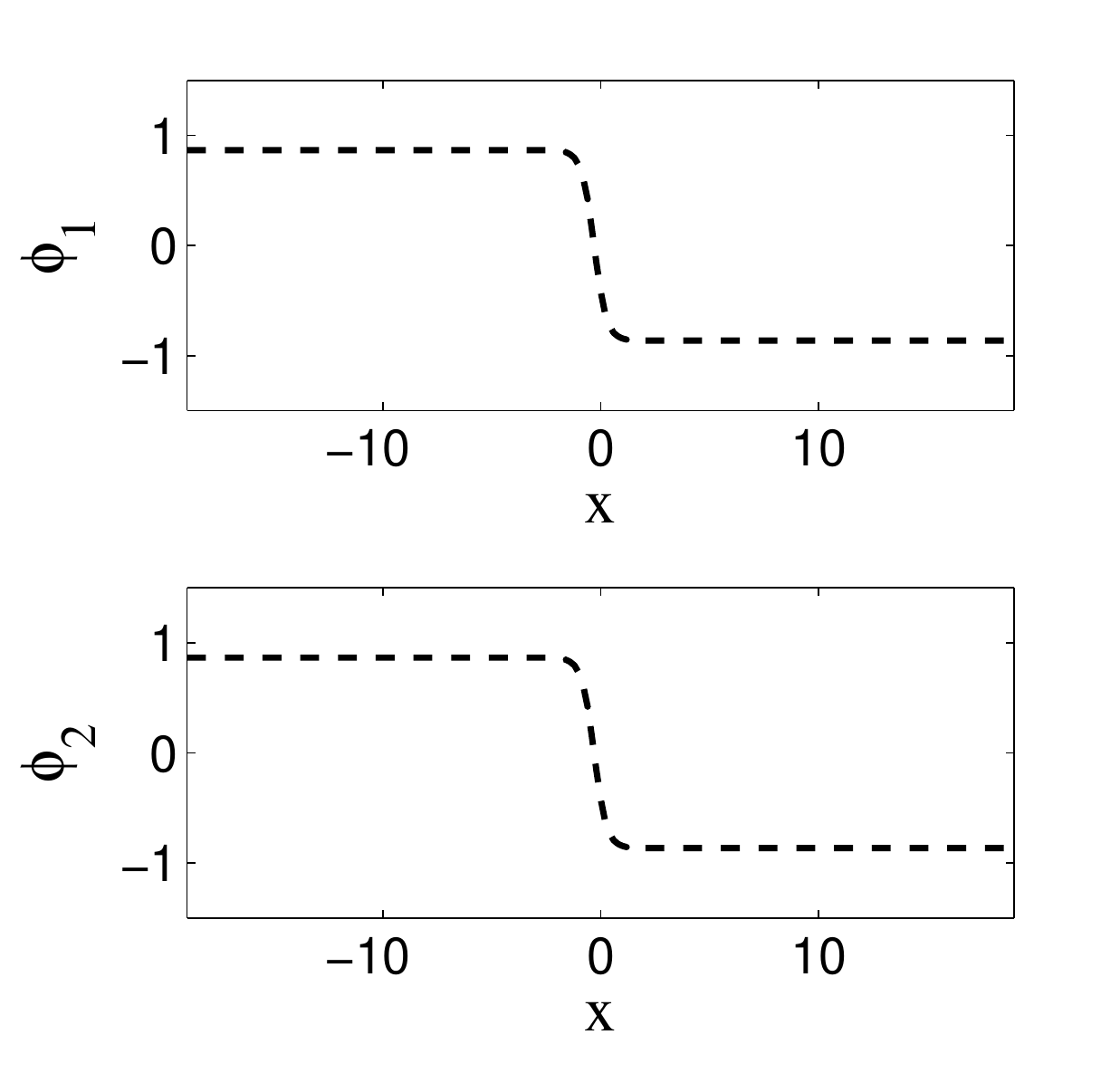}%
\label{Phi12_profiles-d}} 
\caption{Typical examples of the domain walls (a, b) and paired dark
solitons (c, d), found for $g=3$, $\protect\sigma =$ $\protect\kappa =1$,
and $\protect\mu =2$. The four panels represent the patterns of the four
types defined in Eq. (\protect\ref{types}). Stable and unstable solutions
are depicted by the continues and dashed lines, respectively. }
\label{Phi12_profiles}
\end{figure}

DWs are built as transient layers fusing together CW states of different
types. Obviously, such patterns are possible at $\mu ^{2}>\mu _{\mathrm{cr}%
}^{2}$ [see Eq. (\ref{mu-crit})], where asymmetric CW solutions exist, as
given by Eqs. (\ref{phi12_asymm}). Figure~\ref{Phi12_profiles} shows that,
in this case, one can find four different types of the transient layers, if
dark solitons are counted too. These types may be classified by values of
coupled fields $\left( \phi _{1},\phi _{2}\right) $ in the uniform states
connected by the DWs:
\begin{align}
& \left\{ \left( \phi _{1}(x=-\infty ),\phi _{2}\left( x=-\infty \right)
\right) ,\left( \phi _{1}(x=+\infty ),\phi _{2}\left( x=+\infty \right)
\right) \right\}  \notag \\
& =\left\{ \left( A_{1},A_{2}\right) ,\left( A_{2},A_{1}\right) \right\}
;~\left\{ \left( A_{1},A_{2}\right) ,\left( -A_{2},-A_{1}\right) \right\} ;
\notag \\
& \left\{ \left( A_{1},A_{2}\right) ,\left( -A_{1},-A_{2}\right) \right\}
;~\left\{ \left( A_{0},A_{0}\right) ,\left( -A_{0},-A_{0}\right) \right\}
\label{types}
\end{align}%
[recall that $A_{0}$ is given by Eq. (\ref{symm}), and $A_{1,2}$ are given
by Eqs. (\ref{phi12_asymm})]. In fact, only the patterns of the first and
second types in Eq. (\ref{types}) [Figs. \ref{Phi12_profiles-a} and %
\subref{Phi12_profiles-b}] are true DWs, while Figs. \ref{Phi12_profiles-c}
and \subref{Phi12_profiles-d}, corresponding to the profiles of the third
and fourth types in Eq. (\ref{types}), are paired dark solitons. It is also
clear that the dark-soliton pairs of the latter type, displayed in Fig. \ref%
{Phi12_profiles-d}, are unstable past the SBB point, as symmetric CW state (%
\ref{symm}) is unstable in this case.

It is relevant to mention that DWs connecting the symmetric and asymmetric
states -- for instance, $\left( A_{1},A_{2}\right) $ and $\left(
A_{0},A_{0}\right) $ -- are impossible, because a stationary DW may only
exist between two asymptotic flat states with equal Hamiltonian densities
\cite{me}. Comparing the respective densities (\ref{Hsymm}) and (\ref{Hasymm}%
), one can immediately conclude that they coincide solely at the bifurcation
point (\ref{mu-crit}).
\begin{figure}[tbp]
\subfigure[]{\includegraphics[width=3in]{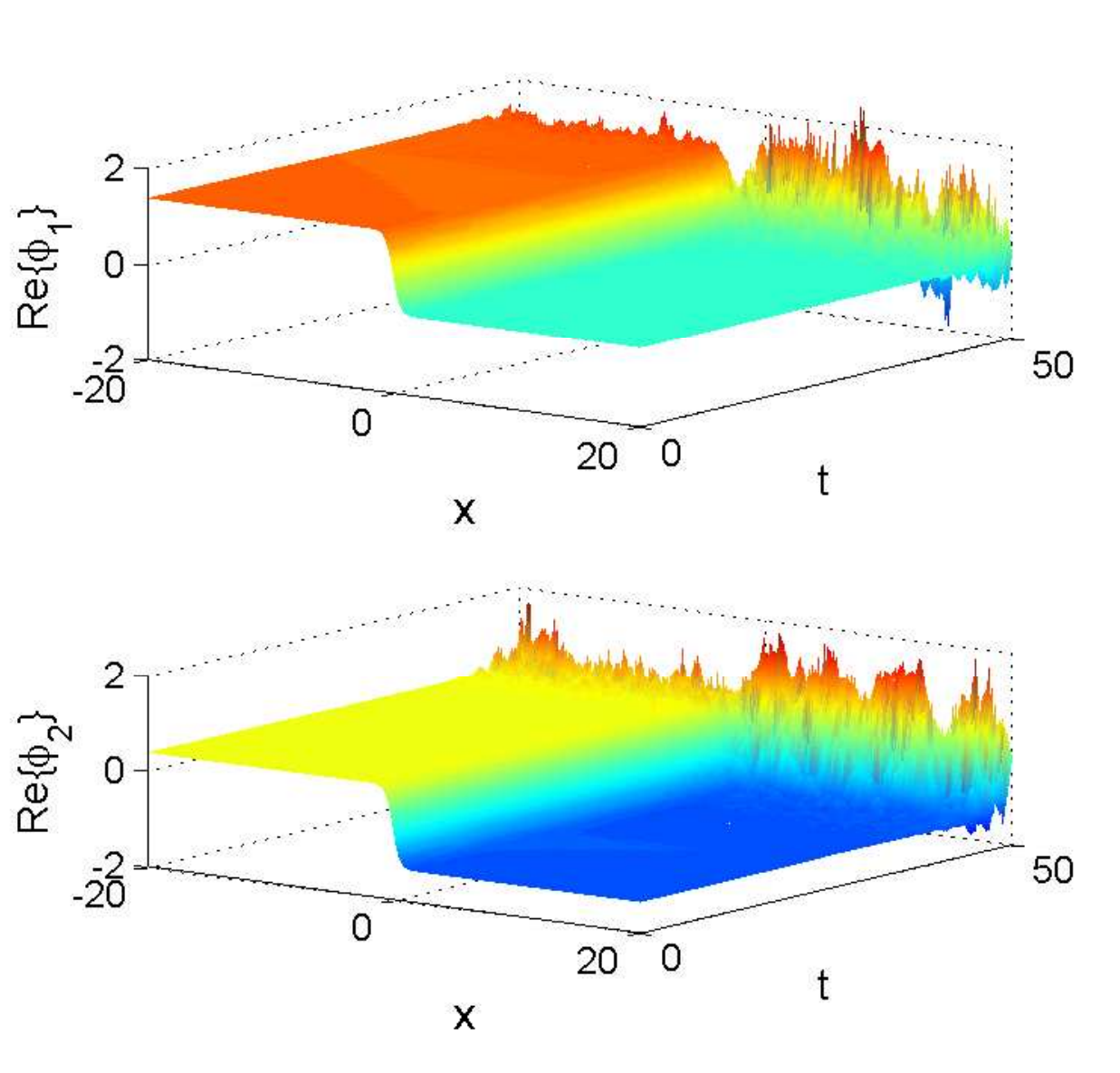}%
\label{Evolution_1DW_phi1mphi2-a}} 
\subfigure[]{\includegraphics[width=3in]{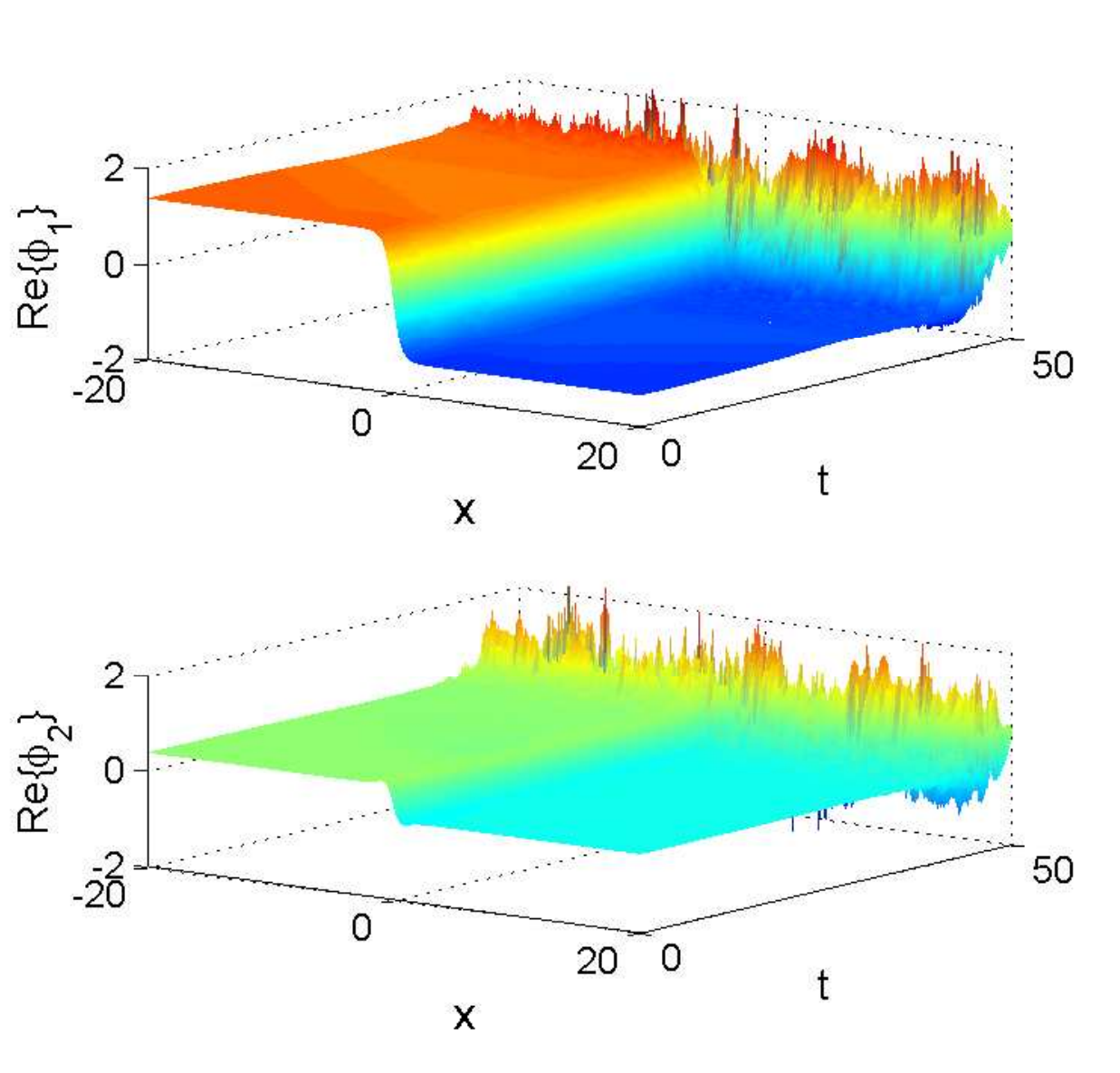}%
\label{Evolution_1DW_phi1mphi1-b}} 
\caption{(Color online) Examples for the evolution of unstable patterns of
the second (a) and third (b) types from Eq. (\protect\ref{types}). The
initial conditions and the parameters are as in Fig. \protect\ref%
{Phi12_profiles-b} and \subref{Phi12_profiles-c}, for (a) and (b),
respectively.}
\label{Phi12_Unstable_Evolution}
\end{figure}

To investigate the stability of the DW patterns, small perturbations were
added to the stationary solutions:%
\begin{eqnarray}
\widetilde{\phi _{1}}(x,t) &=&\phi _{1}(x)+v_{1}(x)e^{-i\gamma
t}+u_{1}^{\ast }(x)e^{i\gamma ^{\ast }t},  \notag \\
\widetilde{\phi _{2}}(x,t) &=&\phi _{2}(x)+v_{2}(x)e^{-i\gamma
t}+u_{2}^{\ast }(x)e^{i\gamma ^{\ast }t},  \label{Perturbed_solution}
\end{eqnarray}%
where $v_{1}$, $u_{1}$ and $v_{2}$, $u_{2}$ constitute eigenmodes of the
infinitesimal perturbation, and $\gamma $ is the corresponding
eigenfrequency that, in general, may be complex. Substituting expressions (%
\ref{Perturbed_solution}) into Eqs. (\ref{Phi12_model}) and linearizing
around the stationary solutions leads to the following eigenvalue problem,
\begin{equation}
\left(
\begin{array}{cccc}
-\hat{L_{1}} & \sigma \phi _{1}^{2} & g\phi _{1}\phi _{2}-\kappa & g\phi
_{1}\phi _{2} \\
-\sigma \phi _{1}^{2} & \hat{L_{1}} & -g\phi _{1}\phi _{2} & -g\phi _{1}\phi
_{2}+\kappa \\
g\phi _{2}\phi _{1}-\kappa & g\phi _{2}\phi _{1} & -\hat{L_{2}} & \sigma
\phi _{2}^{2} \\
-g\phi _{2}\phi _{1} & -g\phi _{2}\phi _{1}+\kappa & -\sigma \phi _{2}^{2} &
\hat{L_{2}}%
\end{array}%
\right) \left(
\begin{array}{c}
v_{1} \\
u_{1} \\
v_{2} \\
u_{2}%
\end{array}%
\right) =\gamma \left(
\begin{array}{c}
v_{1} \\
u_{1} \\
v_{2} \\
u_{2}%
\end{array}%
\right) ,  \label{Eigenvalue_problem}
\end{equation}
\begin{eqnarray}
\hat{L_{1}} &\equiv &\mu +(1/2)d^{2}/dx^{2}-2\sigma \phi _{1}^{2}-g\phi
_{2}^{2},  \notag \\
\hat{L_{2}} &\equiv &\mu +(1/2)d^{2}/dx^{2}-2\sigma \phi _{2}^{2}-g\phi
_{1}^{2}.  \label{L_operator}
\end{eqnarray}%
This eigenvalue problem can be solved using a simple finite-difference
scheme. Accordingly, the solution is identified as a stable one if all the
eigenfrequencies are real.

The stability analysis outlined above demonstrates that the DW family of the
first type in Eq. (\ref{types}), which is represented by Fig. \ref%
{Phi12_profiles-a}, is \emph{completely stable} (past the SBB\ point, where
it exists, along with the asymmetric CW states). In this case, all the other
types of the DW and dark-soliton solutions [i.e., all of them which cross
zero at least in one component, see Fig. \ref{Phi12_profiles}(b,c,d)] are
\emph{completely unstable}, due to the presence of imaginary
eigenfrequencies in the spectrum of small perturbations. We stress that,
unlike the trivial background instability of the pattern of the last type in
Eq. (\ref{types}), those of the second and third types, which are
represented by Figs. \ref{Phi12_profiles-b} and \subref{Phi12_profiles-c},
are destabilized by perturbations localized around the transient layer,
while the CW background is stable. Direct simulations demonstrate that these
unstable DWs decay into expanding turbulent patterns (see Fig. \ref%
{Phi12_Unstable_Evolution}). The stability of the DW of the first type in
Eq. (\ref{types}) was also verified by direct simulations (not shown here).

\subsection{An approximate solution for the DW near the bifurcation point}

An analytical solution for the DW can be constructed in an approximate
asymptotic form near the bifurcation point (\ref{mu-crit}), (\ref{crit}),
i.e., for $\mu $ taken as in Eq. (\ref{over}). In this case, the approximate
solution can be sought for as%
\begin{eqnarray}
\phi _{1}(x) &=&A_{\mathrm{cr}}+\delta \phi _{1}(x)+\delta \phi _{2}(x),
\notag \\
\phi _{2}(x) &=&A_{\mathrm{cr}}-\delta \phi _{1}(x)+\delta \phi _{2}(x),
\label{phi12}
\end{eqnarray}%
where it is implied that $\delta \phi _{1}\sim \sqrt{\delta \mu }$ and $%
\delta \phi _{2}\sim \delta \mu $, cf. Eq. (\ref{delta}). Substituting
expressions (\ref{phi12}) into Eqs. (\ref{Phi12_model}) and expanding the
result in powers of $\delta \mu $ yields a relation between $\delta \phi
_{1} $ and $\delta \phi _{2}$ at order $\delta \mu $,%
\begin{equation}
\delta \phi _{2}\left( x\right) =\frac{A_{\mathrm{cr}}}{2\left( \mu +\kappa
\right) }\left[ \delta \mu +\left( g-3\sigma \right) \left( \delta \phi
_{1}(x)\right) ^{2}\right] .  \label{delta_phi2}
\end{equation}%
Next, at order $\delta \mu ^{3/2}$ the expansion yields the equation for $%
\delta \phi _{1}(x)$:%
\begin{equation}
\left( \delta \phi _{1}\right) ^{\prime \prime }+4\frac{g-\sigma }{g+\sigma }%
\delta \mu \cdot \delta \phi _{1}-16\sigma \frac{g-\sigma }{g+\sigma }\left(
\delta \phi _{1}\right) ^{3}=0.  \label{delta-delta}
\end{equation}%
An exact solution of the DW type to Eq. (\ref{delta-delta}) is%
\begin{equation}
\delta \phi _{1}(x)=\frac{1}{2}\sqrt{\frac{\delta \mu }{\sigma }}\tanh
\left( \sqrt{2\frac{g-\sigma }{g+\sigma }\delta \mu }x\right) .  \label{tanh}
\end{equation}

With regard to the CW-stability condition (\ref{<}), we conclude that
solution (\ref{tanh}) exists if the sign of $\delta \mu $ coincides with the
sign of $\sigma ,$ and only for $\delta \mu >0$, which means that the
solution exists solely for $\sigma >0$. In fact, the numerical solution
demonstrates too that DWs cannot be found at $\sigma <0$, even if asymmetric
CW states (\ref{phi12_asymm}) exist in this case, with $\mu <0$.

In what follows below, we will also need an expression for the full density
of the DW solution. The above formulas yield%
\begin{eqnarray}
&&\left[ \phi _{1}(x)\right] ^{2}+\left[ \phi _{2}(x)\right] ^{2}  \notag \\
&=&\frac{2\kappa }{g-\sigma }+\frac{\delta \mu }{\sigma }-\frac{g-\sigma }{%
g+\sigma }\frac{\delta \mu }{\sigma }\mathrm{sech}^{2}\left( 2\sqrt{\frac{%
g-\sigma }{g+\sigma }\delta \mu }x\right) .  \label{dens0}
\end{eqnarray}

Examples of the analytically predicted profiles, as given by Eqs. (\ref%
{phi12}), (\ref{delta_phi2}) and (\ref{tanh}), together with their
numerically found counterparts, are displayed in Fig.~\ref%
{Approximation_compare-a}, for parameters $g=3$, $\sigma =1$, $\kappa =1$
[i.e., $\mu _{\mathrm{cr}}=1$, see Eq. (\ref{mu-crit})] and $\delta \mu
=0.1,0.5,1$ and $2$. For the same examples, the difference between the
numerical and approximate results is shown in Fig. \ref%
{Approximation_compare-b}. The results presented in Fig. \ref%
{Approximation_compare-a} demonstrate that the prediction loses its accuracy
with the increase of $\delta \mu $. On the other hand, comparing the core of
the analytical and numerical solutions (the transient layer), one can see
that the analytical results are not necessarily most accurate for small $%
\delta \mu $. Actually, this approximation is most suitable for intermediate
values of $\delta \mu $.
\begin{figure}[tbp]
\subfigure[]{\includegraphics[width=2.8in]{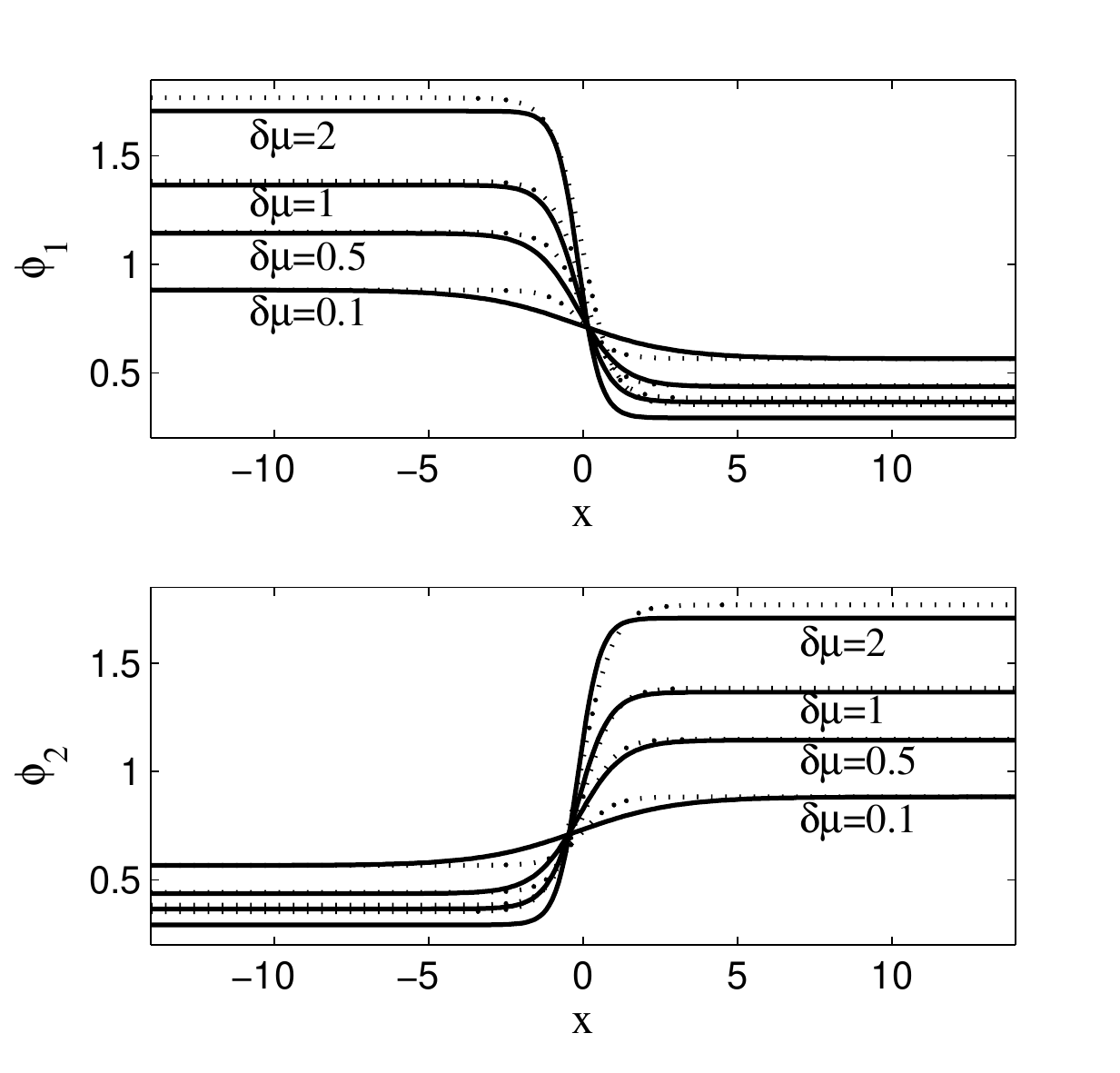}
\label{Approximation_compare-a}} 
\subfigure[]{\includegraphics[width=2.8in]{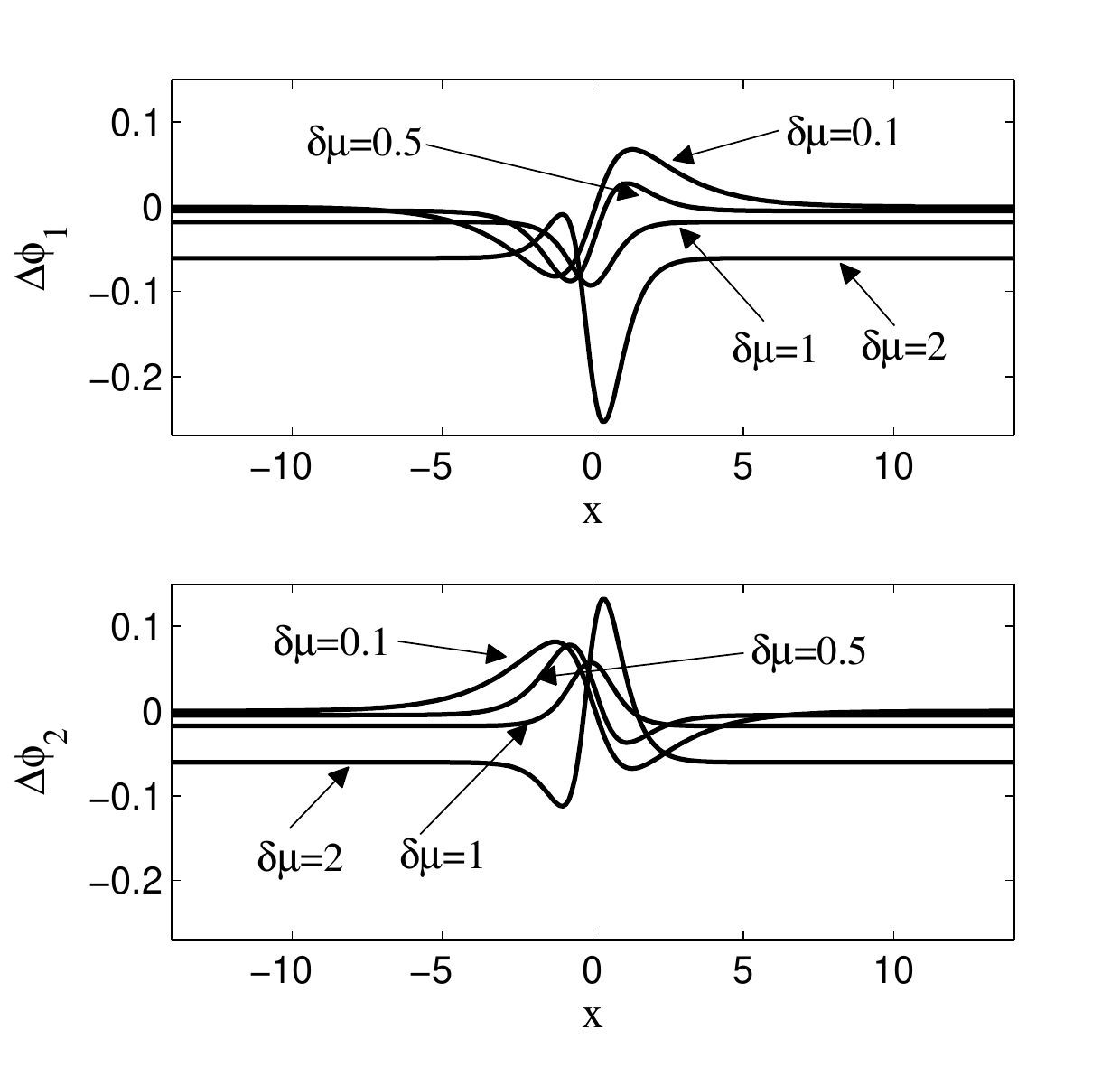}
\label{Approximation_compare-b}}
\caption{(a) Comparison between the analytical approximation given by Eq. (%
\protect\ref{tanh}), which is shown by dotted lines, and numerically found
profiles of the domain walls (solid lines), for $g=3$, $\protect\sigma =1$, $%
\protect\kappa =1$ ($\protect\mu _{\mathrm{cr}}=1$) and $\protect\delta
\protect\mu =0.1,0.5,1$ and $2$. (b) The differences between the numerical
and approximate analytical results, $\Delta \protect\phi _{1,2}=(\protect%
\phi _{1,2})_{\mathrm{numer}}-(\protect\phi _{1,2})_{\mathrm{approx}}$, for
the examples presented in panel (a).}
\label{Approximation_compare}
\end{figure}

\subsection{The exact solution for $g=3\protect\sigma $}

A particular \emph{exact }solution to Eqs. (\ref{Phi12_model}) for the DW
can be found by means of the following ansatz:%
\begin{eqnarray}
\phi _{1}(x) &=&U_{0}+U_{1}\tanh \left( \lambda x\right) ,  \notag \\
&&  \label{ansatz} \\
\phi _{2}(x) &=&U_{0}-U_{1}\tanh \left( \lambda x\right) .  \notag
\end{eqnarray}%
The substitution of the ansatz into Eqs. (\ref{Phi12_model}) demonstrates
that it yields an exact solution at $g=3\sigma $, for chemical potential $%
\mu =\kappa +\lambda ^{2}$, with coefficients%
\begin{eqnarray}
\lambda ^{2} &=&\mu -\kappa ,~U_{0}^{2}=\left( 4\sigma \right) ^{-1}\left(
2\kappa +\lambda ^{2}\right) \equiv \left( 4\sigma \right) ^{-1}\left( \mu
+\kappa \right) ,~  \notag \\
U_{1}^{2} &=&\left( 4\sigma \right) ^{-1}\lambda ^{2}\equiv \left( 4\sigma
\right) ^{-1}\left( \mu -\kappa \right)  \label{exact}
\end{eqnarray}%
(under condition $\mu -\kappa >0$). In fact, this particular exact
stationary solution is similar to the one that was found, also in
the exact form, for the stationary version of coupled
Ginzburg-Landau equations (but without the linear coupling) in Ref.
\cite{we-old}. Note that relation $g=3\sigma $, which is necessary
for the existence of the exact solution, complies with the
CW-stability condition (\ref{<}). In terms of this exact solution,
$\mu $ may be considered as a free parameter, i.e., we
actually have a one-parameter solution family, under constraint $g=3\sigma $%
. An example of the exact DW solution can be seen in Fig.~\ref%
{Phi12_profiles-a}, where the parameters were chosen to comply with the
exact solution. Finally, the total local density of solution (\ref{ansatz})
is%
\begin{equation}
\phi _{1}^{2}(x)+\phi _{2}^{2}(x)=2\left[ \left( U_{0}^{2}+U_{1}^{2}\right)
-U_{1}^{2}\mathrm{sech}^{2}\left( \lambda x\right) \right] ,  \label{dens1}
\end{equation}%
cf. Eq. (\ref{dens0}).

It may also be relevant to mention that, under condition $g=3\sigma $, Eqs. (%
\ref{Phi12_model}) has another formal DW-like solution, given by Eqs. (\ref%
{ansatz}) and (\ref{exact}) with $\tanh $ replaced by $\coth $. However,
this additional solution is singular, therefore it has no physical meaning.

\section{Domain walls in the presence of external potentials}

\label{DWwithPotential}

The model based on Eqs. (\ref{Phi12_model}) can be naturally extended to
include an external potential, $W(x)$:
\begin{eqnarray}
\mu \phi _{1}+\left( 1/2\right) \phi _{1}^{\prime \prime }-\sigma \phi
_{1}^{3}-g\phi _{2}^{2}\phi _{1}+\kappa \phi _{2}-W(x)\phi _{1} &=&0,  \notag
\\
&&  \label{AA} \\
\mu \phi _{2}+\left( 1/2\right) \phi _{2}^{\prime \prime }-\sigma \phi
_{2}^{3}-g\phi _{1}^{2}\phi _{2}+\kappa \phi _{1}-W(x)\phi _{2} &=&0.  \notag
\end{eqnarray}%
In this section we report the analysis performed for the DW patterns
supported by this setting, with $W(x)$ representing single- or double-peak
potentials, with the intention to predict pinning of the DWs by such
potentials. Stable pinning by potential maxima (rather than minima) may be
possible because the DW's core features a minimum of the total density, see
Eqs. (\ref{dens0}) and (\ref{dens1}), hence the total energy of the system
may be minimized by placing the core around a local maximum of the potential.

\subsection{The exact solution in the presence of a single-peak potential}

\label{subsec:ExactPotential}

It is possible to find an exact solution to Eqs. (\ref{AA}) for the DW if
the potential is chosen as%
\begin{equation}
W(x)=W_{0}~\mathrm{sech}^{2}\left( \lambda x\right)  \label{A}
\end{equation}%
where $W_{0}$ and $\lambda $ are considered as given parameters. The
corresponding solution can be looked for in the form of the same ansatz (\ref%
{ansatz}) as used for finding the exact solution in the free-space setting.
The substitution of the ansatz into Eqs. (\ref{AA}) demonstrates that it
yields an exact solution in the present case under the following condition
imposed on parameters of the system:%
\begin{equation}
g=\frac{3\lambda ^{2}+2W_{0}}{\lambda ^{2}+2W_{0}}\sigma ,  \label{g}
\end{equation}%
which goes over into the above relation, $g=3\sigma $, in the limit of $%
W_{0}=0$. Further, coefficients $U_{0}$, $U_{1}$, and $\mu $ of the exact
solution are given by the following expressions:%
\begin{eqnarray}
\mu &=&\lambda ^{-2}\left( \lambda ^{2}+2W_{0}\right) \left( \lambda
^{2}+\kappa \right) ,  \notag \\
U_{0}^{2} &=&\left( 4\sigma \lambda ^{2}\right) ^{-1}\left( \lambda
^{2}+2W_{0}\right) \left( \lambda ^{2}+2\kappa \right) ,  \label{new} \\
U_{1}^{2} &=&\left( 4\sigma \right) ^{-1}\left( \lambda ^{2}+2W_{0}\right) .
\notag
\end{eqnarray}%
It is easy to see that, in the limit of $W_{0}=0$, Eqs. (\ref{new}) carry
over into the above exact solution, given by Eqs. (\ref{exact}). Thus, this
exact solution is an extension of the previous one, although it has no free
parameters [note that $\lambda $, which was an adjustable parameter of the
free-space solution, is now fixed by the given form of potential (\ref{A})].

As argued above, the present exact solution is expected to be stable if $%
W(x) $ represents a repulsive potential barrier (peak), with $W_{0}>0$, and
the solution should be unstable, against spontaneous escape from the pinned
state, in the case of the attractive potential well, with $W_{0}<0$ [cf. Eq.
(\ref{pot}) below]. In turn, Eq. (\ref{g}) with $W_{0}>0$ gives $g<3\sigma $%
. Note also that, unlike its counterpart in the free space, this exact
solution may exist at $\sigma <0$: In the case of the potential well ($%
W_{0}<0$), with $\lambda ^{2}<-2W_{0}<3\lambda ^{2}$, Eq. (\ref{g}) yields $%
\sigma <0$. However, this solution should be unstable according to the above
argument.

\subsection{Analysis of the interaction between DWs and the interaction of
the DW with the external potential}

\label{subsec:AnalyticPotential} The interaction of the DW with an external
potential can be investigated in an approximate form. In fact, a similar
problem which also admits an approximate analytical treatment is the
interaction between two broadly separated DWs with opposite polarities
(i.e., mirror images of each other) in the free space, therefore we start
with this case.

Assuming that the two DWs are set at distance $L$ which is large in
comparison with the inner width of each DW, the interaction between them can
be analyzed by means of the method elaborated in Ref. \cite{interaction}. To
this end, the approximate expression for the nearly flat fields in the
region between the far separated solitons is taken as%
\begin{equation}
\phi _{n}(x)=A_{n}+2U_{n}\left[ \exp \left( -2\lambda \left\vert x-\xi
_{1}\right\vert \right) +\exp \left( -2\lambda \left\vert x-\xi
_{2}\right\vert \right) \right] ,  \label{flat}
\end{equation}%
where $n=1,2$, and $A_{1,2}$ are given by Eqs. (\ref{phi12_asymm}), $\xi
_{1,2}$ are coordinates of the centers of the two DWs, so that $L\equiv \xi
_{2}-\xi _{1}$, and the exponential terms represent small decaying tails of
the DWs on top of the flat background. The decay rate $\lambda >0$ can be
found in the general case, but the expression for it is cumbersome;
amplitude $U_{1,2}$ are not known in an exact form in the general case, as
they may only be found from full solutions for individual DWs. In the
special case of the exact solution given by Eqs. (\ref{ansatz}), at $%
g=3\sigma $), coefficients $\lambda $ and $U_{1,2}$ in Eq. (\ref{flat}) are
actually given by Eq. (\ref{exact}):
\begin{equation}
\lambda =\sqrt{\mu -\kappa },~U_{1}=U_{2}=(1/2)\sqrt{\left( \mu -\kappa
\right) /\sigma }.  \label{lUU}
\end{equation}

Then, using Hamiltonian produced by density (\ref{H}) and the method
developed in Ref. \cite{interaction} (identifying the term accounting for
the interaction energy in the expression for the full Hamiltonian), the
effective potential of the interaction between the two separated DWs is
found in the following form:%
\begin{equation}
U_{\mathrm{int}}(L)=-8\lambda \left( U_{1}^{2}+U_{2}^{2}\right) \exp \left(
-2\lambda L\right) .  \label{Uint0}
\end{equation}%
In the case of the exact DW solution given by Eqs. (\ref{ansatz}) and (\ref%
{exact}), this expression takes an explicit form,%
\begin{equation}
U_{\mathrm{int}}(L)=-4\left( \mu -\kappa \right) ^{3/2}\exp \left( -2\sqrt{%
\mu -\kappa }L\right) .  \label{Uint}
\end{equation}%
A simple but essential property of expressions (\ref{Uint0}) and (\ref{Uint}%
) is that they obviously predict \emph{attraction} between the two DWs.

With the external potential $W(x)$ included into Eqs. (\ref{AA}),
Hamiltonian density (\ref{H}) is modified as%
\begin{gather}
\mathcal{H}=\frac{1}{2}\left[ \left( \phi _{1}^{\prime }\right) ^{2}+\left(
\phi _{2}^{\prime }\right) ^{2}\right] +\frac{\sigma }{2}\left( \phi
_{1}^{4}+\phi _{2}^{4}\right)  \notag \\
+g\phi _{1}^{2}\phi _{2}^{2}-2\kappa \phi _{1}\phi _{2}+2W(x)\left( \phi
_{1}^{2}+\phi _{2}^{2}-A_{1}^{2}-A_{2}^{2}\right) ,  \label{HW}
\end{gather}%
where constants $A_{1,2}^{2}$ are subtracted from $\phi _{1,2}^{2}(x)$ for
convenience (to cancel a formally diverging constant term in the
Hamiltonian), $A_{1,2}$ being the same as in Eqs. (\ref{phi12_asymm}). This
means that the energy of the interaction of the DW with the external
potential is%
\begin{equation}
U_{\mathrm{pot}}=2\int_{-\infty }^{+\infty }W(x)\left[ \phi _{1}^{2}(x)+\phi
_{2}^{2}(x)-A_{1}^{2}-A_{2}^{2}\right] dx.  \label{pot}
\end{equation}

Assuming that $W(x)$ represents a broad potential barrier or well with a
width much larger than the thickness of the DW, and treating the external
potential as a perturbation\emph{\ }(i.e., neglecting the distortion of the
solution under the action of the potential), the substitution of the exact
DW solution given by Eqs. (\ref{ansatz}) and (\ref{exact}) into Eq. (\ref%
{pot}) readily yields:%
\begin{equation}
U_{\mathrm{pot}}\left( \xi \right) \approx -8U_{1}^{2}\lambda ^{-1}W(\xi
)\equiv -2\sqrt{\mu -\kappa }\sigma ^{-1}W\left( \xi \right) ,  \label{Upot}
\end{equation}%
where $\xi $ is the coordinate of the center of the DW. A straightforward
consequence of Eq. (\ref{Upot}) is that, as conjectured above, the DW tends
to be trapped at local \emph{maxima} of the external potential $W(x)$, as,
due to sign minus in Eq. (\ref{Upot}), they correspond to \emph{minima} of
the effective potential (\ref{Upot}).

If two DWs are trapped at two particular maxima of $W(x)$, $\xi _{1}$ and $%
\xi _{2}$, separated by large distance $L$, the equilibrium condition for
each DW is the vanishing of the total force produced by the interaction of
the DW with its counterpart and with the external potential:%
\begin{eqnarray}
&&\frac{\partial }{\partial \xi _{1}}\left[ \frac{2\sqrt{\mu -\kappa }}{%
\sigma }W\left( \xi _{1}\right) +4\left( \mu -\kappa \right) ^{3/2}\exp
\left( -2\sqrt{\mu -\kappa }\left( \xi _{2}-\xi _{1}\right) \right) \right]
\notag \\
&=&\frac{\partial }{\partial \xi _{2}}\left[ \frac{2\sqrt{\mu -\kappa }}{%
\sigma }W\left( \xi _{2}\right) +4\left( \mu -\kappa \right) ^{3/2}\exp
\left( -2\sqrt{\mu -\kappa }\left( \xi _{2}-\xi _{1}\right) \right) \right]
=0.  \label{equil}
\end{eqnarray}%
For example, if the potential is periodic, $W(x)=\epsilon \cos \left( 2\pi
x/\Lambda \right) $, with large period $\Lambda $, one may consider the pair
of DWs trapped at two adjacent potential maxima. Substituting this potential
into Eqs. (\ref{equil}), it is easy to check that its minimum strength,
necessary for holding the DW pair (i.e., preventing it from merger due to
the mutual attraction) is $\epsilon _{\min }=\left( 2/\pi \right) \sigma
\Lambda \left( \mu -\kappa \right) ^{3/2}\exp \left( -\sqrt{\mu -\kappa }%
\Lambda \right) .$

\subsection{Numerical results for the model with an external potential}

\label{subsec:NumericalPotential} The above analysis predicted that two DWs
created in the free space attract each other. Direct simulations confirm the
prediction, see a typical example in Fig.~\ref{Evolution_NoPotential}.
Eventually, the two DWs annihilate into the stable uniform asymmetric state.
Naturally,\ for larger initial values of the separation between the DWs, the
attraction force is weaker, and a considerably longer time is required for
the DW pair to manifest the interaction.

\begin{figure}[tbp]
\begin{minipage}[t]{5.3in}
\includegraphics[width=2.6in]{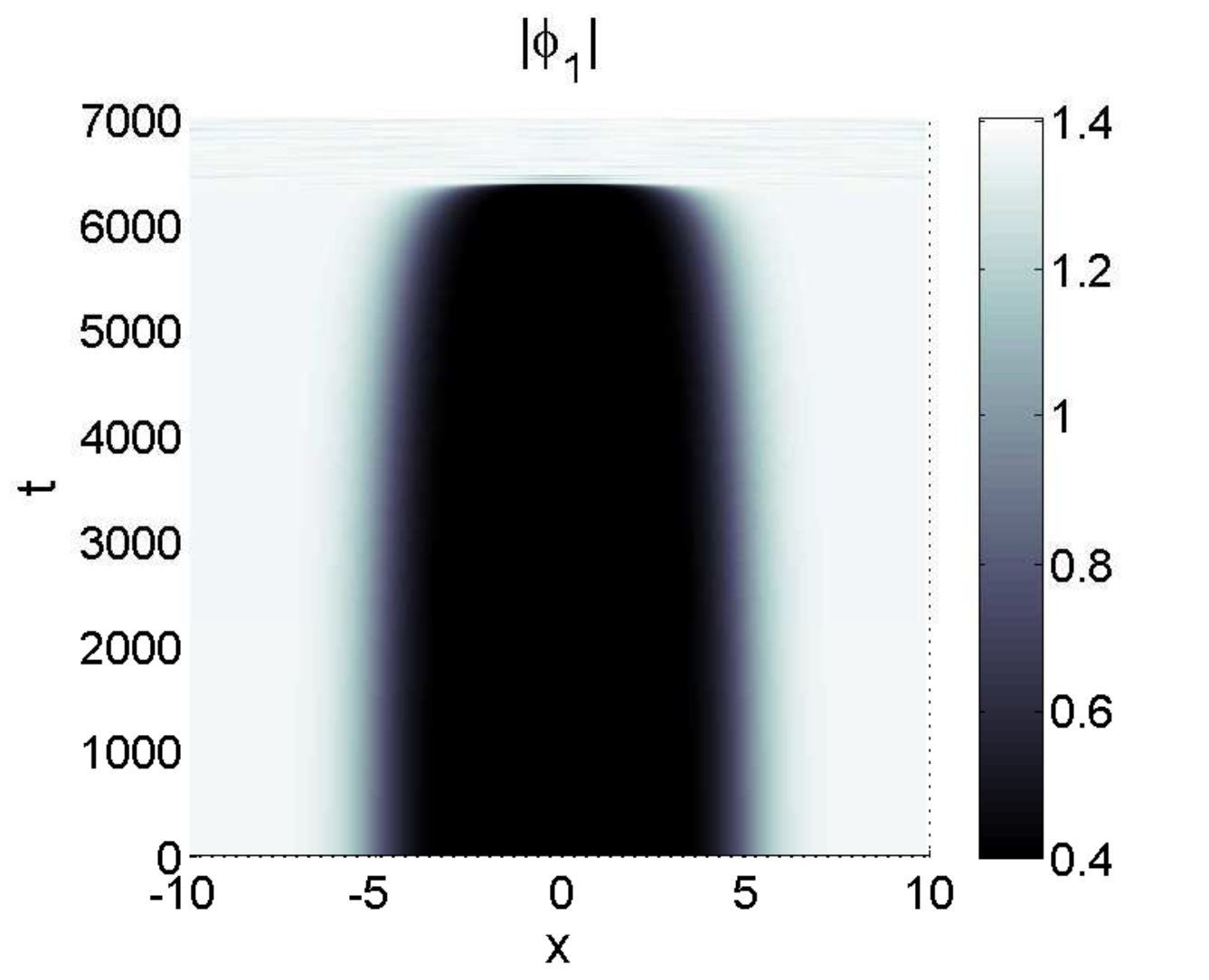}
\begin{minipage}[t]{2.6in}
\includegraphics[width=2.6in]{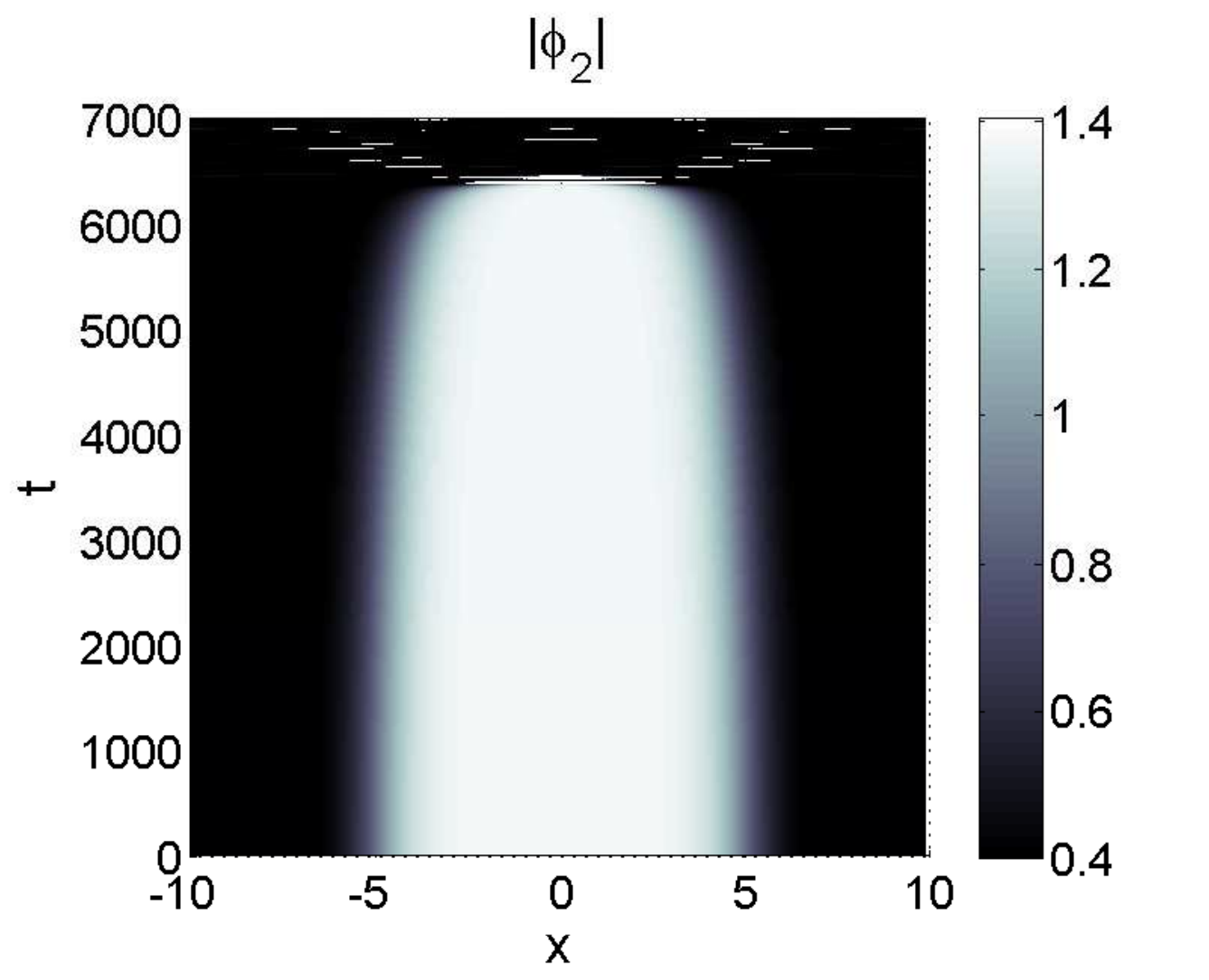}
\label{Evolution_NoPotential}
\end{minipage}
\end{minipage}
\caption{The simulated evolution of a pair of identical
domain walls, initially separated by distance $2\Delta x=10$, is displayed
by means of density contour plots, for $g=3$, $\protect\sigma =1$, $\protect%
\kappa =1$, and $\protect\mu =2$.}
\label{Evolution_NoPotential}
\end{figure}

\begin{figure}[tbp]
\subfigure[]{\includegraphics[width=2.25in]{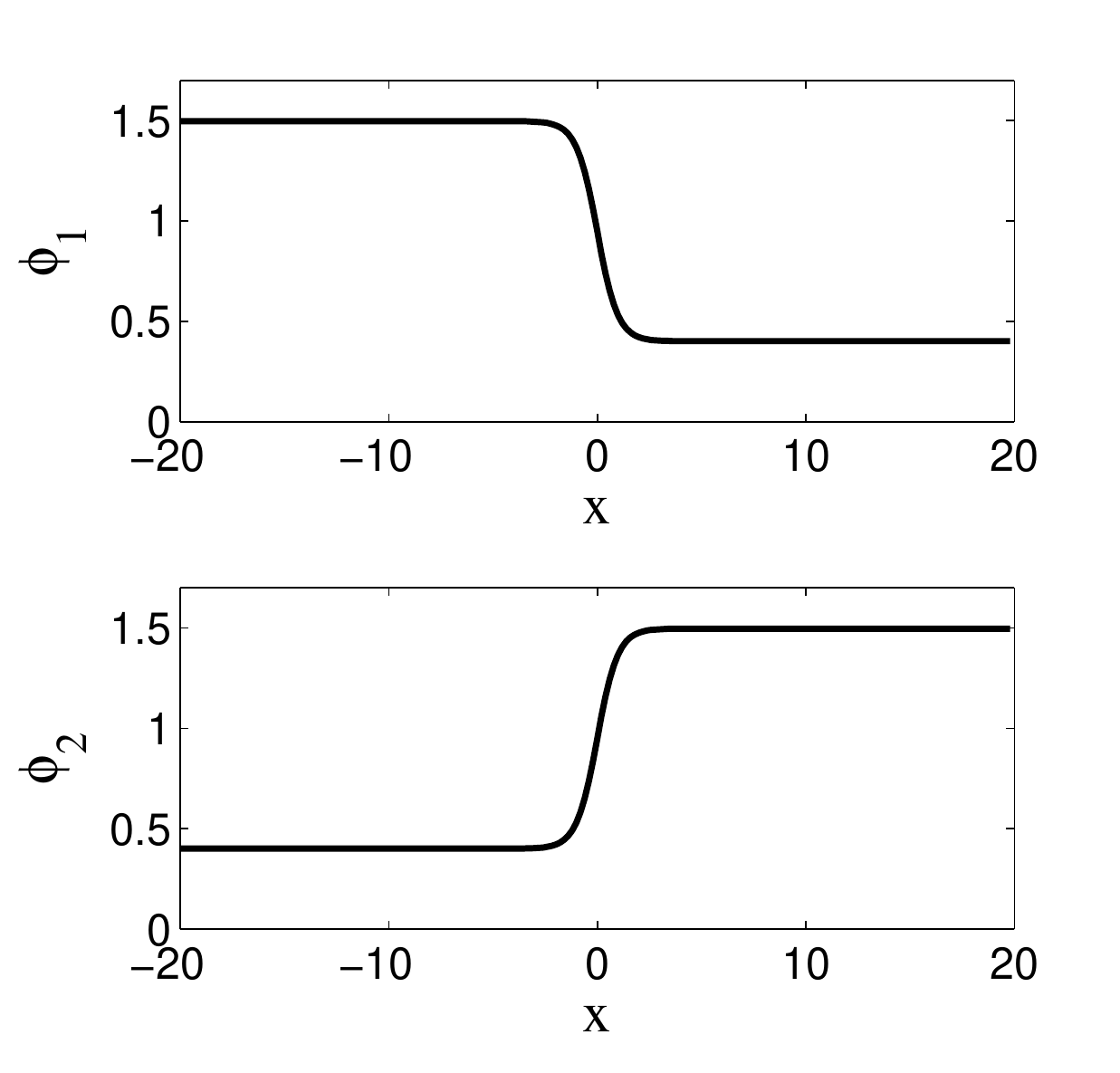}
\label{Phi12_SinglePeak-a}} 
\subfigure[]{\includegraphics[width=2.25in]{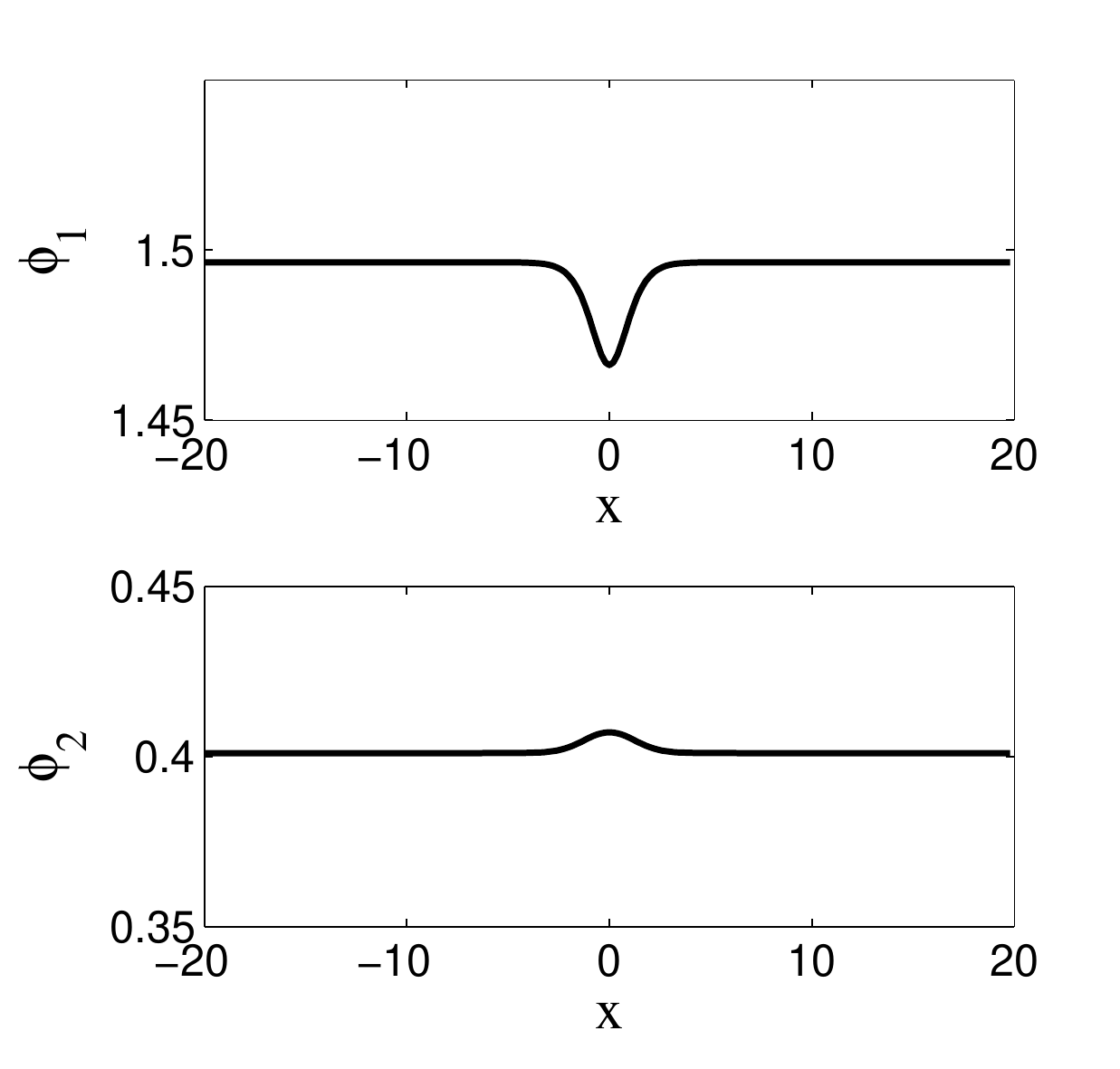}
\label{Phi12_SinglePeak-b}} 
\subfigure[]{\includegraphics[width=2.25in]{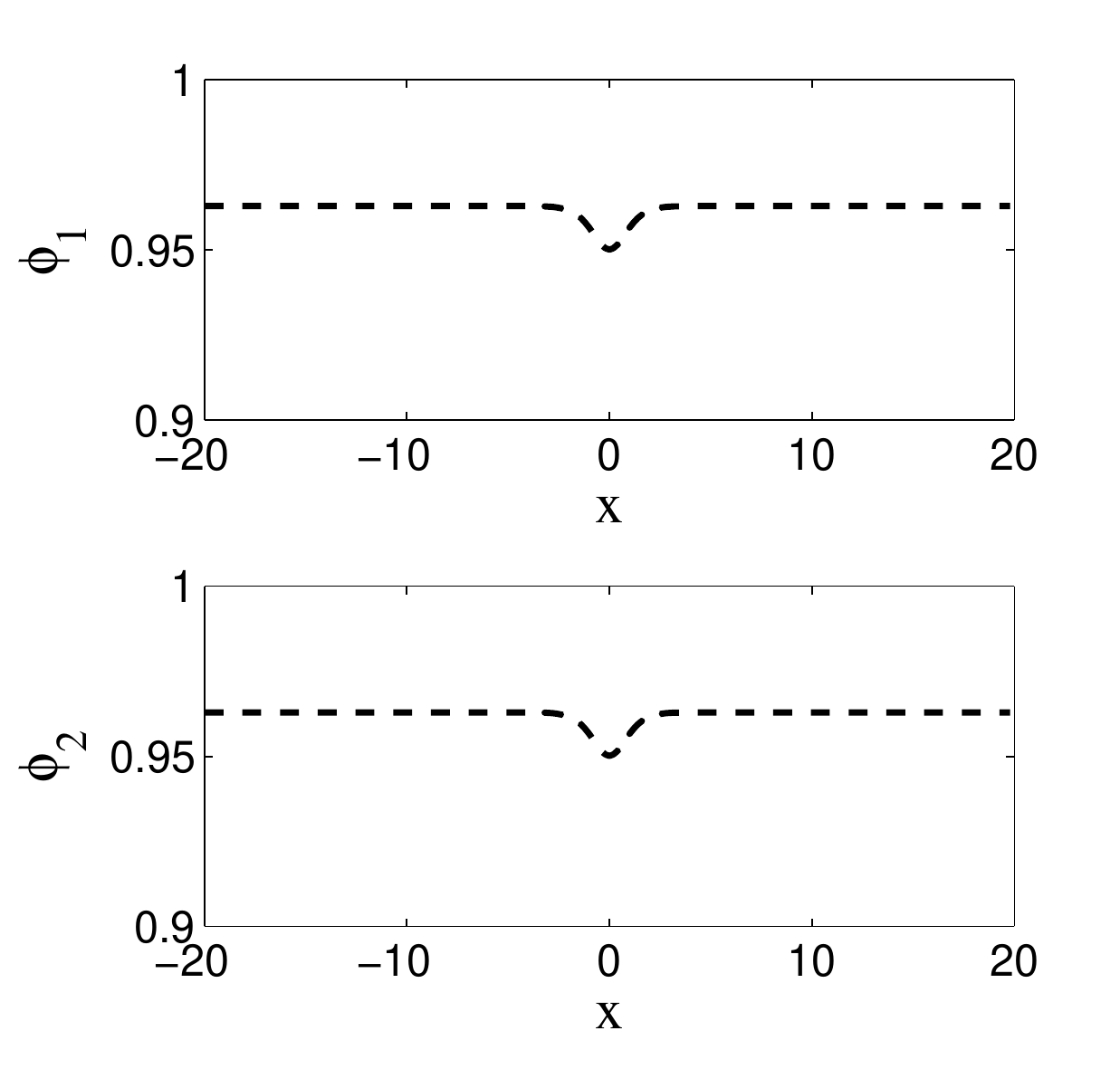}
\label{Phi12_SinglePeak-c}}
\caption{Typical profiles of the domain-wall (a) and ``bubble" (b,c)
stationary solutions found in the presence of the potential peak (\protect
\ref{A}), for $g=8/3$, $\protect\sigma =1$, $\protect\kappa =1$, $\protect%
\lambda =1$, $W_{0}=0.1$, and $\protect\mu =2.4$ [these values of the
parameters admit the existence of the exact DW solution (\protect\ref{g})-(%
\protect\ref{new}), which is actually displayed in (a)]. The stable and
unstable solutions are depicted by continuous and dashed curves,
respectively.}
\label{Phi12_SinglePeak}
\end{figure}

In the presence of the external potential in Eqs. (\ref{AA}), the numerical
investigation was carried out for two different shapes of $W(x)$. First, the
single potential barrier was considered, taken in the same form (\ref{A})
which was used to obtain the exact solution. The corresponding stationary DW
solutions are similar to their counterparts found in the free space.
Specifically, out of the four types of the DWs categorized by the values of $%
\left( \phi _{1}(x),\phi _{2}(x)\right) $ at $x=\pm \infty $ as per Eq. (\ref%
{types}), only the first one is stable, see a typical example in Fig. \ref%
{Phi12_SinglePeak}(a). In addition to the DWs, the potential barrier
supports states in the form of ``bubbles" (defined as per Ref. \cite{bubbles}%
), i.e., patterns with identical values of $\left( \phi _{1}(x),\phi
_{2}(x)\right) $ at $x=\pm \infty $, and localized perturbations of $\phi
_{1,2}(x)$ around the barrier, see Figs.~\ref{Phi12_SinglePeak-b}--%
\subref{Phi12_SinglePeak-c}. Naturally, the bubbles supported by the
asymmetric and symmetric CWs are stable and unstable, respectively.

The expectation that the DW is stably trapped by the potential peak is
confirmed by direct simulations displayed in Fig.~\ref%
{Evolution_SinglePeak-a}, for the DW of the first type, in terms of Eq. (\ref%
{types}), i.e. $\{(A_{1}A_{2}),(A_{2}A_{1})\}$ [it is the same DW which is
displayed in Fig. \ref{Phi12_SinglePeak}(a)]. The DW, if shifted from the
potential maximum by $\Delta x=5$, performs decaying oscillations around the
peak. The decay of the oscillations is explained by emission of radiation
waves into the background by the oscillating DW.

The attraction of DWs to the potential peak is further illustrated by Fig. %
\ref{Phi12_SinglePeak}(b), which displays a pair of identical DWs
symmetrically placed, at $t=0$, at distances $\Delta x=\pm 5$ from the peak.
Due to the attraction of both DWs to the peak, in this case their
annihilation happens much sooner than for the same pair in the free space,
cf. Fig. \ref{Evolution_NoPotential} for the same value of separation $%
2\Delta x$ between the DWs.
\begin{figure}[tbp]
\subfigure[]{
\begin{minipage}{2.6in}
\includegraphics[width=2.6in]{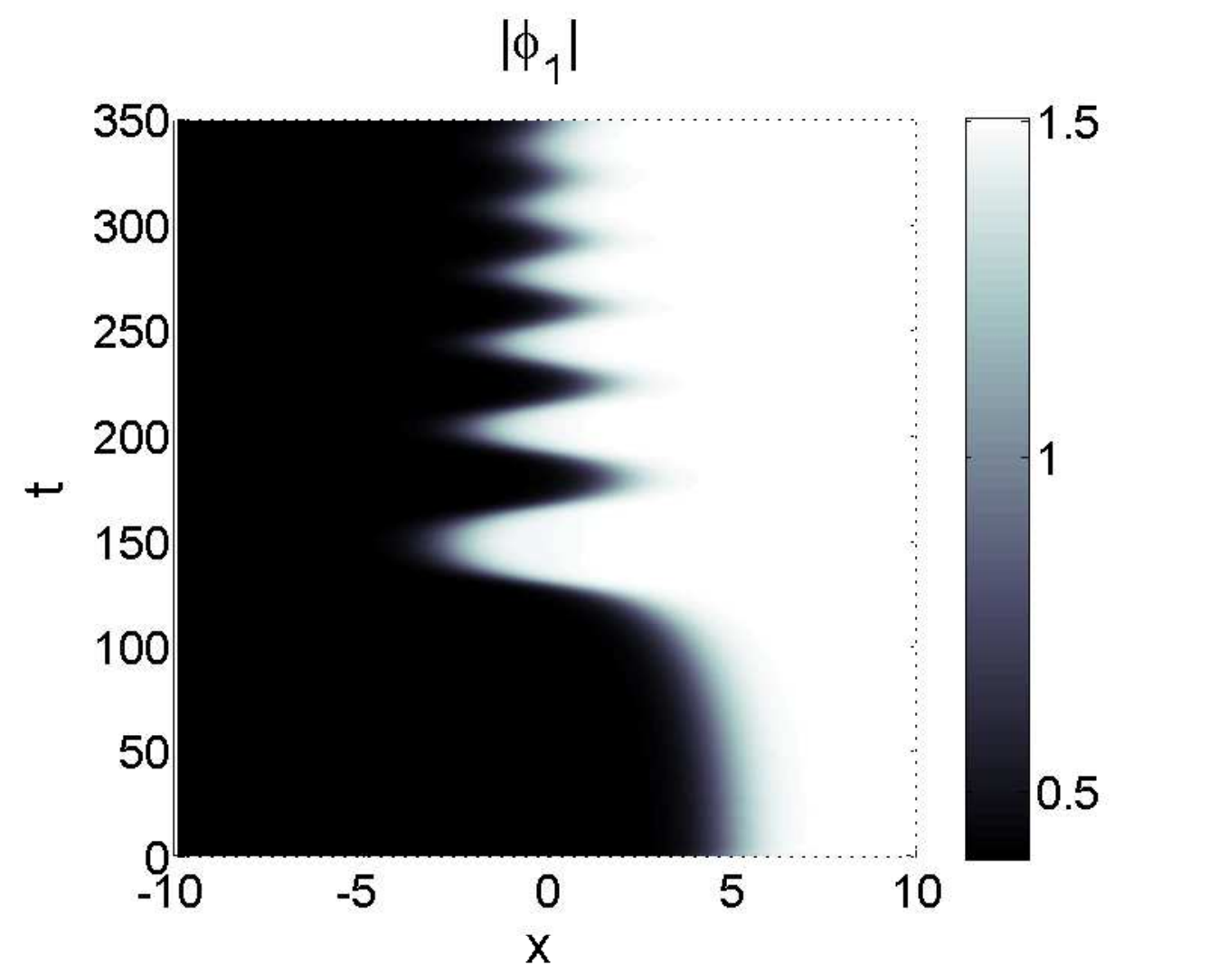}\\
\begin{minipage}{2.6in}
\includegraphics[width=2.6in]{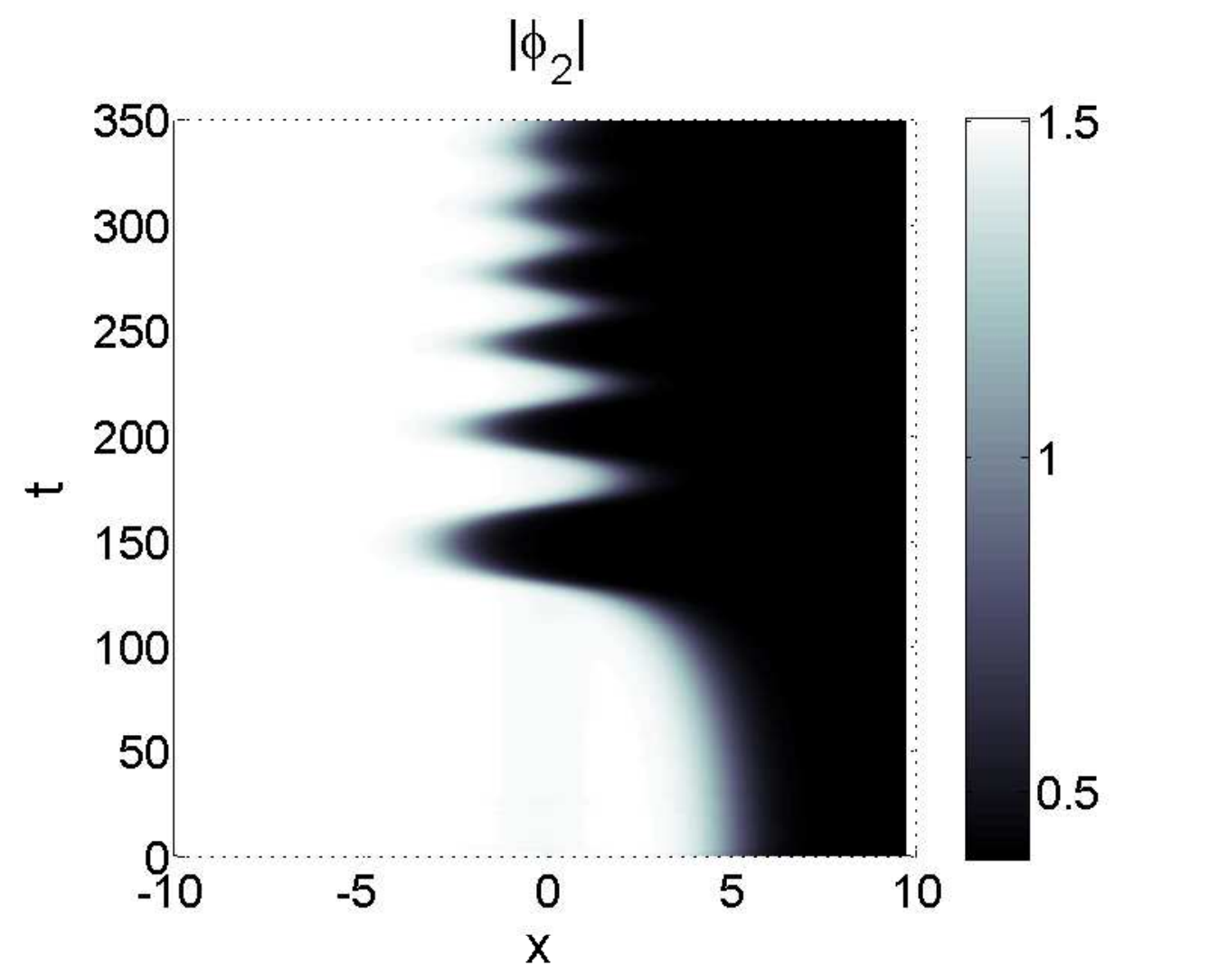}
\label{Evolution_SinglePeak-a}
\end{minipage}
\end{minipage}} 
\subfigure[]{
\begin{minipage}{2.6in}
\includegraphics[width=2.6in]{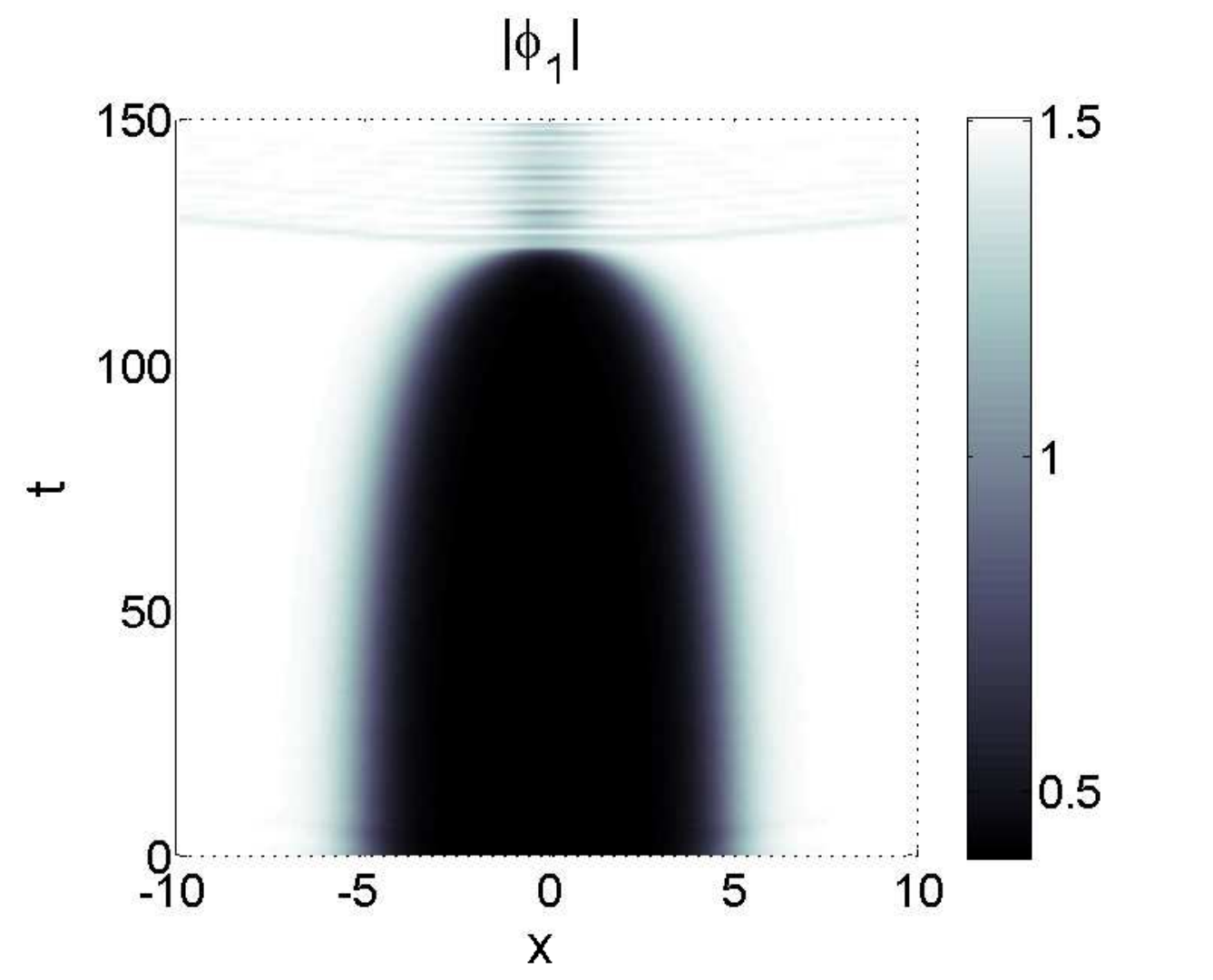}\\
\begin{minipage}{2.6in}
\includegraphics[width=2.6in]{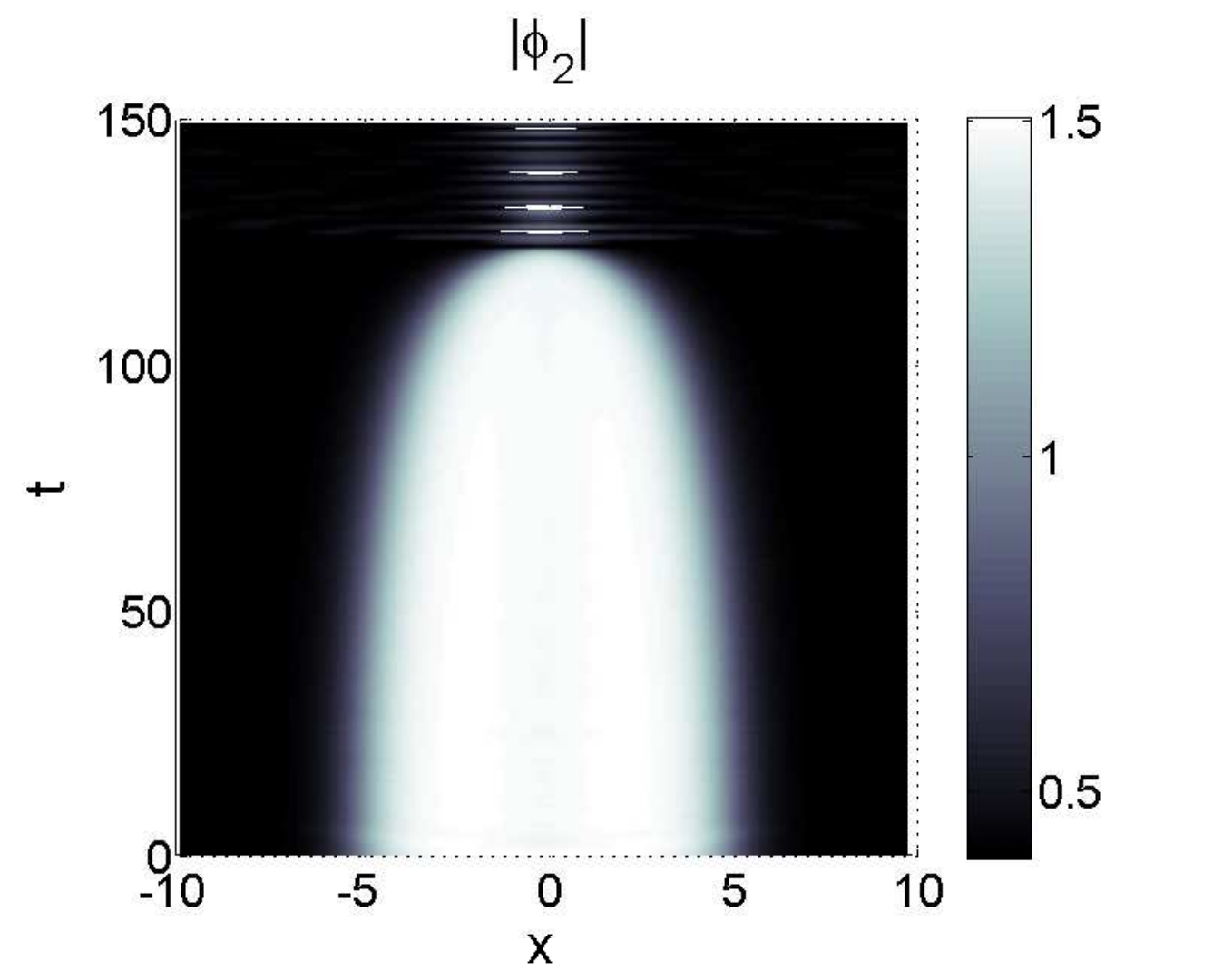}
\label{Evolution_SinglePeak-b}
\end{minipage}
\end{minipage}} 
\caption{(a) The evolution of the domain wall shifted by $%
\Delta x=5$ from the maximum of potential barrier (\protect\ref{A}), for the
same parameters as in Fig.~\protect\ref{Phi12_SinglePeak}(a). (b) Similar to
panel (a), but for the pair of DWs, placed symmetrically on both sides of
the potential maximum, at $x=\pm 5$.}
\label{Evolution_SinglePeak}
\end{figure}

The numerical analysis was also carried out for the double-peak potential,
taken in the form of
\begin{equation}
W(x)=W_{0}\left( 2x/L\right) ^{2}\exp \left( 1-\left( 2x/L\right)
^{2}\right) ,  \label{2PeakPotential}
\end{equation}%
where $W_{0}$ is the height of the two peaks and $L$ is the distance between
them. The numerical solution of Eqs. (\ref{AA}) with potential (\ref%
{2PeakPotential}) has revealed multiple structures, which may be considered
as transient layers between CW states filling the three regions separated by
the two potential barriers. These structures can be categorized into two
groups, built of symmetric or asymmetric CWs. The structures of the former
type are arranged as the following sets of the CW states in the three
regions [cf. Eq. (\ref{types})]:%
\begin{gather}
\left\{ \left( A_{0},A_{0}\right) ,\left( A_{0},A_{0}\right) ,\left(
A_{0},A_{0}\right) \right\} ;~\left\{ \left( A_{0},A_{0}\right) ,\left(
A_{0},A_{0}\right) ,\left( -A_{0},-A_{0}\right) \right\} ;  \notag \\
\left\{ \left( A_{0},A_{0}\right) ,\left( -A_{0},-A_{0}\right) ,\left(
A_{0},A_{0}\right) \right\} .  \label{types-S}
\end{gather}%
These three varieties may be interpreted, respectively, as containing none,
one, or two dark solitons trapped in each component.

\begin{figure}[tbp]
\subfigure[]{\includegraphics[width=2.3in]{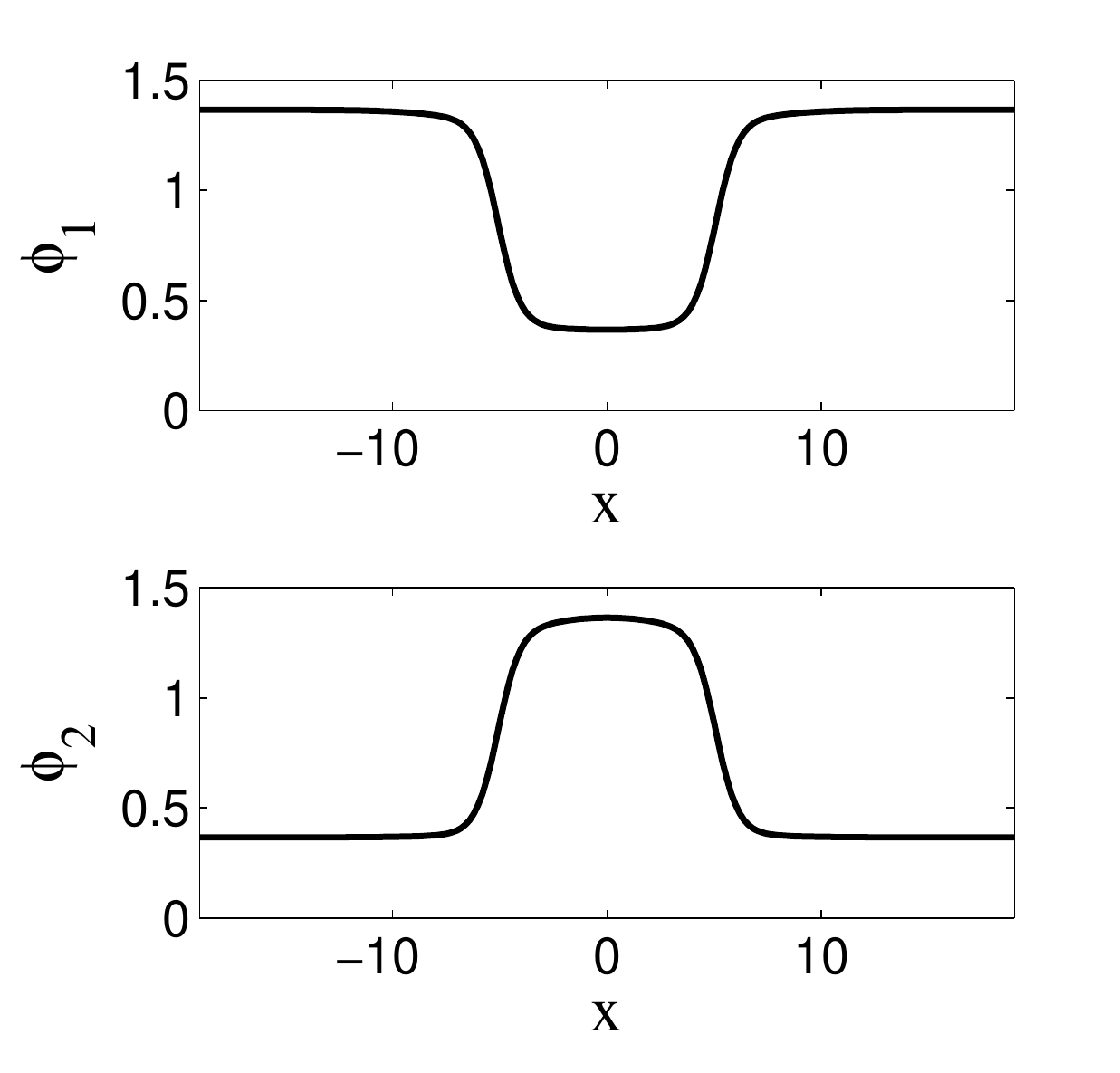}
\label{Phi12_2lePeaks-a}} 
\subfigure[]{\includegraphics[width=2.3in]{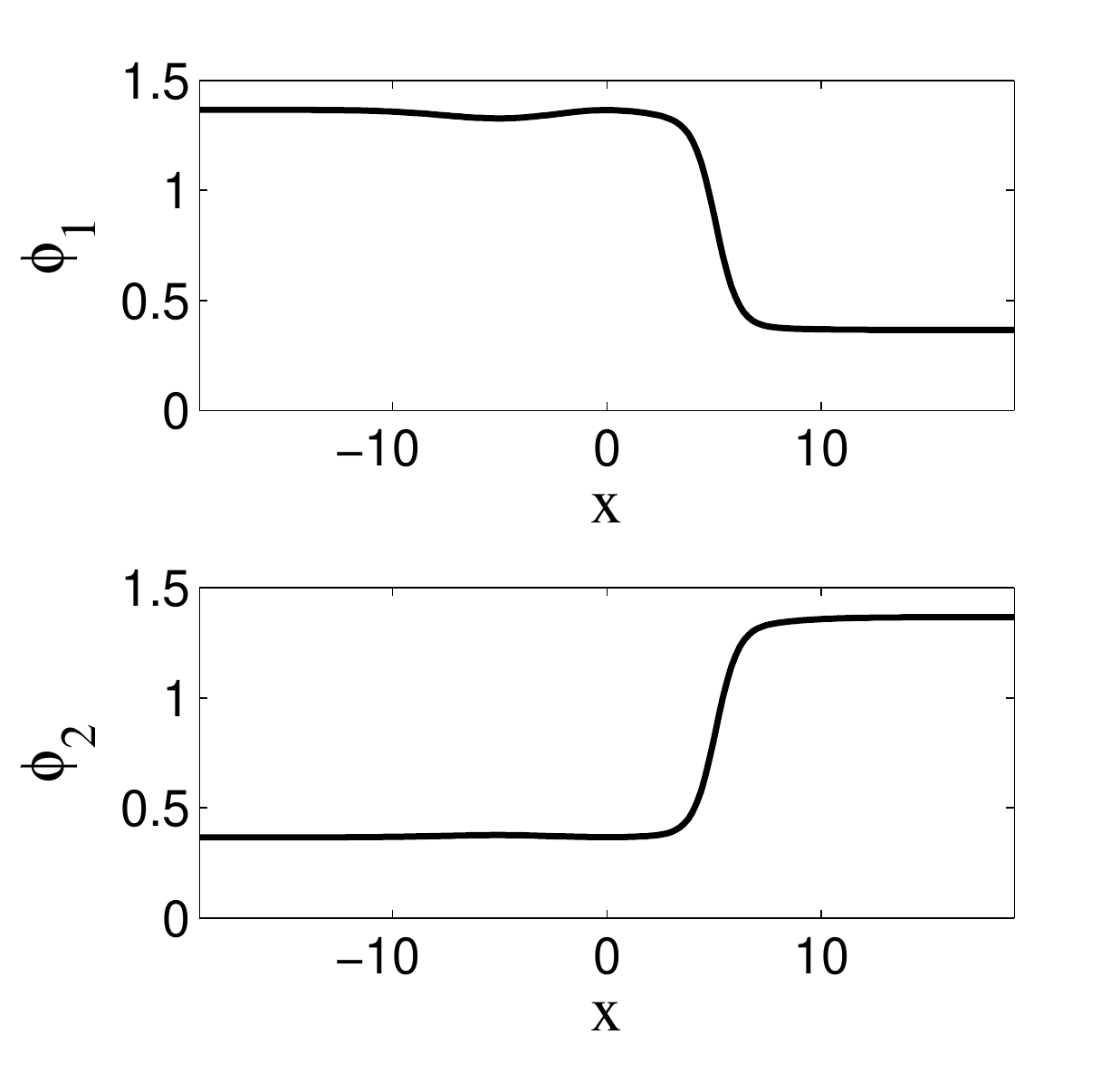}
\label{Phi12_2lePeaks-b}} \\
\subfigure[]{\includegraphics[width=2.3in]{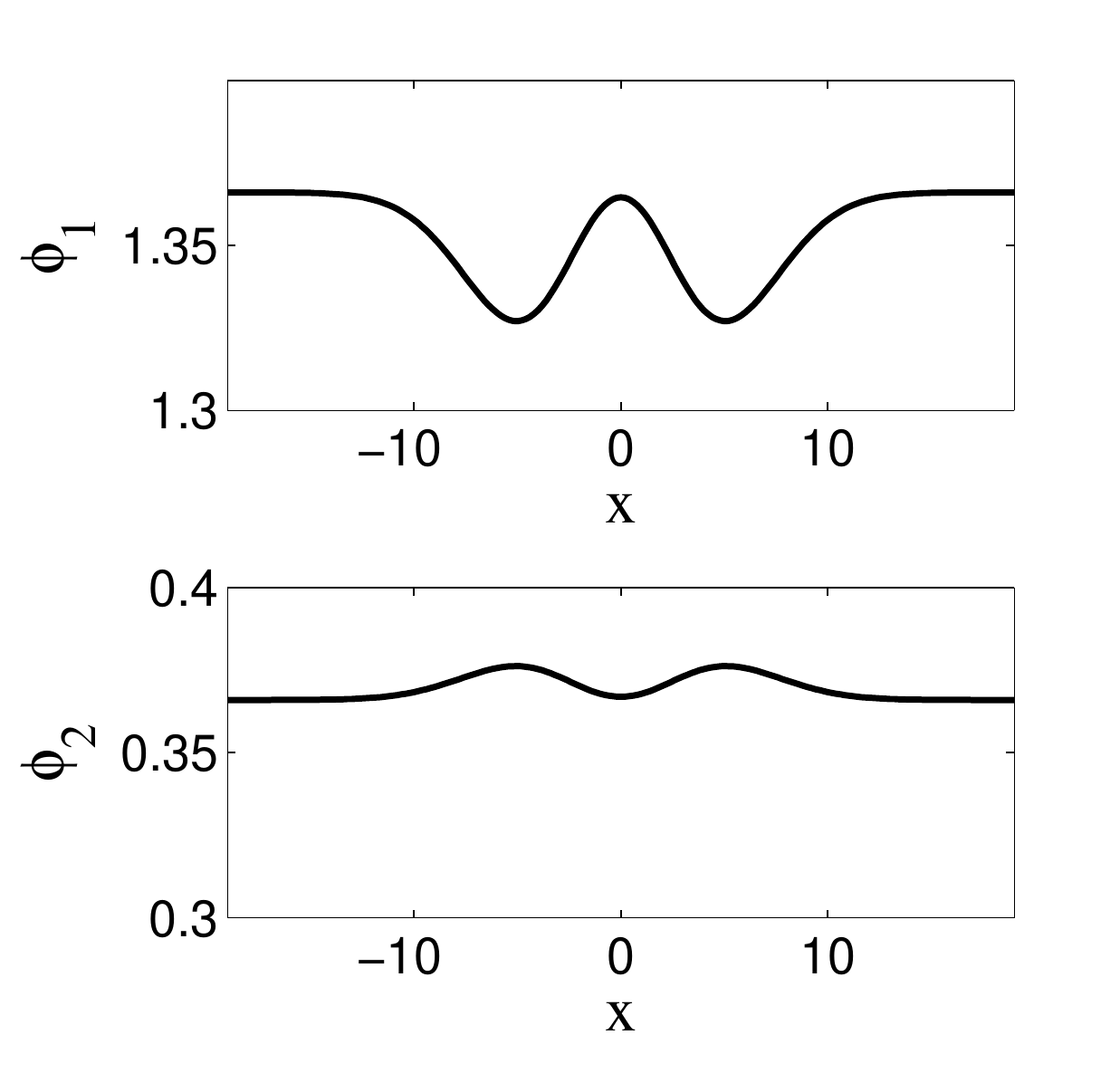}
\label{Phi12_2lePeaks-c}} 
\subfigure[]{\includegraphics[width=2.3in]{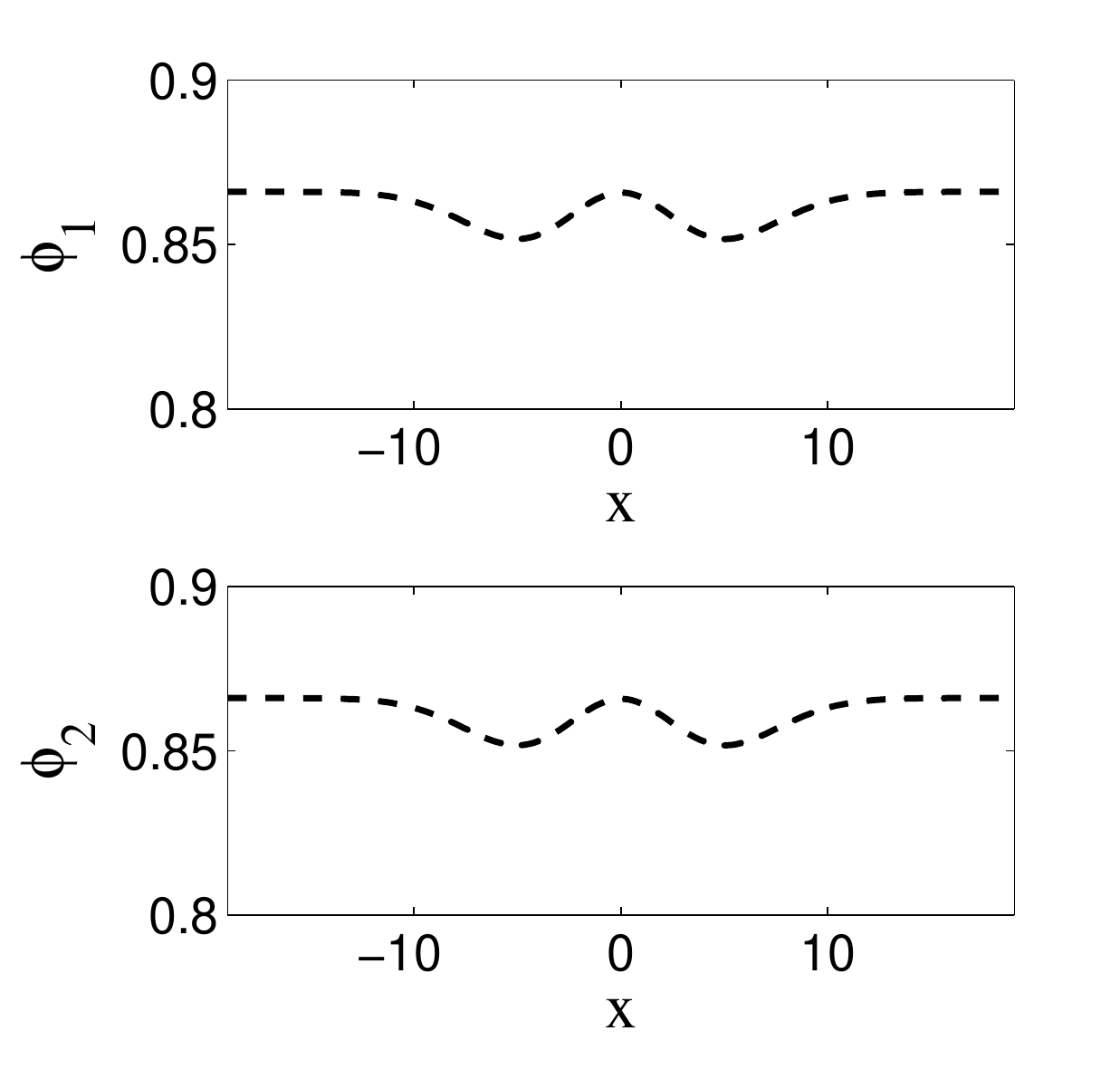}
\label{Phi12_2lePeaks-d}}
\caption{Typical examples of patterns supported by the double-barrier
potential (\protect\ref{2PeakPotential}), with $W_{0}=0.1$ and $L=10$,
constants $g$, $\protect\sigma $, $\protect\kappa $ and $\protect\mu $ being
the same as in Fig.~\protect\ref{Evolution_NoPotential}. Panels (a), (b) and
(c) demonstrate the stable patterns of the following types, in terms of Eq. (%
\protect\ref{types-As}): $\left\{ \left( A_{1},A_{2}\right) ,\left(
A_{2},A_{1}\right) ,\left( A_{1},A_{2}\right) \right\} $, $\left\{ \left(
A_{1},A_{2}\right) ,\left( A_{1},A_{2}\right) ,\left( A_{2},A_{1}\right)
\right\} $, and $\left\{ \left( A_{1},A_{2}\right) ,\left(
A_{1},A_{2}\right) ,\left( A_{1},A_{2}\right) \right\} $, respectively. The
unstable pattern of type $\left\{ \left( A_{0},A_{0}\right) ,\left(
A_{0},A_{0}\right) ,\left( A_{0},A_{0}\right) \right\} $, from set (\protect
\ref{types-S}), is presented in panel (d).}
\label{Phi12_2lePeaks}
\end{figure}

Using various combinations of the asymmetric CW states, we have built the
following patterns supported by the double-peak potential, cf. Eqs. (\ref%
{types}) and (\ref{types-S}):
\begin{gather}
\left\{ \left( A_{1},A_{2}\right) ,\left( A_{1},A_{2}\right) ,\left(
A_{1},A_{2}\right) \right\} ;\left\{ \left( A_{1},A_{2}\right) ,\left(
A_{1},A_{2}\right) ,\left( A_{2},A_{1}\right) \right\} ;  \notag \\
\left\{ \left( A_{1},A_{2}\right) ,\left( A_{2},A_{1}\right) ,\left(
A_{1},A_{2}\right) \right\} ;  \label{types-As}
\end{gather}%
\begin{gather}
\left\{ \left( A_{1},A_{2}\right) ,\left( A_{1},A_{2}\right) ,\left(
-A_{1},-A_{2}\right) \right\} ;~\left\{ \left( A_{1},A_{2}\right) ,\left(
-A_{1},-A_{2}\right) ,\left( A_{1},A_{2}\right) \right\} ;  \notag \\
\left\{ \left( A_{1},A_{2}\right) ,\left( A_{1},A_{2}\right) ,\left(
-A_{2},-A_{1}\right) \right\} ;~\left\{ \left( -A_{1},-A_{2}\right) ,\left(
A_{1},A_{2}\right) ,\left( A_{2},A_{1}\right) \right\} ;  \notag \\
\left\{ \left( A_{1},A_{2}\right) ,\left( -A_{1},-A_{2}\right) ,\left(
A_{2},A_{1}\right) \right\} ;\left\{ \left( A_{1},A_{2}\right) ,\left(
A_{2},A_{1}\right) ,\left( -A_{1},-A_{2}\right) \right\} ;  \notag \\
~\left\{ \left( A_{1},A_{2}\right) ,\left( -A_{2},-A_{1}\right) ,\left(
A_{1},A_{2}\right) \right\} .  \label{types-As2}
\end{gather}%
The three patterns (\ref{types-As}) feature, respectively, none, one, or two
trapped DWs, while seven patterns (\ref{types-As2}) include dark solitons or
zero-crossing (sign-changing) DWs. As before, those patterns which do not
cross zero in any component are stable [in the case of the first pattern in (%
\ref{types-S}), this is, naturally, true prior to the SBB], while all the
solutions featuring at least one zero crossing are unstable. Examples of the
patterns of these types are presented in Fig.~\ref{Phi12_2lePeaks}, for $%
W_{0}=0.1$ and $L=10$.

\begin{figure}[tbp]
\subfigure[]{
\begin{minipage}{2.25in}
\includegraphics[width=2.25in]{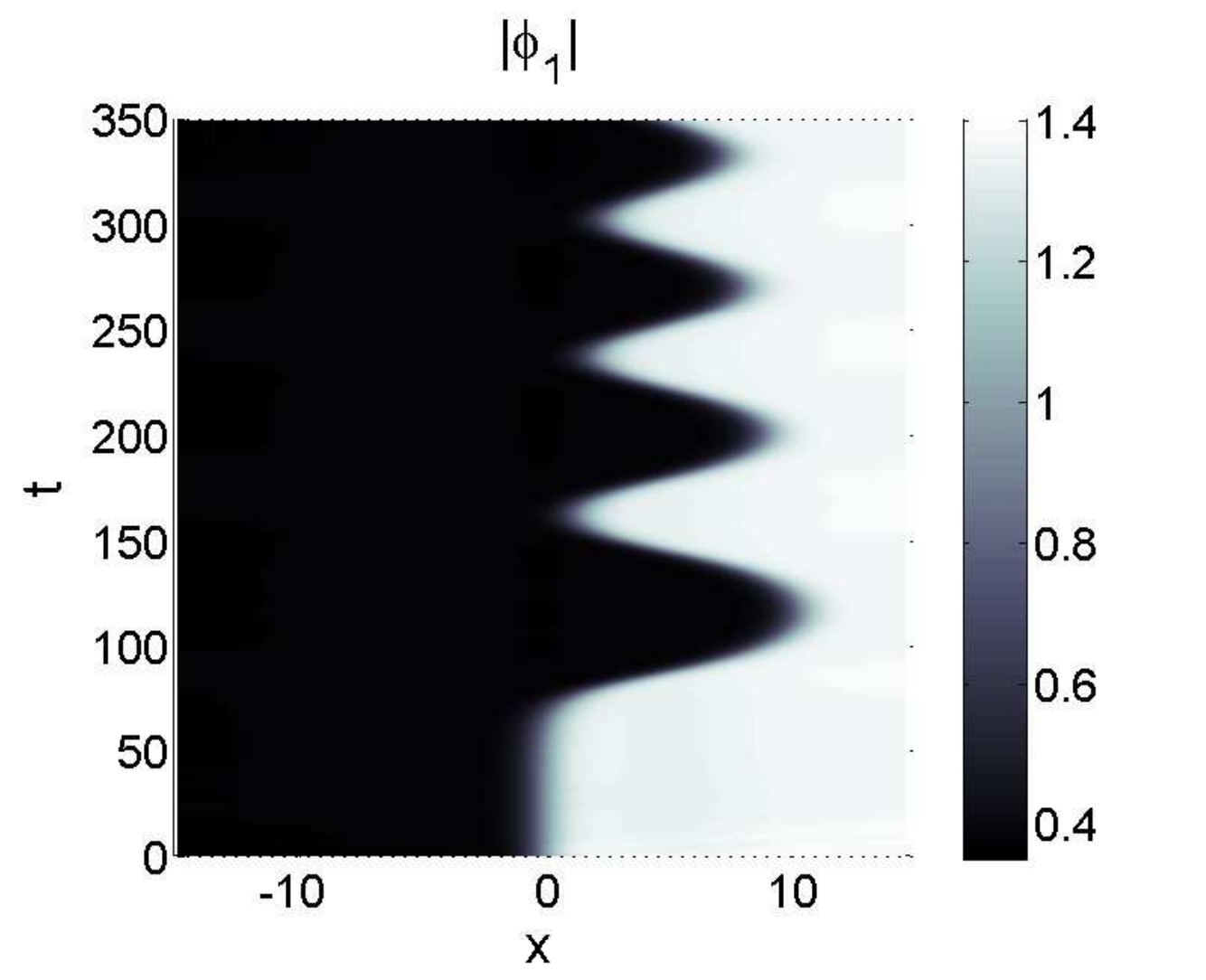}\\
\begin{minipage}{2.25in}
\includegraphics[width=2.25in]{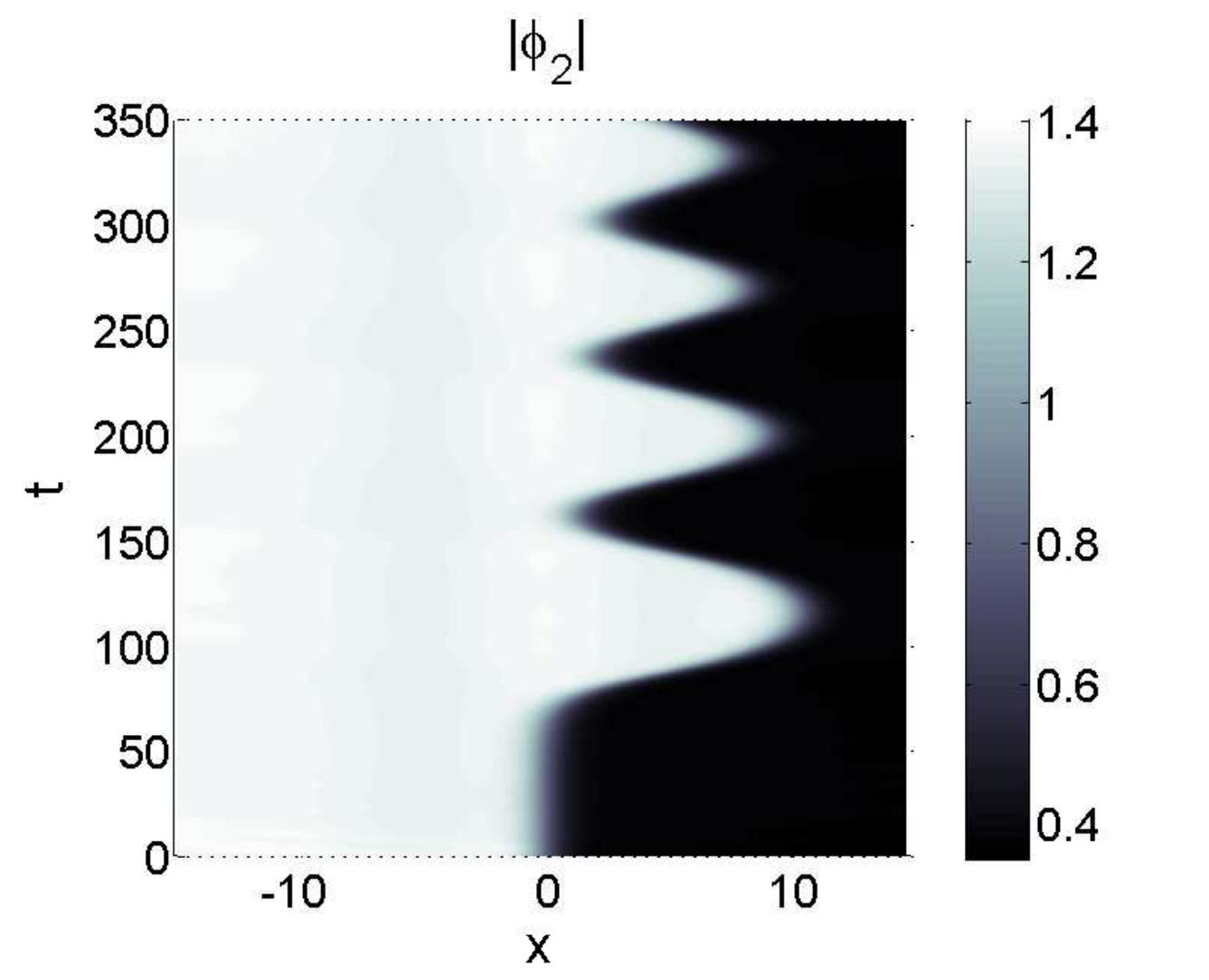}
\label{Evolution_2Peaks_1DW}
\end{minipage}
\end{minipage}} 
\subfigure[]{
\begin{minipage}{2.25in}
\includegraphics[width=2.25in]{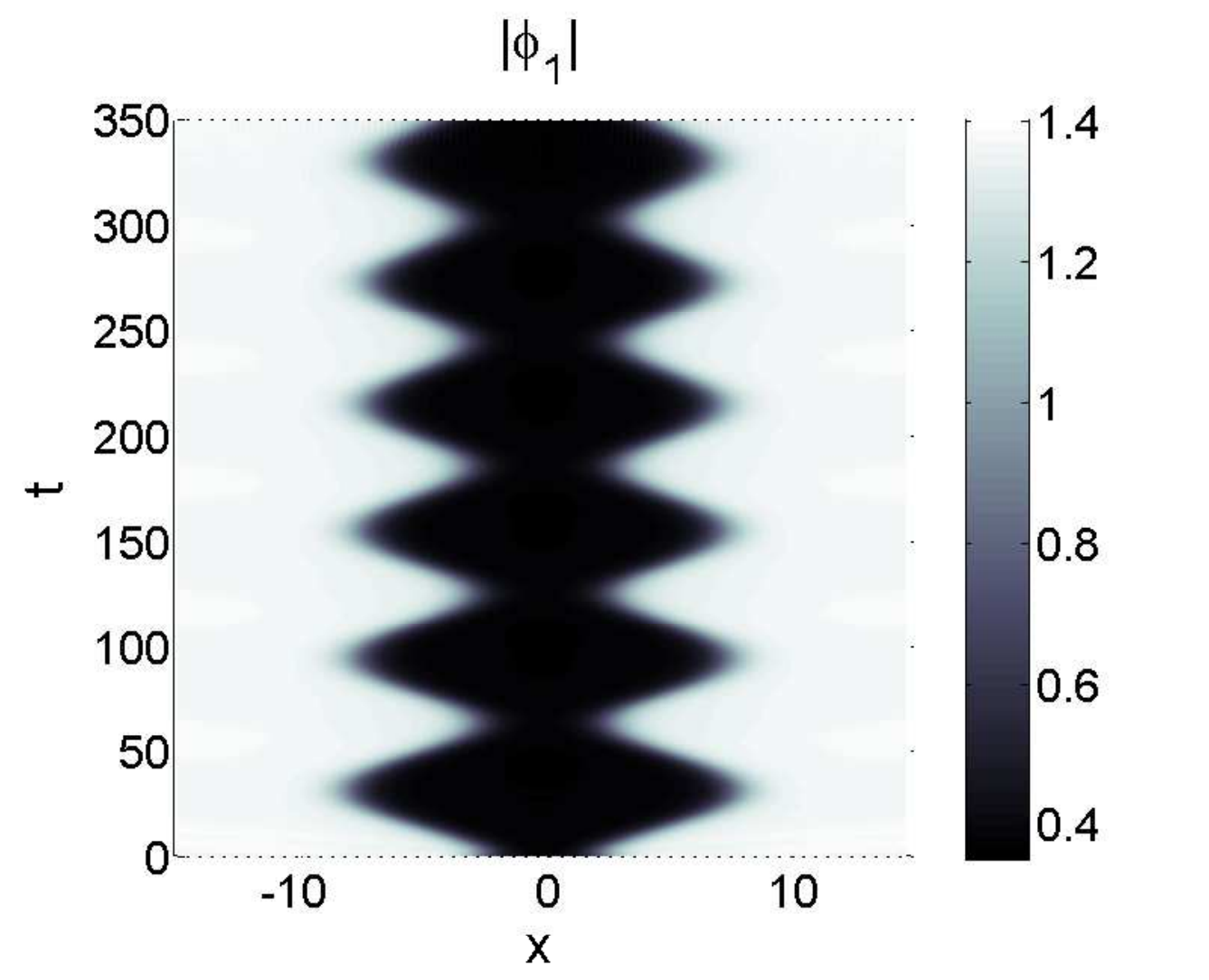}\\
\begin{minipage}{2.25in}
\includegraphics[width=2.25in]{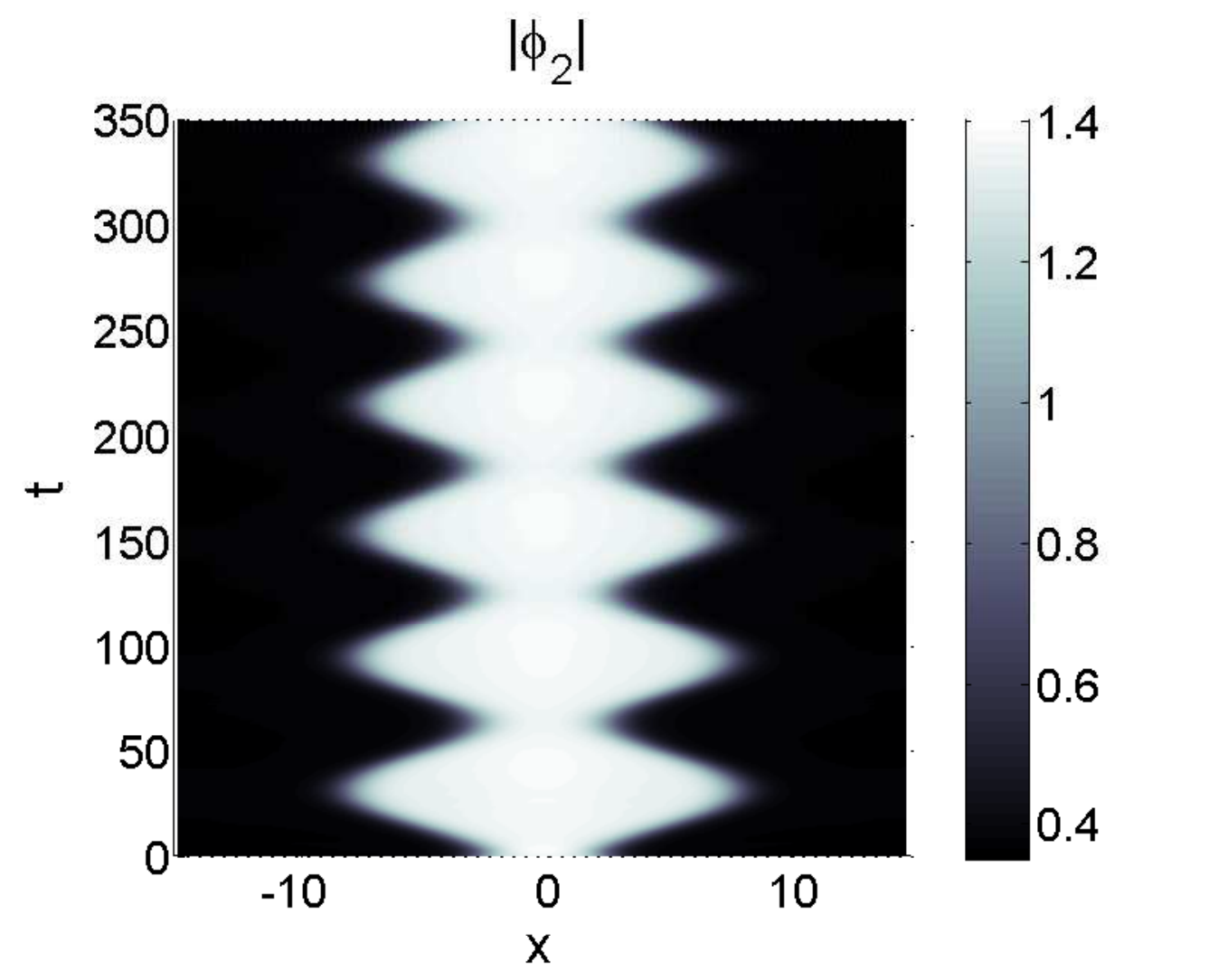}
\label{Evolution_2Peaks_1DW_L5}
\end{minipage}
\end{minipage}} 
\subfigure[]{
\begin{minipage}{2.25in}
\includegraphics[width=2.25in]{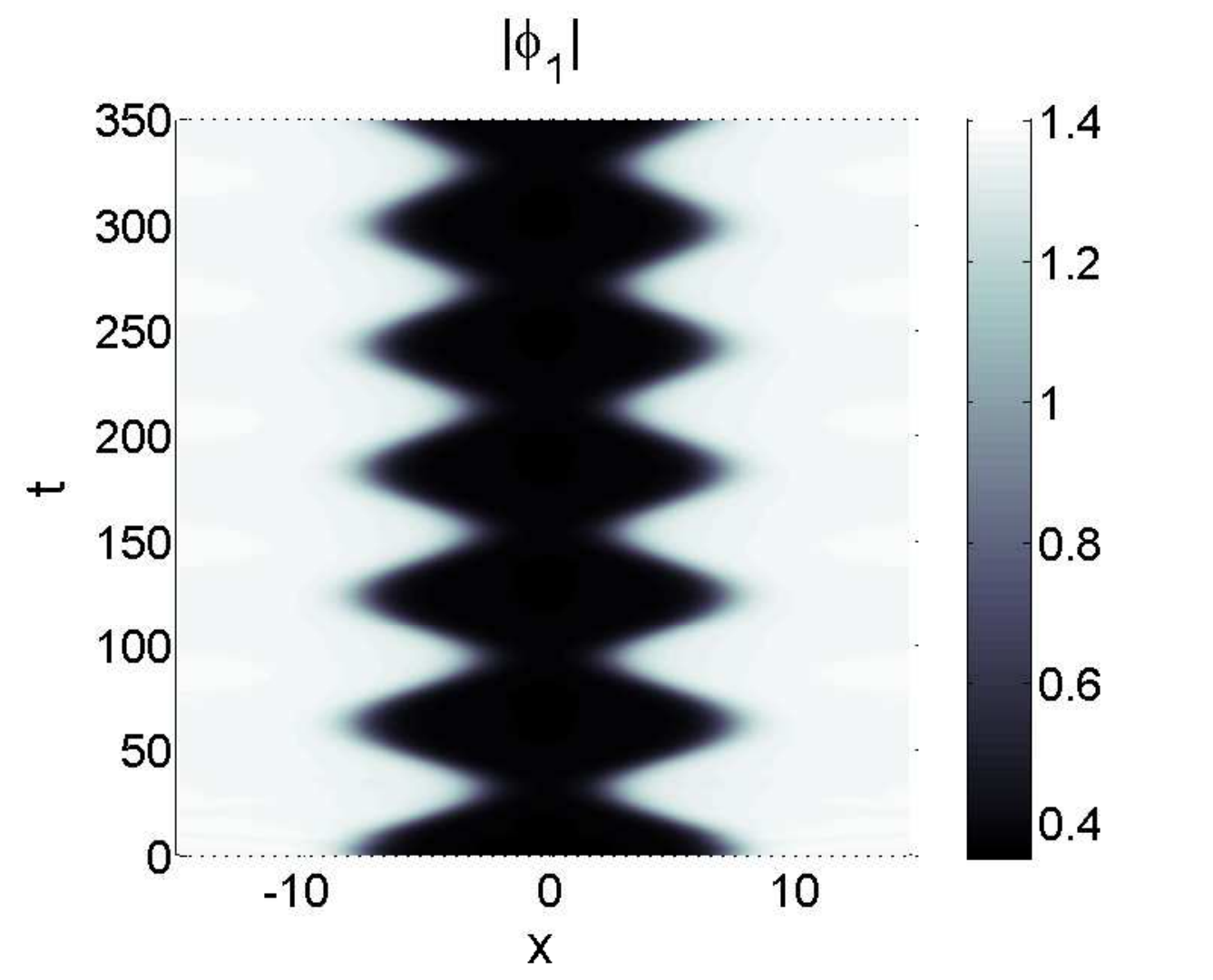}\\
\begin{minipage}{2.25in}
\includegraphics[width=2.25in]{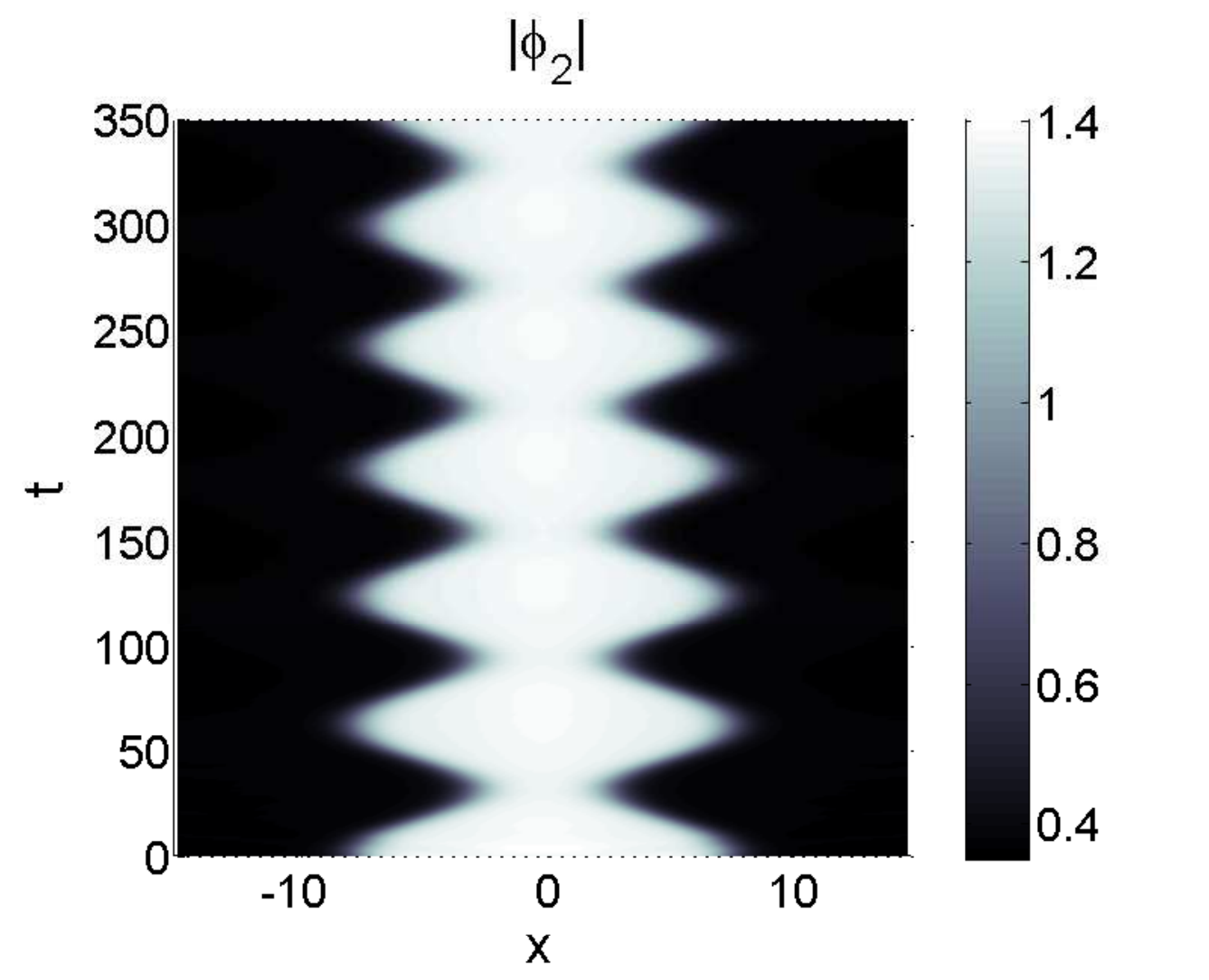}
\label{Evolution_2Peaks_1DW_L15}
\end{minipage}
\end{minipage}}
\caption{The evolution of a single DW and pairs of DWs in the
presence of the double-peak external potential, (\protect\ref{2PeakPotential}%
), with $W_{0}=0.1$ and $L=10$. The other parameters are as in Figs.~\protect
\ref{Evolution_NoPotential} and \protect\ref{Phi12_2lePeaks}. In panel (a),
the single DW is initially positioned exactly between the two peaks. The
evolution of pairs of DWs, symmetrically placed at $x=\pm 2.5$ and $x=\pm
7.5 $, is displayed in panels (b) and (c), respectively.}
\label{Evolution_TwoPeaks}
\end{figure}

The attractive interaction of the DWs with the potential peaks is
additionally illustrated by simulations displayed in Fig. \ \ref%
{Evolution_TwoPeaks}. In particular, if a single DW is initially placed at
the midpoint between the peaks, the evolution of this obviously unstable
configuration leads to damped oscillations of the DW around either peak to
which it is pulled, see Fig.~\ref{Evolution_2Peaks_1DW}. On the other hand,
symmetrically placing pairs of identical DWs either between (Fig.~\ref%
{Evolution_2Peaks_1DW_L5})\ or outside (Fig.~\ref{Evolution_2Peaks_1DW_L15})
of the potential peaks, we observe coherent oscillations with the apparently
unbroken symmetry. In the latter case, the symmetric oscillations feature
very slow damping, which is explained by the small value of $W_{0}=0.1$ in
this case. Similar simulations for larger $W_{0}$ demonstrate oscillations
which converge much faster to a stable symmetric configuration, with each DW
trapped by one potential peak.

\section{Vortices and two-dimensional circular domain walls}

\label{sec:2DModel} The 2D version of model (\ref{GPE}) is
\begin{eqnarray}
i(\psi _{1})_{t} &=&-\left( 1/2\right) [(\psi _{1})_{xx}+(\psi
_{1})_{yy}]+\sigma |\psi _{1}|^{2}\psi _{1}+g|\psi _{2}|^{2}\psi _{1}-\kappa
\psi _{2},  \notag \\
i(\psi _{2})_{t} &=&-\left( 1/2\right) [(\psi _{2})_{xx}+(\psi
_{2})_{yy}]+\sigma |\psi _{2}|^{2}\psi _{2}+g|\psi _{1}|^{2}\psi _{2}-\kappa
\psi _{1}.  \label{2DModelPsi1Psi2}
\end{eqnarray}%
In terms of BEC, Eqs. (\ref{2DModelPsi1Psi2}) admit the straightforward
interpretation as the GP equations for the two-component condensate in a 2D
pancake-shaped configuration. In terms of optics, these equations, with $t$
replaced by propagation coordinate $z$, govern the transmission of a
stationary beam through the bulk nonlinear medium, with functions $\psi
_{1,2}$ representing two circular polarizations. In the latter case, the
linear mixing between the polarizations can be induced by the linear
birefringence, which, in turn, may be imposed by mechanical stress applied
perpendicular to the propagation axis \cite{Agrawal}, or, alternatively, by
dc magnetic field applied in the same direction \cite{magneto-optics} (the
birefringence imposed by the transverse magnetic field leads to the
classical Cotton-Mouton/Voigt effects \cite{Boyd}).

General axisymmetric solutions to Eqs. (\ref{2DModelPsi1Psi2}) are looked
for in the form of the optical vortices (alias 2D dark solitons) \cite%
{Swartzlander,Agrawal}, which also correspond to the vortex modes in BEC
\cite{Pit}:%
\begin{equation}
\psi _{1,2}(x,y,t)=\phi _{1,2}(r,t)\exp (is\theta )\exp (-i\mu t ),
\label{AxisymmetricSolutions}
\end{equation}%
where $r$ and $\theta $ are the polar coordinates in the $(x,y)$ plane, and
integer $s$ is the topological charge (vorticity, alias ``spin").

The substitution of expressions (\ref{AxisymmetricSolutions}) into Eqs. (\ref%
{2DModelPsi1Psi2}) leads to the equations for radial wave functions $\phi
_{1,2}(r,t)$:
\begin{gather}
i(\phi _{1})_{t}+\mu \phi _{1}+(1/2)\left[ (\phi _{1})_{rr}+r^{-1}(\phi
_{1})_{r}-s^{2}r^{-2}\phi _{1}\right]  \notag \\
-\sigma (\phi _{1})^{3}-g(\phi _{2})^{2}\phi _{1}+\kappa \phi _{2}=0,  \notag
\\
i(\phi _{2})_{t}+\mu \phi _{2}+(1/2)\left[ (\phi _{2})_{rr}+r^{-1}(\phi
_{2})_{r}-s^{2}r^{-2}\phi _{2}\right]  \notag \\
-\sigma (\phi _{2})^{3}-g(\phi _{1})^{2}\phi _{2}+\kappa \phi _{1}=0.
\label{2DModelPhi1Phi2}
\end{gather}%
Obviously, stationary CW solutions in the 2D model are identical to those
obtained for the 1D model, see Eqs. (\ref{symm}) and (\ref{phi12_asymm}). In
particular, the bifurcation described by Eqs. (\ref{mu-crit}) and (\ref{crit}%
) is relevant in the 2D case too.

The study of the existence, stability and dynamics of 2D axisymmetric
patterns, generated by Eqs. (\ref{2DModelPhi1Phi2}), is presented below in
two steps. First, stationary vortices, supported by the asymmetric CW
background, are obtained for spins $s=0,1,2$ and $3$, and their stability is
determined. Then, the evolution of the \textit{pulsons}, i.e., circular DWs
oscillating in the radial direction,\textit{\ }is investigated by means of
direct simulations. In this connection, it is relevant to mention that
patterns in the form of circular DWs are well known in various magnetic
media \cite{circularDW}

\subsection{Stationary vortices}

\label{sec:StationarySolutions} 
\begin{figure}[tbp]
\subfigure[]{\includegraphics[width=2.2in]{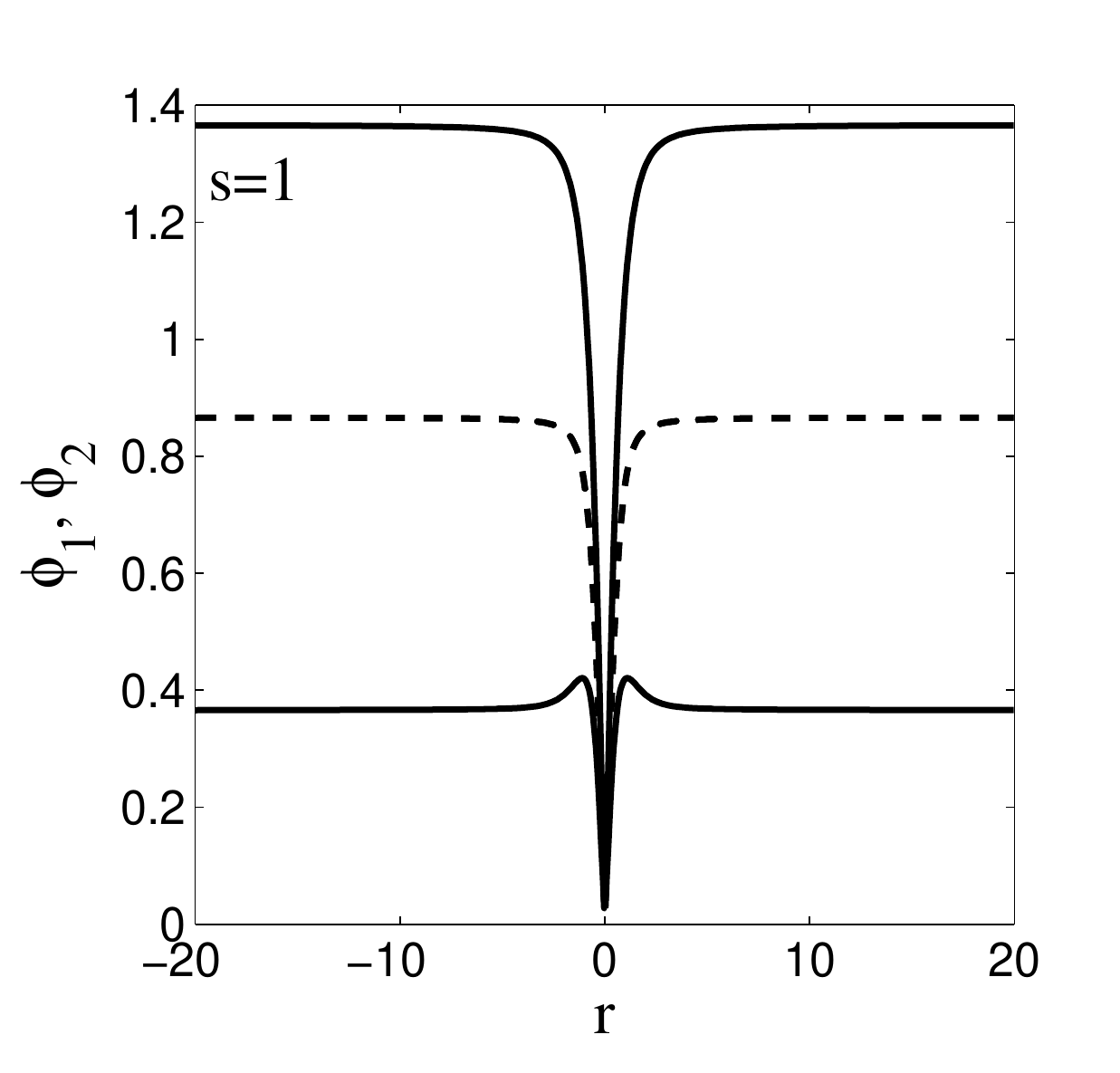}}
\subfigure[]{\includegraphics[width=2.2in]{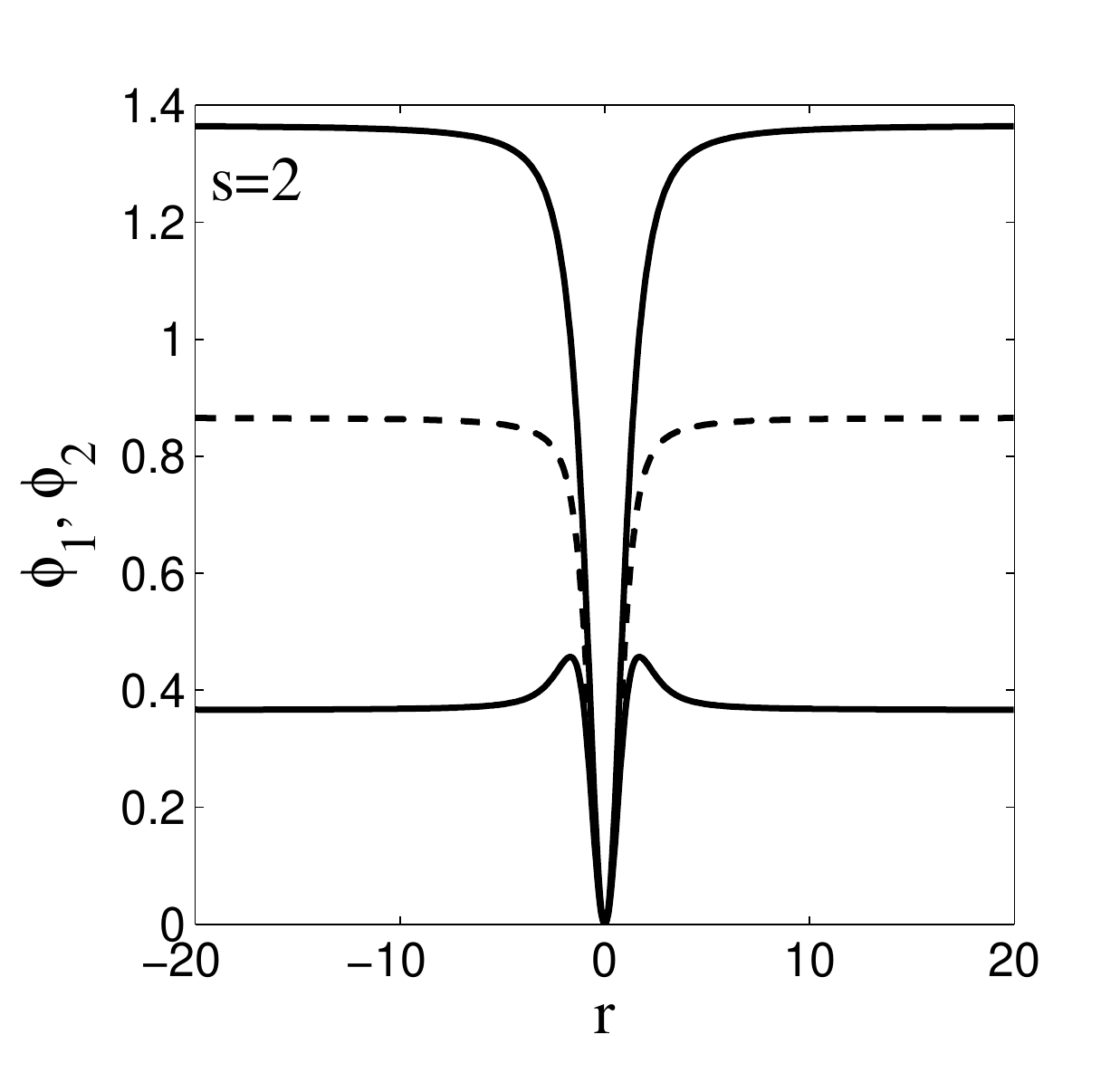}}
\subfigure[]{\includegraphics[width=2.2in]{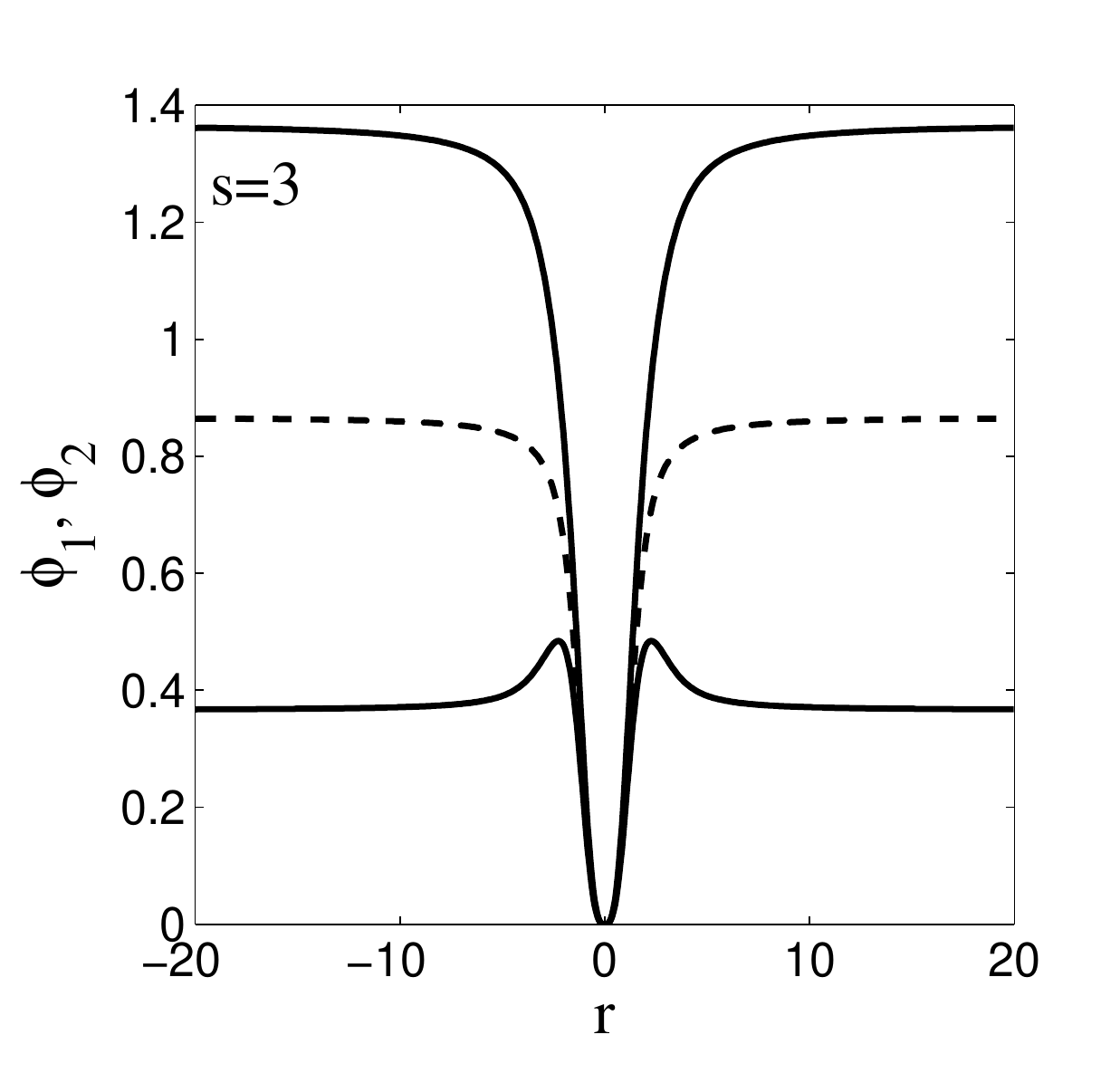}}
\caption{(a,b,c): Radial profiles of the symmetric and asymmetric stationary
vortices, with spin $s=1,~2$, and $3$, respectively (no non-flat stationary
2D solutions were found for $s=0$). The parameters are $g=3,\protect\sigma %
=1,\protect\kappa =1$, and $\protect\mu =2$. Solid and dashed curves depict,
respectively, two components of the stable asymmetric solutions, and
unstable symmetric ones.}
\label{Stationary_s123}
\end{figure}

To find stationary solutions of Eqs. (\ref%
{2DModelPhi1Phi2}), we proceed to the time-independent version of these
equations and apply the Newton-Raphson method to the respective nonlinear
boundary-value problem. The boundary conditions demand that $\phi
_{1,2}(r)\rightarrow r^{s}$ at $r\rightarrow 0$, while, at $r\rightarrow
\infty $, $\phi _{1,2}(r)$ asymptotically approach the symmetric or
asymmetric CW solutions, (\ref{symm}) or (\ref{phi12_asymm}). Examples of
the radial profiles of such modes are displayed in Fig.~\ref{Stationary_s123}%
. For the $s=0$, no stationary solutions exist, apart from the obvious flat
states (which are not shown in Fig.~\ref{Stationary_s123}). For each
non-zero value of the spin that was examined, $s=1,2$ and $3$, two families
of nontrivial stationary solutions were obtained, supported [past
bifurcation point (\ref{mu-crit})] by the symmetric and asymmetric CW states
at $r\rightarrow \infty $, hence they demonstrate precisely the same
existence and bifurcation features as the corresponding flat states, (\ref%
{symm}) and (\ref{phi12_asymm}).

To explore the stability of the 2D stationary solutions, we seek for
perturbed solutions as [cf. Eqs. (\ref{Perturbed_solution})]
\begin{eqnarray}
\widetilde{\phi _{1}}(r,\theta ,t) &=&U(r)+\left[ U_{+}(r)\exp (in\theta
)+U_{-}(r)\exp (-in\theta )\right] \exp (\gamma _{n}t),  \notag \\
\widetilde{\phi _{2}}(r,\theta ,t) &=&V(r)+\left[ V_{+}(r)\exp (in\theta
)+V_{-}(r)\exp (-in\theta )\right] \exp (\gamma _{n}t),
\label{PerturbedSolution}
\end{eqnarray}%
where integer $n>0$ is an arbitrary azimuthal index of the perturbation, and
$\gamma _{n}$ is the corresponding instability growth rate. After
substituting expressions (\ref{PerturbedSolution}) into Eqs. (\ref%
{2DModelPhi1Phi2}) and linearizing, the following system of equations is
obtained:
\begin{gather}
\mu U_{+}+i\gamma _{n}U_{+}+(1/2)U_{+}^{\prime \prime }+\left( 2r\right)
^{-1}U_{+}^{\prime }-(s+n)^{2}\left( 2r^{2}\right) ^{-1}U_{+}  \notag \\
-gUV(V_{-}^{\ast }+V_{+})-\sigma U^{2}(U_{-}^{\ast
}+2U_{+})-gV^{2}U_{+}+\kappa V_{+}=0;  \notag \\
\mu U_{-}+i\gamma _{n}U_{-}+(1/2)U_{-}^{\prime \prime }+\left( 2r\right)
^{-1}U_{-}^{\prime }-(s-n)^{2}\left( 2r^{2}\right) ^{-1}U_{-}  \notag \\
-gUV(V_{+}^{\ast }+V_{-})-\sigma U^{2}(U_{+}^{\ast
}+2U_{-})-gV^{2}U_{-}+\kappa V_{-}=0;  \notag \\
\mu V_{+}+i\gamma _{n}V_{+}+(1/2)V_{+}^{\prime \prime }+\left( 2r\right)
^{-1}V_{+}^{\prime }-(s+n)^{2}\left( 2r^{2}\right) ^{-1}V_{+}  \notag \\
-gUV(U_{-}^{\ast }+U_{+})-\sigma V^{2}(V_{-}^{\ast
}+2V_{+})-gU^{2}U_{+}+\kappa U_{+}=0;  \notag \\
\mu V_{-}+i\gamma _{n}V_{-}+(1/2)V_{-}^{\prime \prime }+\left( 2r\right)
^{-1}V_{-}^{\prime }-(s-n)^{2}\left( 2r^{2}\right) ^{-1}V_{-}  \notag \\
-gUV(U_{+}^{\ast }+U_{-})-\sigma V^{2}(V_{+}^{\ast
}+2V_{-})-gU^{2}V_{-}+\kappa U_{-}=0,  \label{UVpmEquations}
\end{gather}%
where the prime stands for $d/dr$. We treat Eqs. (\ref{UVpmEquations}) as an
algebraic eigenvalue problem for $\gamma _{n}$ and solve it directly, using
a finite-difference method. The largest instability-growth rate is
identified as the real part of the most unstable eigenvalue, $\max \{\mathrm{%
Re}(\gamma _{n})\}$. Following this approach, we have confirmed that the
only source of the destabilization is the transition between the symmetric
and asymmetric modes, which, as said above, is actually driven by the SBB in
the flat background at $r\rightarrow \infty $. In particular, no azimuthal
instability, that would break the axial symmetry, was found for $s=1,2,3$,
for all integer values of $n$ considered. The absence of the azimuthal
instability in the case of the self-defocusing nonlinearity is actually a
natural feature \cite{review}. In addition, this finding implies that the
flat (quasi-one-dimensional) DW is stable against corrugations in the 2D
setting.

\subsection{Direct simulations -- shrinking domain walls and pulsons}

Another aspect of the 2D model based on equations (\ref{2DModelPsi1Psi2})
was examined by investigating the development of circular DWs into pulsons,
i.e., annular grain boundaries periodically shrinking and expanding in the
radial directions. It is known that, while exact solutions for such pulsons
do not exist, in some models, such as the 2D single-component sine-Gordon
equation, the pulsons may be remarkably stable, featuring hundreds \cite%
{Geicke0,eccentric} and thousands \cite{Geicke} of radial pulsations with
very little loss.

We have performed simulations of pulsating radial DWs for three values of
the spin, $s=0,1$ and $2$. The simulations were implemented by means of the
linearized Crank-Nicolson scheme. As initial conditions, the following DW
configuration, based on the asymmetric uniform states (\ref{phi12_asymm}),
was used,
\begin{eqnarray}
\phi _{1}(r,t &=&0)=(1/2)\tanh ^{s}(r)\left\{ A_{1}[1+\tanh
(r-R_{0})]+A_{2}[1-\tanh (r-R_{0})]\right\} ,  \notag \\
\phi _{2}(r,t &=&0)=(1/2)\tanh ^{s}(r)\left\{ A_{1}[1-\tanh
(r-R_{0})]+A_{2}[1+\tanh (r-R_{0})]\right\} ,  \label{RadialInitialCon}
\end{eqnarray}%
where $R_{0}$ is the initial radius of the DW. Note that this initial ansatz
complies with the necessary boundary condition at $r\rightarrow 0$, $\phi
_{1,2}(r)\sim r^{s}$. The boundary conditions at the right edge of the
integration interval, $0<r<\rho $, were adopted as $\left( \partial \phi
_{1,2}/\partial r+\partial \phi _{1,2}/\partial t\right) |_{r=\rho }=0$.
These conditions prevent the reflection of the emitted radiation from the
boundary.

For each value of the spin considered here, $s=0,1,2$, the simulations were
run for several values of the initial radius, $R_{0}$. Typical examples of
the initial stage of the observed evolution are demonstrated in Fig.~\ref%
{Initial_Evolution_S012}, for $R_{0}=20$ and for the parameters $g=3$, $%
\sigma =1$, $\kappa =1$ and $\mu =2$. Similar results were obtained for
other values of $R_{0}$.
\begin{figure}[tbp]
\subfigure[]{\includegraphics[width=2.3in]{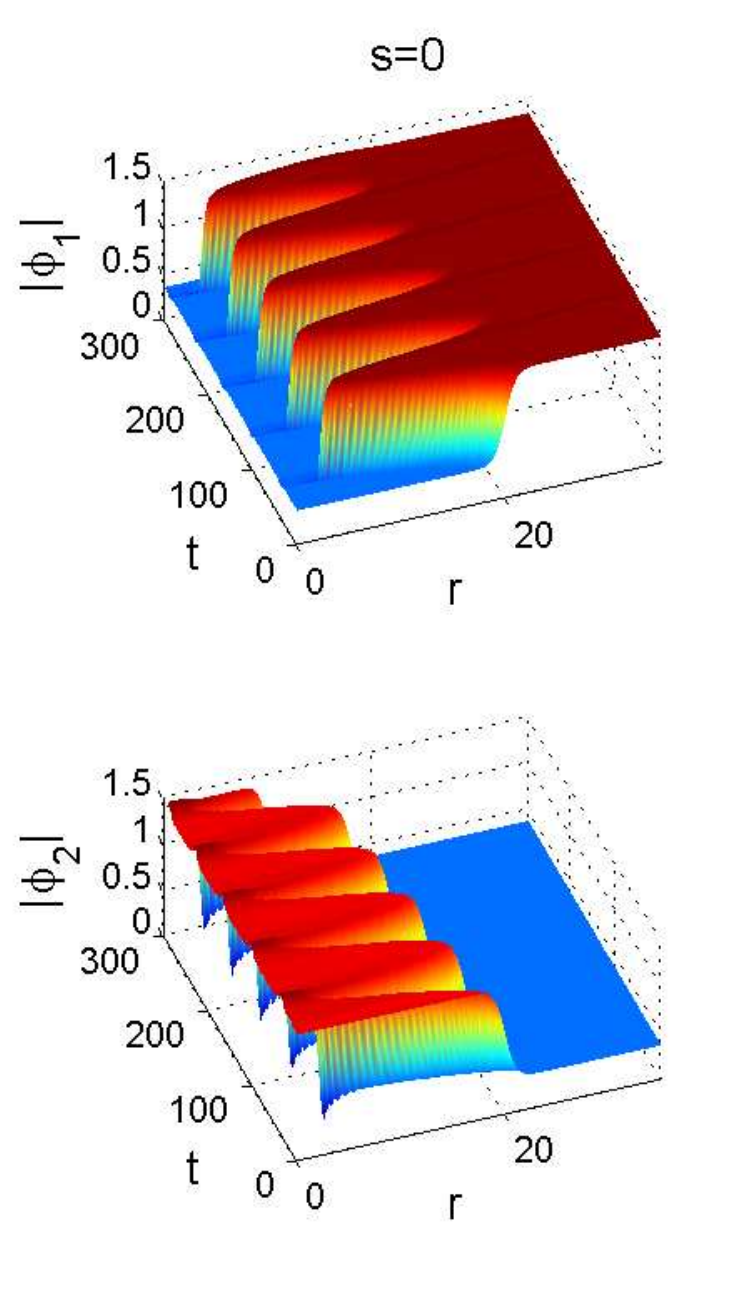}} 
\subfigure[]{\includegraphics[width=2.3in]{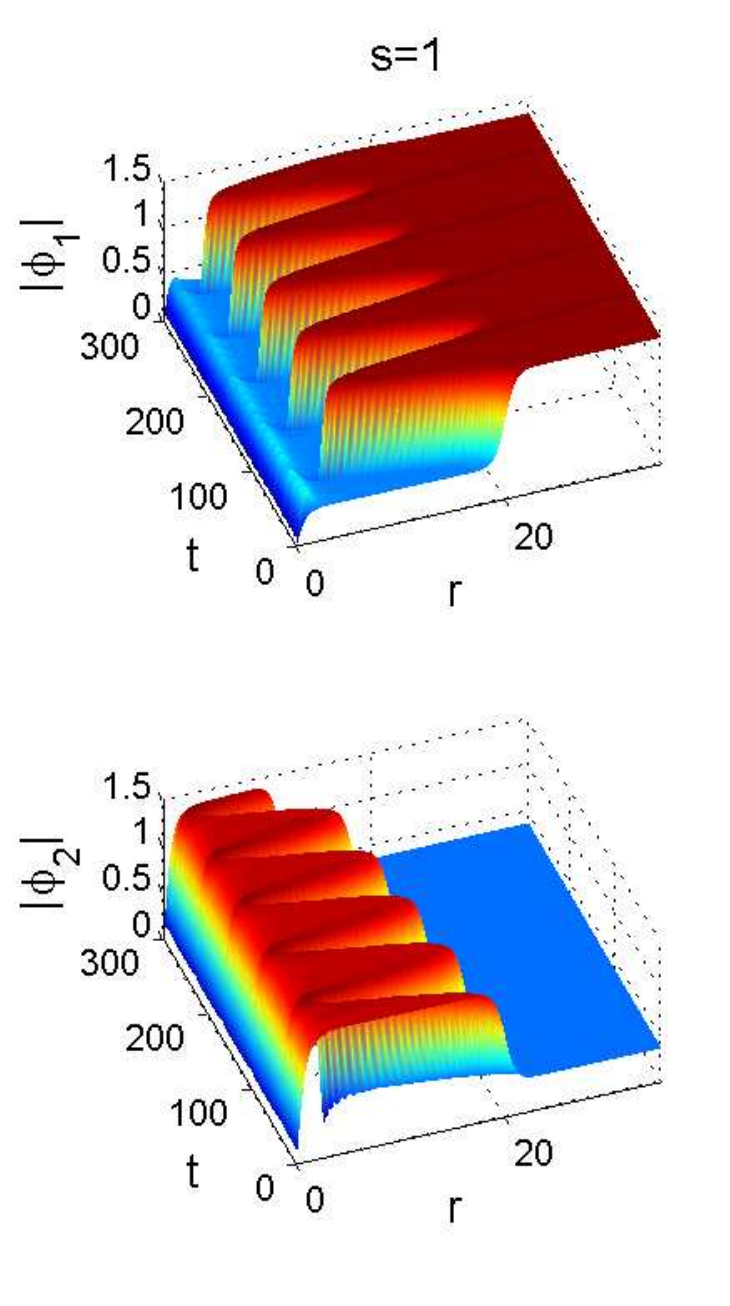}} 
\subfigure[]{\includegraphics[width=2.3in]{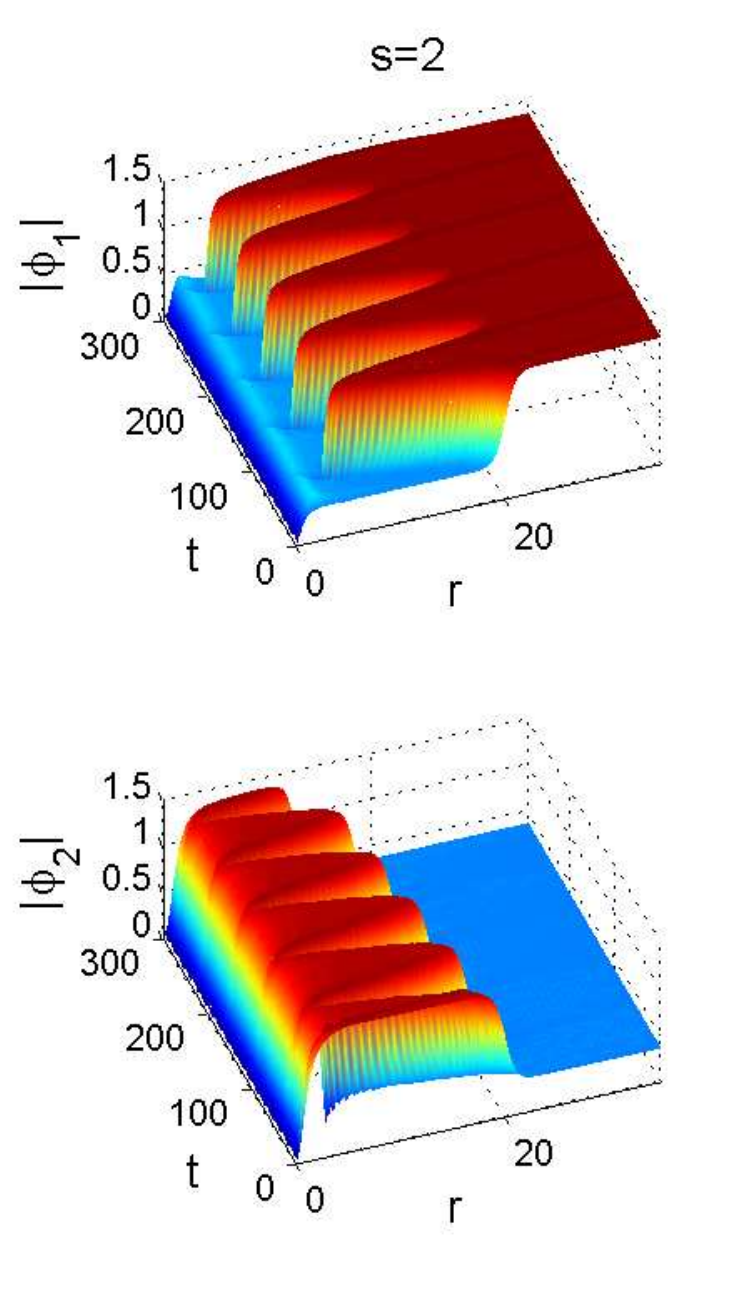}}
\caption{The initial evolution of the pulsons (circular domain walls) for $%
s=0$ (a), $s=1$ (b), and $s=2$ (c). The initial radius of the circular DW is
$R_{0}=20$, the other parameters being $g=3$, $\protect\sigma =1$, $\protect%
\kappa =1$ and $\protect\mu =2$.}
\label{Initial_Evolution_S012}
\end{figure}

\begin{figure}[tbp]
{\includegraphics[width=3in]{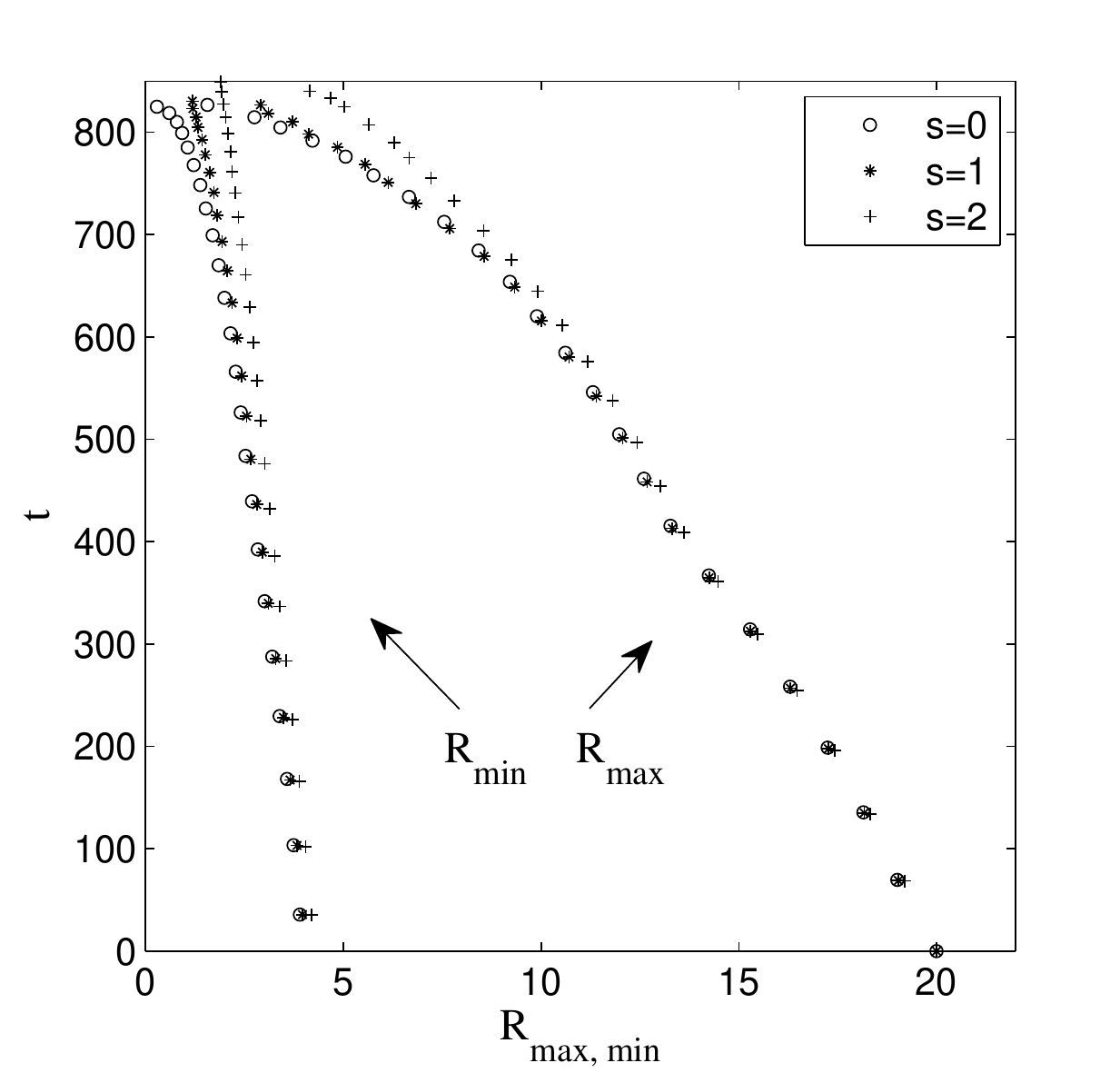}}
\caption{The temporal decay of the largest and smallest radii of the pulson,
in the cases shown in Fig.~\protect\ref{Initial_Evolution_S012}, for $s=0,1$
and $2$. }
\label{Rminmax_s012}
\end{figure}

It was observed that, while the pulsons shrink and expand in a
quasi-periodic manner, their smallest and largest radii, $R_{\min }$ and $%
R_{\max }$, do not remain constant, slowly decreasing from a cycle to a
cycle, as shown in Fig.~\ref{Rminmax_s012}. In particular, the decay of $%
R_{\max }$ is roughly exponential in time, with some irregularities observed
at $t\simeq 400$ and at the final stage of the evolution. The decay is
plausibly caused by the emission of radiation waves by the pulsating DW.

Also decreasing is the oscillation period, $\tau _{l}$, taken as the time
interval between two consecutive points at which the pulson expands to the
largest radius, $\tau _{l}=t(r=R_{\max ,l})-t(r=R_{\max ,l-1})$, where $l$
is the number of the cycle. The relation between the slowly decreasing
period and the effective maximum radius, $R_{\max \mathrm{,eff}}\equiv (R_{%
\mathrm{max},l}+R_{\mathrm{max},l-1})/2$, is presented in Fig.~\ref%
{TauVsReff_s012}. The plots demonstrate a nearly linear dependence for all $%
s $. The same linear dependence was obtained for different values of the
initial radius, including $R_{0}=10,15,30,$ and $40$.
\begin{figure}[tbp]
{\includegraphics[width=3in]{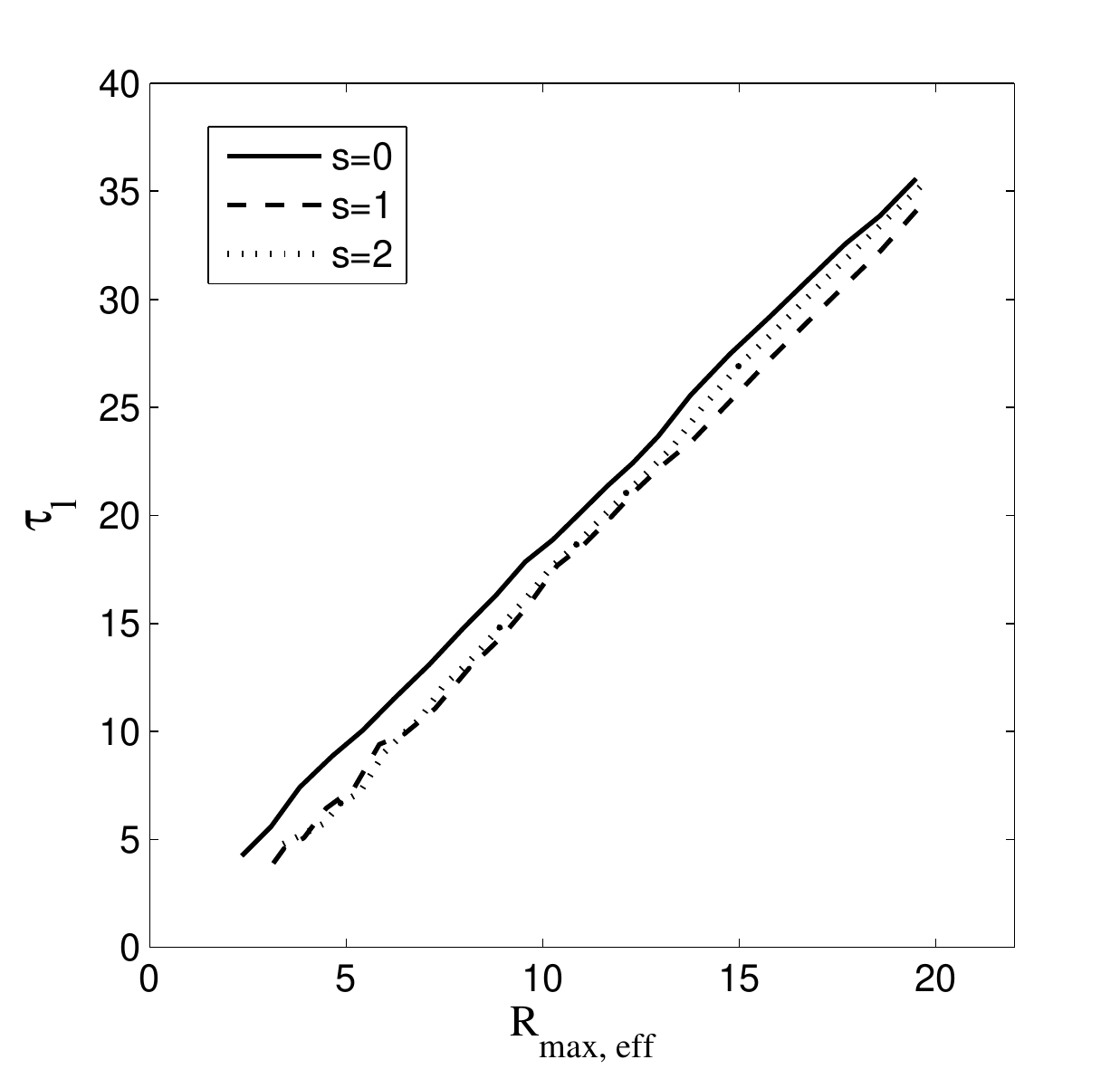}}
\caption{The relation between the gradually decreasing oscillation period of
the pulson, $\protect\tau _{l}$, and the effective maximum radius, $R_{\max
\mathrm{,eff}}$, for $s=0,1$ and $2$. The parameters are the same as in Fig.~%
\protect\ref{Initial_Evolution_S012}.}
\label{TauVsReff_s012}
\end{figure}

The linear relation between $\tau _{l}$ and $R_{\max }$ can be easily
explained. Indeed, considering a large-radius circular DW, with radius $R$
much larger than the thickness of the DW in the radial direction, one can
define the effective mass for the radial pulsations, $M=2\pi Rm$, and the
effective surface-tension energy, $E_{\mathrm{ST}}=2\pi R\alpha $, where $m$
and $\alpha $ are the effective mass and surface-energy densities of the
quasi-one-dimensional DW. Thus, the Newton's equation of motion for the DW
in the radial direction is%
\begin{equation}
\frac{d}{dt}\left( 2\pi mR\frac{dR}{dt}\right) =-\frac{d}{dR}\left( 2\pi
\alpha R\right) ,
\end{equation}%
from where the law of motion follows: $R^{2}(t)=R_{\max }^{2}-\left( \alpha
/m\right) t^{2}$, assuming that the motion starts with the zero initial
velocity and $R=R_{\max }$. According to this result, the shrinking DW ring
will bounce back from the center at the moment of time $\tau _{l}/2=\sqrt{%
\alpha /m}R_{\max }$, which is obviously equal to a half of the period. This
result explains the linear proportionality of $\tau _{l}$ to $R_{\max }$.

\begin{figure}[tbp]
\subfigure[]{\includegraphics[width=2.3in]{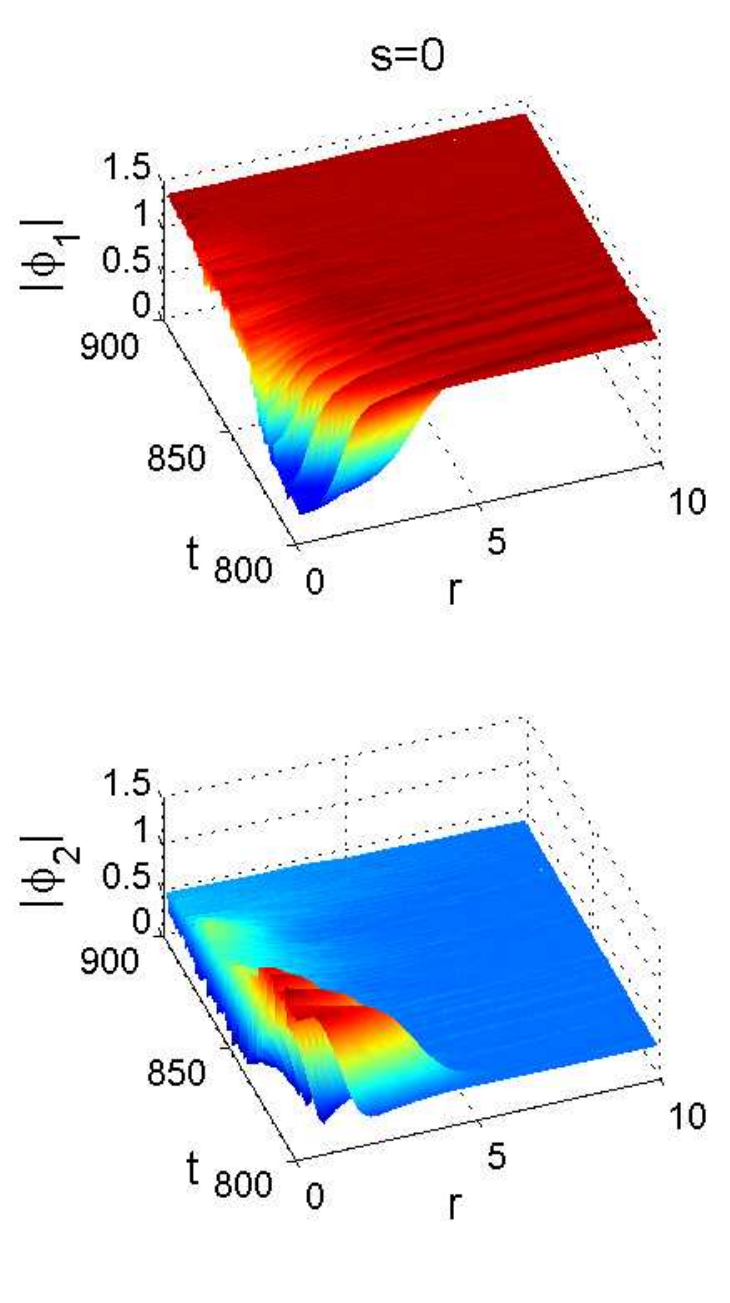}}
\subfigure[]{\includegraphics[width=2.3in]{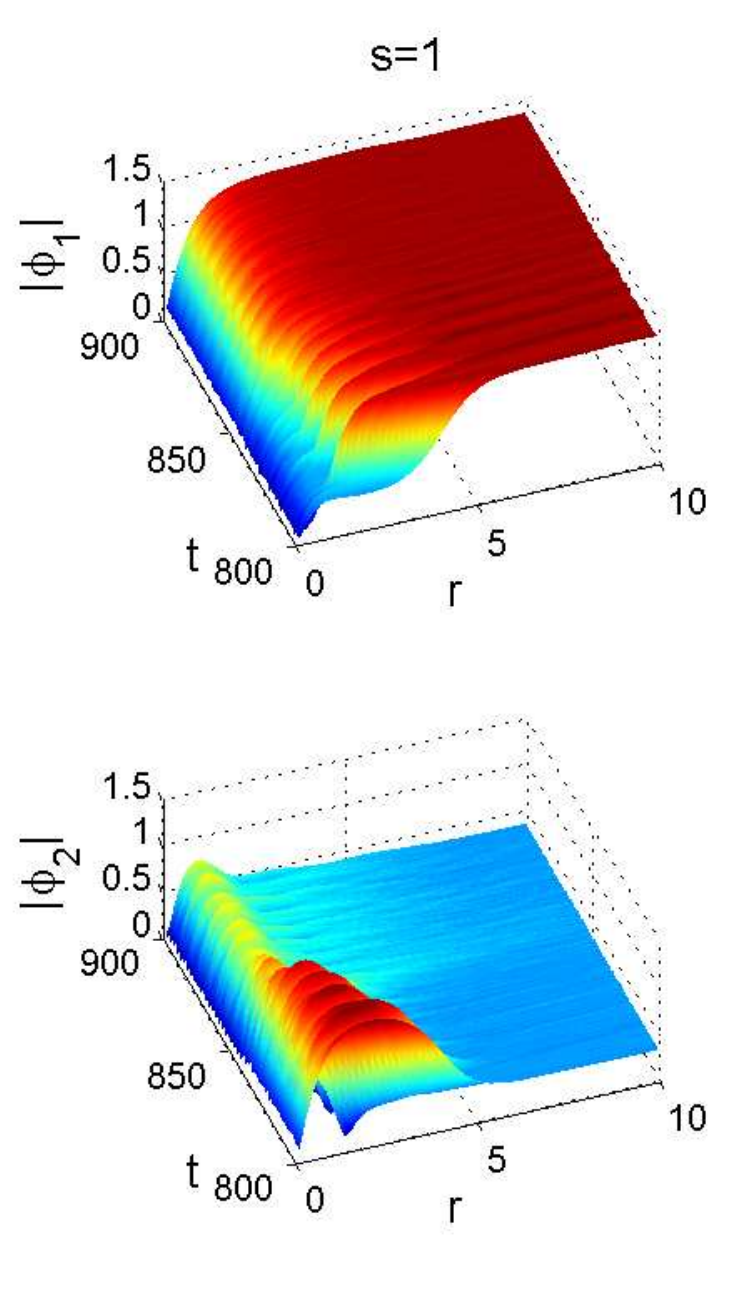}}
\subfigure[]{\includegraphics[width=2.3in]{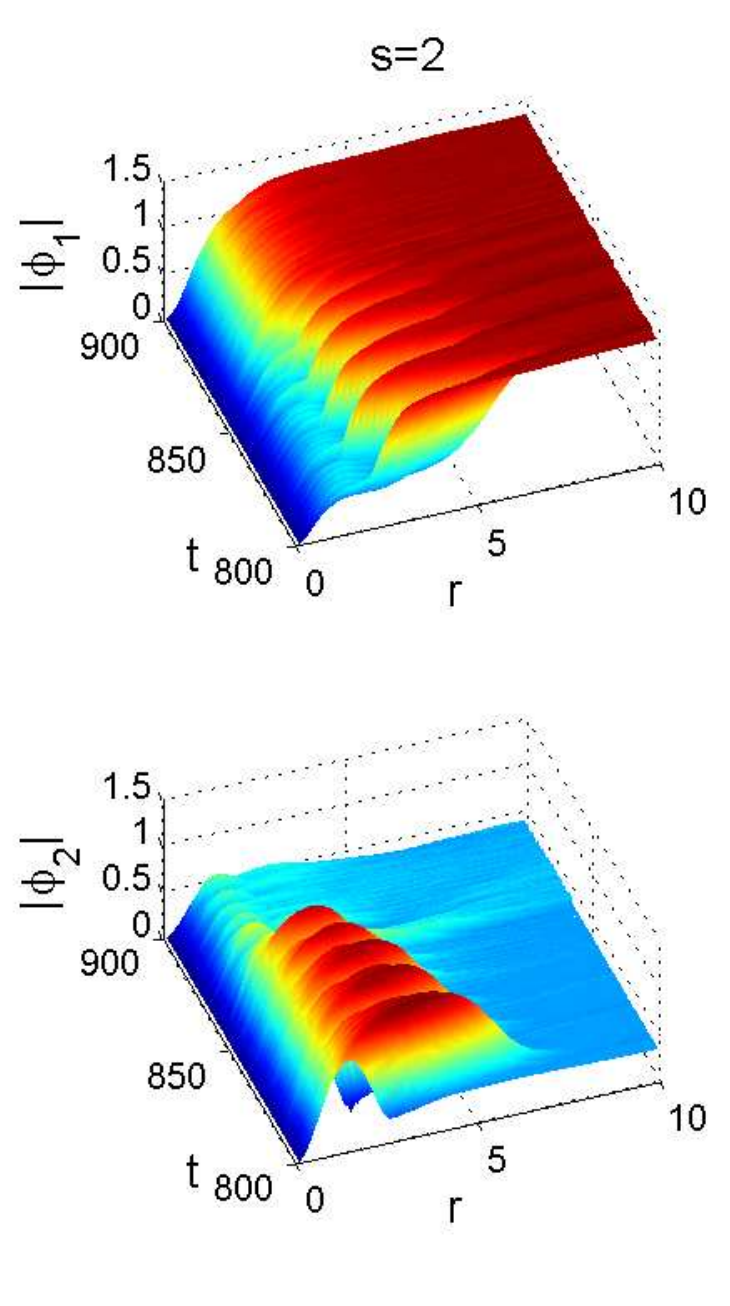}}
\caption{(Color online) The final stage of the evolution of the pulsons with
$s=0,1$ and $2$, whose initial evolution was displayed in Fig.~\protect\ref%
{Initial_Evolution_S012}. The pulsons with $s=1$ and $2$ transform into the
asymmetric stationary (stable) modes shown in Fig.~\protect\ref%
{Stationary_s123}.}
\label{FinalEvolution_s012}
\end{figure}

Finally, Fig.~\ref{FinalEvolution_s012} displays the final stage of the
evolution, observed when the pulson's radius and oscillation period have
been sufficiently reduced. In this case, the oscillatory behavior fades out
and the pulsons transform into the stable asymmetric stationary states which
are presented above in subsection~\ref{sec:StationarySolutions} (in
particular, they are flat for $s=0$, and feature the vortex shape for $s=1$
and $2$). For the case considered above (with $R_{0}=20$), the transition to
the eventual stationary state occurs at $t\approx 830$, for all the values
of the spin.

\section{Conclusions}

\label{sec:Conclusion}

The objective of this work was to investigate fundamental DW
(domain-wall) modes in the system of
nonlinear-Schr\"{o}dinger/Gross-Pitaevskii equations, coupled by the
linear and XPM\ (cross-phase-modulation) terms. The system of
coupled equations has natural realizations in two-component
Bose-Einstein condensate, and in nonlinear optics, where the two
wave functions represent amplitudes of co-propagating signals with
orthogonal polarizations. First, conditions providing for the
stability of the uniform CW (continuous wave) symmetric and
asymmetric bimodal states, which support the DW patterns, were
identified, and then the general families of DW solutions were
constructed in the 1D setting. In particular, approximate analytical
solutions were found near the symmetry-breaking bifurcation point of
the CW states, and an exact solution was found for the XPM/SPM ratio
$3:1$. The DW states connecting asymptotically flat asymmetric
states which are mirror images to each other (without the change of
the overall sign) are stable, while all other types of the DWs,
which feature zero crossings (including dark solitons), are
unstable. Interactions between two DWs with opposite polarities were
also considered. The potential of the attraction between them was
found in the analytical form, and numerical simulations have
demonstrated that the attraction leads to annihilation of the DW
pair.

DWs trapped by the single or double potential peaks were investigated too.
An exact solution for the DW trapped by a single peak was found. In the
general case, it was predicted and corroborated by direct simulations that
the bound state of the DW placed at a maximum of the external potential is
stable. The interaction of two DWs in the presence of the single- and
double-peak potentials was also studied.

The analysis was extended to axisymmetric patterns in the 2D geometry,
including vortices carrying the topological charge $s=1,2,3$ and supported
by the asymmetric flat background at infinity. Stable stationary states for
the vorticity-carrying DW rings were found. The stability of the vortices
against azimuthal perturbations was verified through the computation of the
corresponding eigenvalues. Oscillations of annular DW-shaped pulsons were
studied by means of systematic simulations (the linear relation between the
period of the radial oscillations and largest radius of the annular DW was
obtained in an analytical form). The evolution of the pulsons ends up with
their relaxation into the stationary 2D modes (vortices, for $s\neq 0$, or
simply the flat asymmetric state in the case of $s=0$).

This work may be extended by a more systematic investigation of the 2D
system, without assuming the axial symmetry of the patterns. In particular,
oscillations of eccentricity in elliptically deformed pulsons may be
interesting to study, cf. Ref. \cite{eccentric}. In 1D, it may be also
interesting to study in detail the behavior of DWs against the backdrop of
periodic potentials (optical lattices), as well as in nonstationary systems
with time-dependent parameters.

\section*{Acknowledgment}

The work of J.Z. was supported, in a part, by a postdoctoral fellowship
provided by the Tel Aviv University, and by grant No. 149/2006 from the
German-Israel Foundation.

\end{document}